\documentclass{aa}

\usepackage{graphicx}
\usepackage{txfonts}
\usepackage{upgreek}    
\begin{document}

   \title{Asteroseismology of $Kepler$ Algol type oscillating eclipsing binaries}

   \author{A. Liakos
           }
   \institute{Institute for Astronomy, Astrophysics, Space Applications and Remote Sensing, National Observatory of Athens,\\
            Metaxa \& Vas. Pavlou St., GR-15236, Penteli, Athens, Greece\\\\
              \email{alliakos@noa.gr}
             }

   \date{Received September XX, 2017; accepted March XX, 2017}


\abstract
   {This research paper contains light curve modelling, spectroscopy and detailed asteroseismic studies for four out of five in total semidetached eclipsing binaries with a $\delta$~Scuti component, that have been detected to date through $Kepler$ mission, namely KIC~06669809, KIC~10581918, KIC~10619109 and KIC~11175495.}
   {The goal is to study the pulsational characteristics of the oscillating stars of the systems as well as to estimate their absolute parameters and enrich the so far poor sample of this kind of systems.}
   {Ground based spectroscopic observations provide the means to estimate the spectral types of the primary components and to model the light curves with higher certainty. The photometric data are analysed using eclipsing binary modeling techniques, while Fourier analysis is applied on their residuals in order to reveal the pulsation frequency modes.}
   {The results of analyses show that the primaries are pulsating stars of $\delta$~Scuti type and that all systems belong to the group of Algol type binaries with an oscillating star, namely oEA stars. The primaries of KIC~06669809, KIC~10581918, and KIC~10619109 pulsate in three, two, and five frequencies, respectively and in more than 200 other detected as combinations. The $\delta$~Scuti star of KIC~11175495 is the youngest and the fastest pulsator in binary systems that has ever been found and it oscillates in three main non-radial frequencies, while other 153 are also found as depended ones. Moreover, a comparison of their properties with other systems of the same type as well as with theoretical models of pulsating stars are also presented and discussed.}
   {}

   \keywords{stars:binaries:eclipsing -- stars:fundamental parameters -- (Stars:) binaries (including multiple): close -- Stars: oscillations (including pulsations) -- Stars: variables: delta Scuti -- Stars: individual: KIC~06669809, KIC~10581918, KIC~10619109, KIC~11175495}

   \maketitle
%

\section{Introduction}
\label{sec:intro}

In general, $\delta$~Scuti stars are multiperiodic pulsating variables oscillating in radial and non-radial modes. They are of A-F spectral types and III-V luminosity classes, their mass values range between 1.4-3~M$_{\sun}$, and they are located in the classical instability strip. The radial and the low-order non-radial pulsations are excited due to $\kappa$-mechanism \citep{BAL15}, while recently it was proposed that the turbulent pressure in the Hydrogen convective zone might be the explanation for the high-order non-radial modes \citep{ANT14, GRA15}.

Eclipsing binary systems (hereafter EBs) can be considered as the most essential tools for absolute parameters and evolutionary status estimation of stellar components. The combination of light curves (hereafter LCs) and radial velocities curves (hereafter RVs) provide the means for a very accurate calculation of stellar masses, radii, luminosities etc. The geometrical status of an EB (i.e. Roche geometry) and all its geometrical parameters (e.g. orbital period, inclination) are information than can be extracted directly through the LCs analysis. Moreover, using the `Eclipse Timing Variations' (ETV) method is possible to detect mechanisms (e.g. mass transfer) occurring in the EB. In general, RVs for binaries with orbital periods larger than 1-2 days cannot be easily obtained, mostly because they require large telescopes and different dates of observations, and in many cases, especially for Algol type EBs, the brighter member dominates the spectrum and makes impossible to detect any spectral lines from its fainter companion. In addition, it is also difficult to obtain data for ETV analysis for recently discovered cases, because typical period changes require decades of observations for photometric minima. However, the LCs are generally more easy to be obtained and by using fair assumptions based on stellar evolution theory, it is feasible to derive relatively accurate results and plausible conclusions.

The study of $\delta$~Scuti stars in EBs can be considered as quite important not only because two different stand-alone astrophysical issues (i.e. pulsations and binarity) are combined, but also because by using the binarity the absolute parameters of the pulsator can be derived. The first who proposed this group as a subclass of EBs were \citet{MKR02}. In that paper, a new category of EBs, namely the `$oEA~stars$' (oscillating eclipsing binaries of Algol type), was introduced as the Algol-type mass accreting pulsators of (B)A-F type. Later on, \citet{SOY06a} announced the first connection between orbital ($P_{\rm orb}$) and dominant pulsation ($P_{\rm pul}$) periods, while \citet{SOY06b} published lists with candidate EBs with a $\delta$~Scuti member. \citet{LIA12} after a long-term observational campaign published a catalogue with 74 cases and updated correlations between fundamental parameters. \cite{ZHA13} made the first theoretical attempt for the $P_{\rm pul}-P_{\rm orb}$ correlation. \citet{LIAN15, LIAN16} noticed for the first time that there is a boundary between $P_{\rm pul}-P_{\rm orb}$ beyond that these two quantities do not correlate. Finally, \citet{LIAN17} published the most complete catalogue\footnote{\url{http://alexiosliakos.weebly.com/catalogue.html}} (204 $\delta$~Scuti stars in 199 binary systems) and correlations between fundamental parameters up to date.

The $Kepler$ mission provides the means for high precision asteroseismic models of $\delta$~Scuti stars because of (a) the unprecedented photometric precision of the data (order of tens of mmag), that makes frequencies with amplitudes of tens of $\upmu$mag feasible to be detected \citep[e.g.][]{MUR13}, (b) the continuous monitoring of the targets, that eliminates the effect of alias in frequencies detections \citep{BRE00}, and (c) the relatively high time resolution of the short-cadence data ($\sim1$~min). It should to be noted that the long-cadence data (i.e. time resolution $\sim30$~min) can be also considered extremely valuable for asteroseismology in cases of other pulsating stars such as Cepheids, $\gamma$~Doradus etc., which show pulsation periods of the order of several hours to several days. In addition, long-cadence data for targets observed during many quarters of the mission allow the study of short term pulsation period changes (e.g. order of a few years). Another very important point is that all the archives of the mission are publicly available and the data can be easily obtained and analyzed. Particularly for the EBs, an excellent online database, namely `$Kepler$ Eclipsing Binary Catalog'\footnote{\url{http://keplerebs.villanova.edu/}} \citep[$KEBC$,][]{PRS11}, has been created and provides all the preliminary information (e.g. ephemerides, magnitudes, temperatures) needed for a further and more detailed analysis.

According to the catalogue of \citet{LIAN17} only 17 cases of $Kepler$ EBs with a $\delta$~Scuti component have been published so far. In particular, KIC~06220497 \citep{LEE16} and the four presented herein are oEA stars, nine of these are detached systems, while another three are still unclassified regarding their Roche geometry \citep[see][and references therein]{LIAN17}.

For the present study, KIC~06669809, KIC~10581918, KIC~10619109, and KIC~11175495 have been selected for detailed analyses for first time. Ground based spectroscopic observations allowed the estimation of the spectral types of the systems' primary components and provided the means for a more accurate LCs modelling that led us to estimate the absolute parameters of their components and study in deep the pulsation properties of their $\delta$~Scuti stars. The fifth oEA $Kepler$ system, namely KIC~06220497 \citep{LEE16}, is not part of the present sample because it was not included in our spectroscopic campaign. The present sample consists the $\sim$6\% of the total known (66) oEA systems with $\delta$~Scuti component and the $\sim$7\% of those (56 in total) whose absolute parameters are known. Finally, the present research contributes in the least studied region of the pulsating stars in binaries, namely the $Kepler$ oEA stars.

Preliminary results regarding the dominant pulsation frequency and the Roche geometry of the systems have been published by \citet{LIAN16} for KIC~06669809 and by \citet{LIAN17} for the rest three EBs. In Table~\ref{tab:Obslog} are listed for all studied systems the respective observations logs.

For a more easy reading of the systems' identification names, KIC~06669809 is referred as KIC~066, KIC~10581918 as KIC~105, KIC~10619109 as KIC~106, and KIC~11175495 as KIC~111 in the whole text.

\begin{table}
\centering
\caption{Log of observations from the $Kepler$ mission for all studied systems. $Q$ denotes the quarters of $Kepler$ mission during which long-cadence ($L-C$) and short-cadence data ($S-C$) were obtained. $llc$ denotes the level of light contamination, while the total $points$, the number of consecutive $days$ of the S-C data, and the number of the fully covered $LCs$ are also given in the last three columns.}
\label{tab:Obslog}
\scalebox{0.9}{
\begin{tabular}{l cccccc}
\hline													
KIC &	           L-C Q	    &	S-C Q	&	        llc	     &	Points	&	Days	&	LCs	\\
    &                           &           &  ($\times 10^{-4}$)&  (S-C)     & (S-C)   &  (S-C)  \\
\hline													
066	&	1-4; 6-8; &	    4	   &5	 &	45449	&	31.06	&	42	\\
	&10-12; 14-16 &	    	  &                    &	     	&	     	&	 	\\
105	&	2-4; 6-8; &	    7	   &	       0	     &	42700	&	29.18	&	16	\\
	&10-12; 14-16 &	    	  &                    &	     	&	     	&	 	\\

106	&	0-6; 8-10;&	   3	    &	       0	     &	38004	&	25.97	&	12	\\
	&12-14; 16-17 &	    	  &                    &	     	&	     	&	 	\\
111	& 0-17	      &	   5	   &0.9&	49145	&	33.57	&	15	\\
\hline													
\end{tabular}}
\end{table}

\subsection{KIC 06669809}
\label{sec:introKIC066}
The system (cross ID: TYC~3127-1399-1) has a period of $\sim0.73$~d and a LC of $\beta$~Lyrae type. The first LCs in $V$ and $I$ filters were obtained by the $ASAS$ survey \citep{PIG09}. Its temperature is referred as 7239~K \citep{CHR12,PIN12}, 7150~K in the $HIPPARCOS$ Catalogue \citep{ESA97}, 7452~K by \citet{HUB14}, and 7810~K by \citet{AMM06}. \citet{CON14} presented the ETV diagram of the system, while \citet{BOR16} analyzed it using one periodic and one cubic terms. The periodic term was attributed to a possible Light-Time Travel effect due to the presence of a third companion with an orbital period of $\sim$194~d and a mass function (i.e. eq.~3) of $f(M)\sim0.0062$~M$_{\sun}$.

\subsection{KIC 10581918}
\label{sec:introKIC105}
This binary system (cross IDs: 2MASS~J18521048+4748166; WX~Dra) was discovered by \citet{TSE60} and has an orbital period of $\sim1.8$~d. Except for many timings of minima neither LCs modeling nor spectroscopic studies have been published so far. The only available LCs come from the $ASAS$ survey \citep{PIG09} in $V$ and $I$ filters and from the $Trans-Atlantic~Exoplanet~Survey$ \citep[][T-Lyr1-05887]{DEV08} in $R$ band. \citet{CHR12} referred the temperature of the system as 7252~K, \citet{ARM14} as 7371~K, \citet{HUB14} as 7475~K, but the $LAMOST$ spectroscopic survey \citep{FRA16} classified it as of A7V spectral type (i.e. $\sim 7800$~K). \citet{PET12} based on the long and short cadence data of $Kepler$ concluded that there is a migrating hot spot on the primary's surface and that this component pulsates with a dominant frequency of $\sim 35.8$~cycle~d$^{-1}$. \citet{WOL15} and \citet{ZAS15} analyzed the ETV diagram of the system and they found a 14~yr periodic modulation of the orbital period which they attributed it to the possible existence of a third body in an eccentric orbit with a mass function of $f(M)\sim0.00027$~M$_{\sun}$. In the latter paper, the model of the $Kepler$ LC of the system was presented assuming a mass ratio of 1, but no third light contribution was found.

\subsection{KIC 10619109}
\label{sec:introKIC106}
This system (cross ID: TYC~3562-985-1) was identified as an EB with an orbital period of $\sim 2$~days by the $Trans-Atlantic~Exoplanet~Survey$ \citep[][T-Cyg1-01956]{DEV08}, from which the first LC in $R$ pass band was obtained. The temperature estimation of the system ranges from 7028~K \citep{CHR12} and 7128~K \citep{HUB14} up to 7441~K \citep{ARM14}. \citet{GAO16} announced that among 1049 binaries observed by $Kepler$ mission, they discovered flare activity in 234 of them including also KIC~106. Based on $\sim 3$~yr data, they found 14 flares with a frequency of 0.013~d$^{-1}$ and an amplitude of 45~mmag. \citet{CON14} and \citet{GIE15} performed an ETV analysis but they found very weak evidence for a potential third star orbiting the system.

\subsection{KIC 11175495}
\label{sec:introKIC111}
KIC~111 (cross ID: 2MASS~J18501133+4851229) has a period of $\sim2.19$~d and was discovered as a high-frequency A-type pulsator ($f_{\rm dom}\sim64.44$~cycle~d$^{-1}$) by \citet{HOL14} using data from the $SWASP$ survey. Its temperature is given as 8070~K by \citet{SLA11} and \citet{CHR12}, and as 8300~K by \citet{HUB14}.

\section{Ground based Spectroscopy}
\label{sec:sp}

Spectroscopic observations were obtained with the 2.3~m Ritchey-Cretien `Aristarchos' telescope at Helmos Observatory in Greece on summer 2016 using the \emph{Aristarchos~Transient~Spectrometer}\footnote{\url{http://helmos.astro.noa.gr/ats.html}} (ATS) instrument \citep{BOU04}. ATS is a low/medium dispersion fibre spectrometer that consists of 50 fibres (50~$\upmu$m diameter each) providing a field-of-view of $\sim10$~arcsec on the sky. The U47-MB Apogee CCD camera (Back illuminated, 1024$\times$1024~pixels, 13~$\upmu$m$^{2}$~pixel size) and the low resolution grating (600~lines~mm$^{-1}$) were used for the observations. This setup provided a resolution of $\sim3.2$~{\AA}~pixel$^{-1}$ and a spectral coverage between app. 4000~{\AA}-7260~{\AA}, where the first four Balmer (i.e. H$_{\upalpha}$-H$_{\updelta}$) and many metallic lines (e.g. Mg$_{\rm I}$-triplet, Na$_{\rm I}$-doublet) are present.

Approximately 45 spectroscopic standard stars, suggested by GEMINI Observatory\footnote{\url{http://www.gemini.edu/}}, ranging from B0V to K8V spectral types and the KIC targets were observed with the same instrumental set-up. The objective of these observations was to derive the spectral types of the primary components of the KIC systems. The observations of standard stars took place in 3 nights in July, 5 nights in August and 2 nights in October 2016. In Table~\ref{tab:Standards} are listed the spectroscopic standards stars (the names are taken from the $HIPPARCOS$ catalogue--Hip No) used for the comparisons in the A-F spectral type region. The spectroscopic observations log for the KIC systems is given in Table~\ref{tab:SpecResults}. The technical calibration of all spectra (bias, dark, flat-field corrections) was made using the \textsc{MaxIm DL} software, while the data reduction (wavelength calibration, cosmic rays removal, spectra normalization, sky background removal) was performed using the \textsc{RaVeRe} v.2.2c software \citep{NEL09}.

\begin{figure}
\includegraphics[width=\columnwidth]{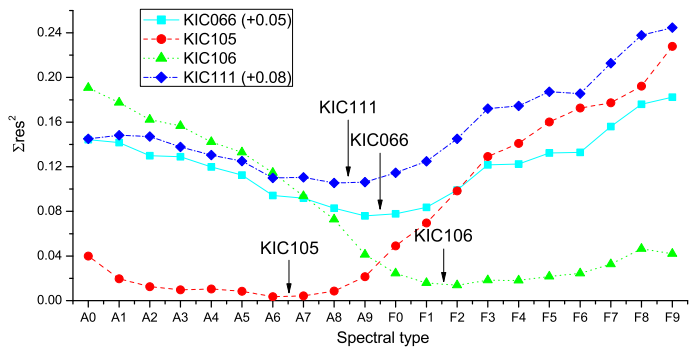}
\caption{Spectral type-search plots for all studied KIC systems. The points of KIC~066 and KIC~111 are shifted vertically for better viewing. The arrows show the adopted spectral type for the primary component of each system.}
\label{fig:STs}
\end{figure}

\begin{table}
\centering
\caption{Spectroscopic observations log of the standard stars used for the comparisons in the A-F spectral types (ST) region.}
\label{tab:Standards}
\scalebox{0.85}{
\begin{tabular}{lc lc lc lc}
\hline															
HIP No	&	ST	&	HIP No	&	ST	&	HIP No	&	ST	&	HIP No	&	ST	\\
\hline															
12640	&	A0V	&	4436	&	A5V	&	194	    &	F0V	&	5833	&	F5V	\\
111169	&	A1V	&	111259	&	A6V	&	19205	&	F1V	&	116505	&	F6V	\\
111242	&	A2V	&	12332	&	A7V	&	13904	&	F2V	&	6711	&	F7V	\\
92405	&	A3V	&	102253	&	A8V	&	90345	&	F3V	&	109572	&	F8V	\\
94620	&	A4V	&	96286	&	A9V	&	10830	&	F4V	&	102611	&	F9V	\\
\hline															
\end{tabular}}
\end{table}

\begin{table}
\centering
\caption{Spectroscopic observations log and results for the primary components of the studied systems. The columns include: the name of the system (KIC), the dates of observations (in DD/MM/YY format), the exposure time used (E.T.), and the orbital phase of the system when the spectrum was obtained ($\Phi_{\rm orb}$). The errors are given in parentheses alongside values and correspond to the last digit(s).}
\label{tab:SpecResults}
\begin{tabular}{l c cc cc}
\hline											
KIC	&	Date	&	E.T.	&	$\Phi_{\rm orb}$	&	Spectral type	&	$T_{\rm eff}$ 	\\
	&		&	(min)	&		&		&	(K)	\\
\hline											
066	&	15/8/16	&	10	&	0.35	&	A9V-F0V	&	7400(100)	\\
105	&	6/10/16	&	15	&	0.49	&	A6V-A7V	&	7900(100)	\\
106	&	5/10/16	&	20	&	0.13	&	F1V-F2V	&	7070(75)	\\
111	&	16/8/16	&	20	&	0.66	&	A8V-A9V	&	7550(100)	\\
\hline											
\end{tabular}
\end{table}

\begin{figure*}
\includegraphics[width=18cm]{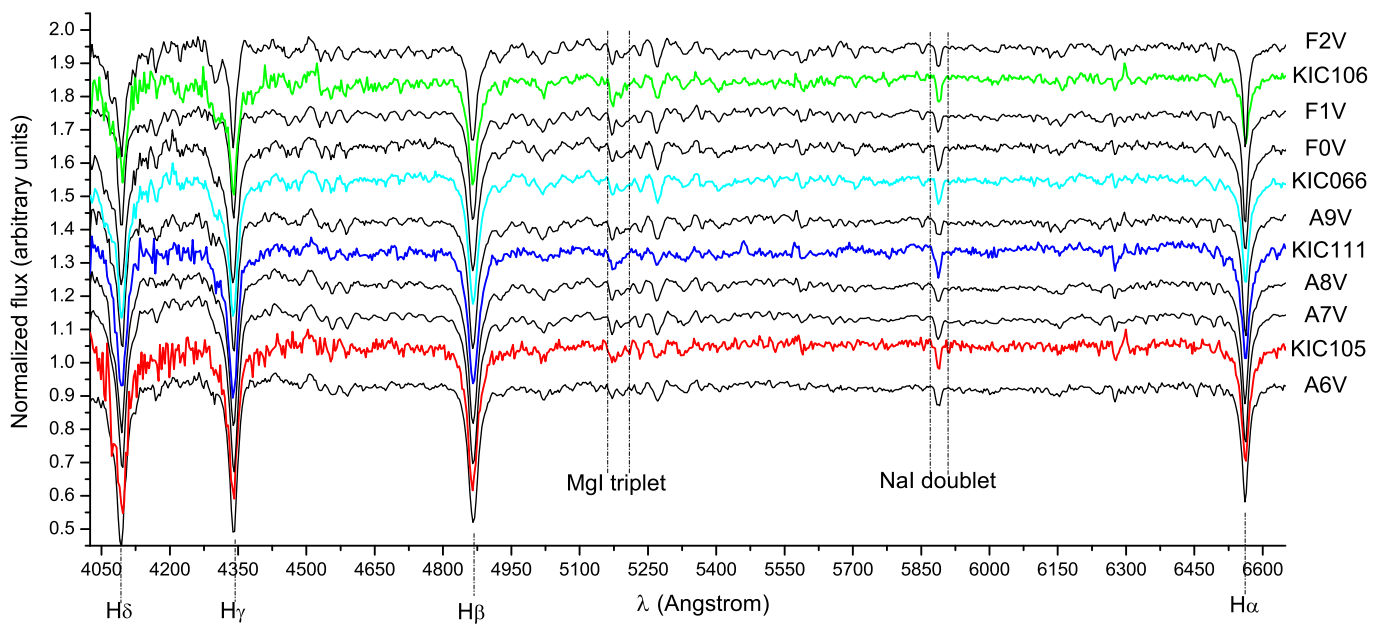}
\caption{Comparison spectra of the KIC targets and standard stars with the closest spectral types. The Balmer and some strong metallic lines are also indicated.}
\label{fig:spectra}
\end{figure*}

All spectra were calibrated and normalized to enable direct comparisons. The spectra were then shifted, using the Balmer lines as reference, to compensate for the relative Doppler shifts of each standard and the variables. The spectral region between 4000~{\AA} and 6800~{\AA}, where the Balmer and numerous metallic lines are strong, was used for the spectral classification. The remaining spectral regions were ignored because they generally lacked sufficient metallic lines with significant signal-to-noise ratios (S/N). The depths of each Balmer and other metallic lines, that are sensitive to the temperature, were calculated in all spectra of the standard stars and the variables. The differences of spectral line depths between each standard star and the variables derive sums of squared residuals in each case. These least squares sums guided us to the closest match between the spectra of variables and standards. In Fig.~\ref{fig:STs} is shown the $\sum \rm res^{2}$ against spectral type plot for each variable. The comparison is shown only between A-F spectral types due to scaling reasons. In the case of KIC~066 the best match was found with the spectrum of an A9V standard star and as shown in Fig.~\ref{fig:STs}, the F0V spectral type provides better match than that of an A8V. So, in conclusion, KIC~066 is between A9V-F0V spectral types with a formal error of half subclass. Following the same method, it is found that KIC~105 lays between A6V-A7V, KIC~106 between F1V-F2V, and KIC~111 between A8V-A9V spectral types with the same formal error. In Fig.~\ref{fig:spectra} are plotted the spectra of all systems along with those of best-fit standard stars. The last two columns of Table~\ref{tab:SpecResults} host the spectral classification for each system using the aforementioned spectral comparisons and the assigned temperature values ($T_{\rm eff}$ ) according to the relations between effective temperatures-spectral types of \citet{COX00}.

It should to be noted that the spectra of all systems, except for KIC~105, were obtained during out-of-eclipses phases, thus, typically, they correspond to the combined spectra of the components. However, in Section~\ref{sec:LCmdl} is shown that the secondary components of the systems have a very small contribution to the total light and, in addition, the temperature differences between them and their companions are relatively large. Therefore, since the secondaries were found to be cool stars, it can be plausibly assumed that the spectrum of each system practically reflects the spectral type of its primary. For KIC~066 and KIC~106 the present spectral classifications are in relatively good agreement with the results of previous photometry-based methods (see Sections~\ref{sec:introKIC066} and \ref{sec:introKIC106}). The primary of KIC~105 is found hotter than the photometry-based methods estimated, but the present result is in excellent agreement with other spectroscopic ones (see Section~\ref{sec:introKIC105}). Finally, for KIC~111 the present spectroscopic results reveal a cooler spectral type in comparison with those suggested by other photometry-based methods (see Section~\ref{sec:introKIC111}).

\section{Light curve modelling and absolute parameters calculation}
\label{sec:LCmdl}
\begin{figure*}
\begin{tabular}{c}
\includegraphics[width=17cm]{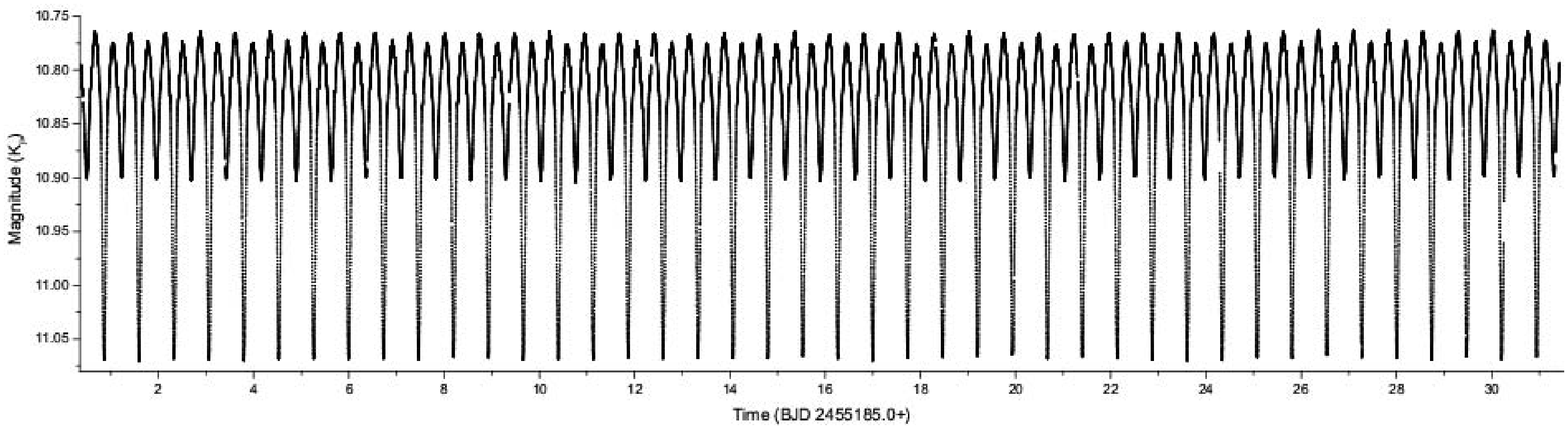}\\
\includegraphics[width=17cm]{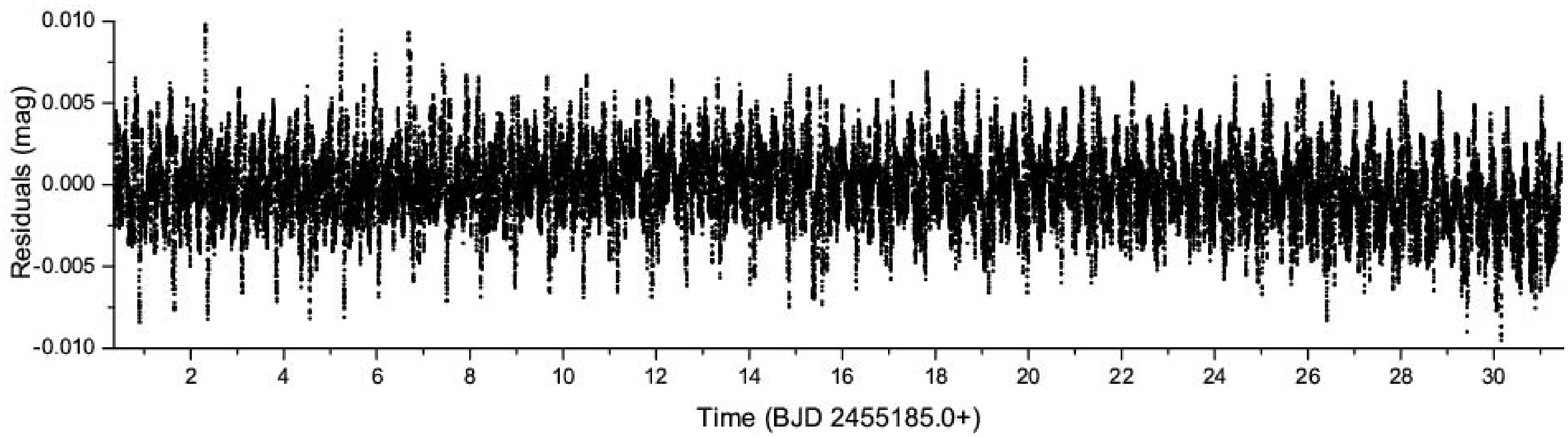}
\end{tabular}
\caption{Upper panel: Short cadence LCs for KIC~06669809. Lower panel: LCs residuals after the subtraction of the binary models.}
\label{fig:KIC066LCs_Res}
\end{figure*}

\begin{figure*}
\begin{tabular}{c}
\includegraphics[width=17cm]{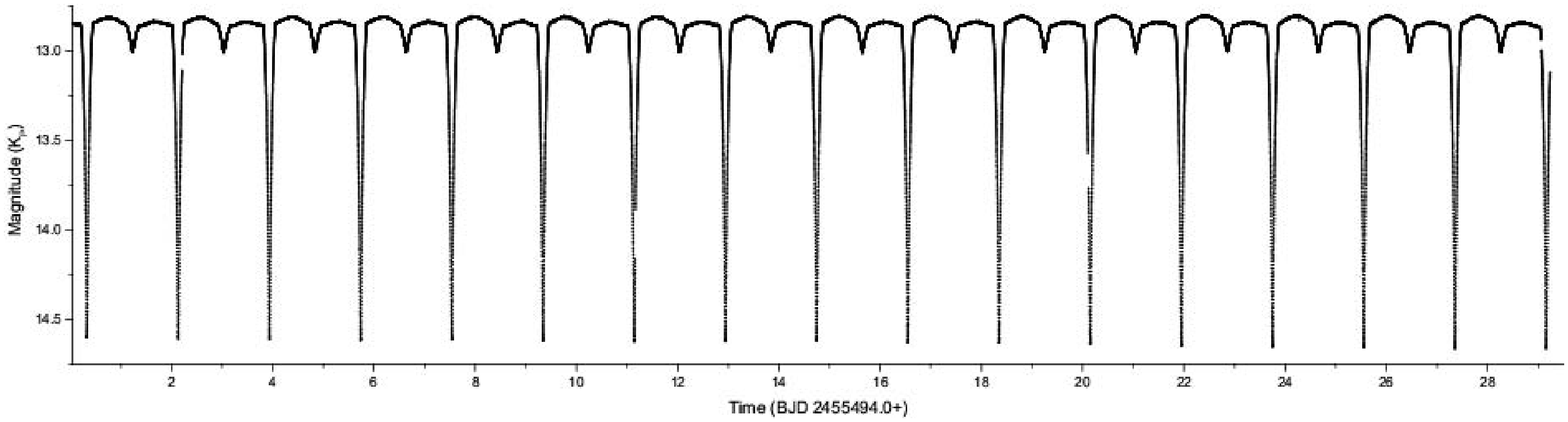}\\
\includegraphics[width=17cm]{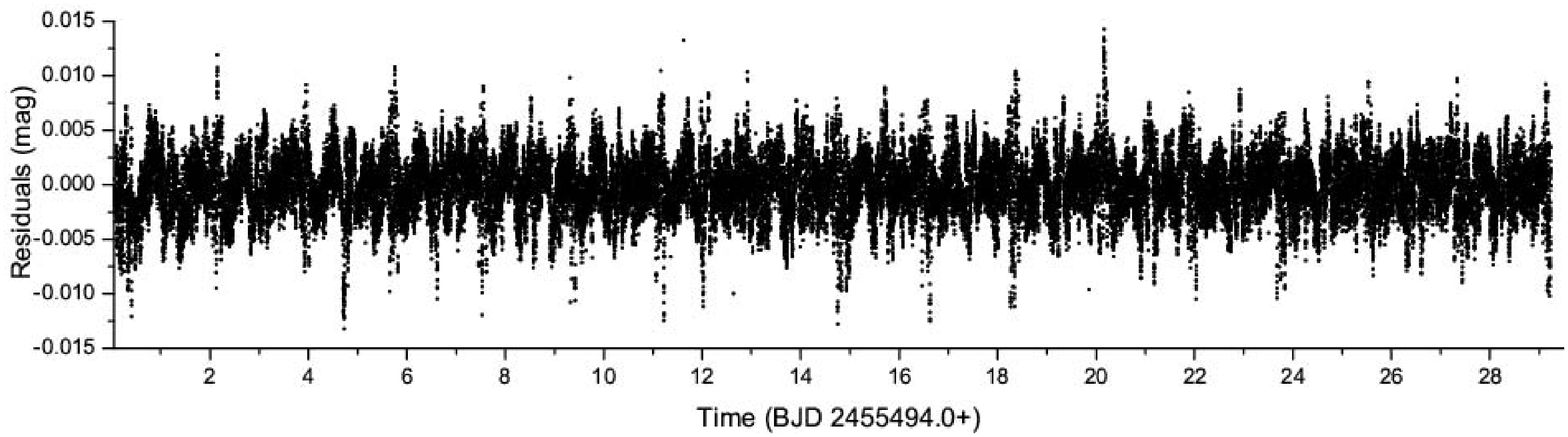}
\end{tabular}
\caption{The same as Fig.~\ref{fig:KIC066LCs_Res}, but for KIC~10581918.}
\label{fig:KIC105LCs_Res}
\end{figure*}

\begin{figure*}
\begin{tabular}{c}
\includegraphics[width=17cm]{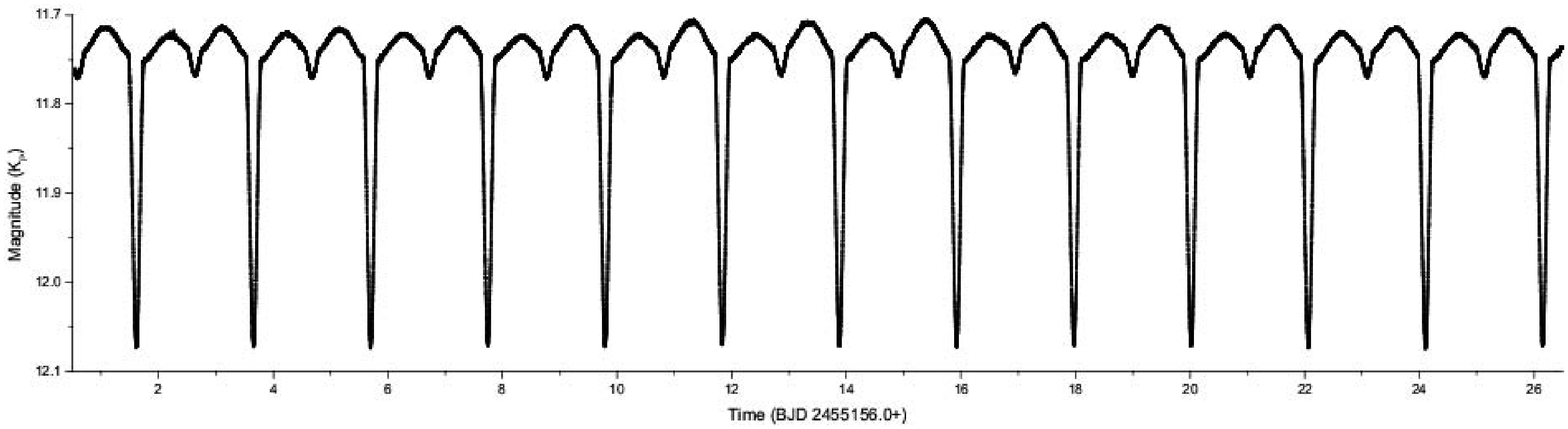}\\
\includegraphics[width=17cm]{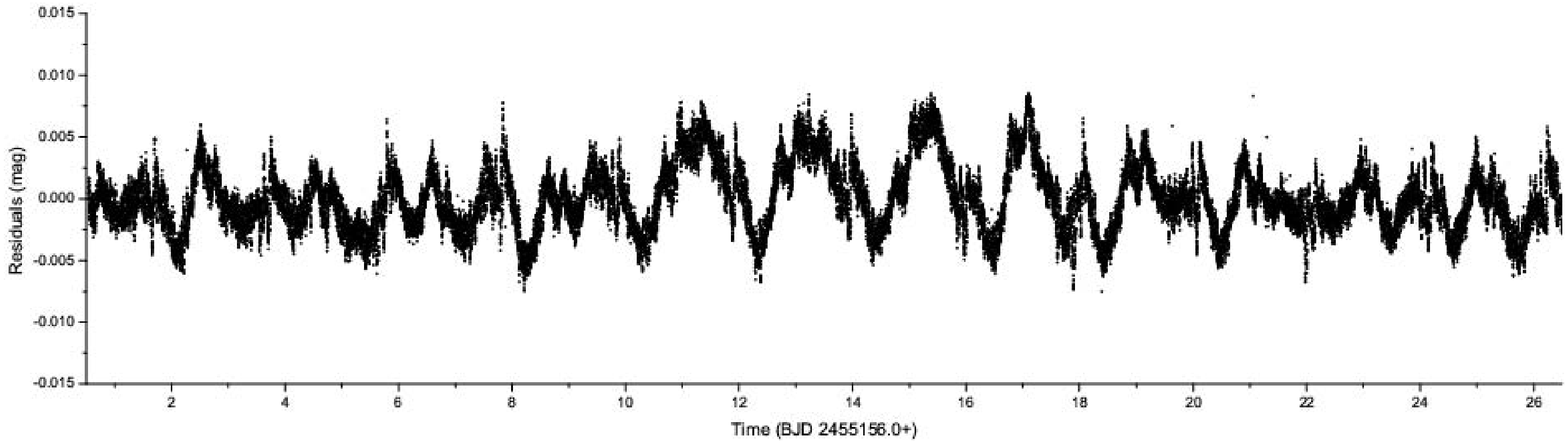}
\end{tabular}
\caption{The same as Fig.~\ref{fig:KIC066LCs_Res}, but for KIC~10619109.}
\label{fig:KIC106LCs_Res}
\end{figure*}

\begin{figure*}
\begin{tabular}{c}
\includegraphics[width=17cm]{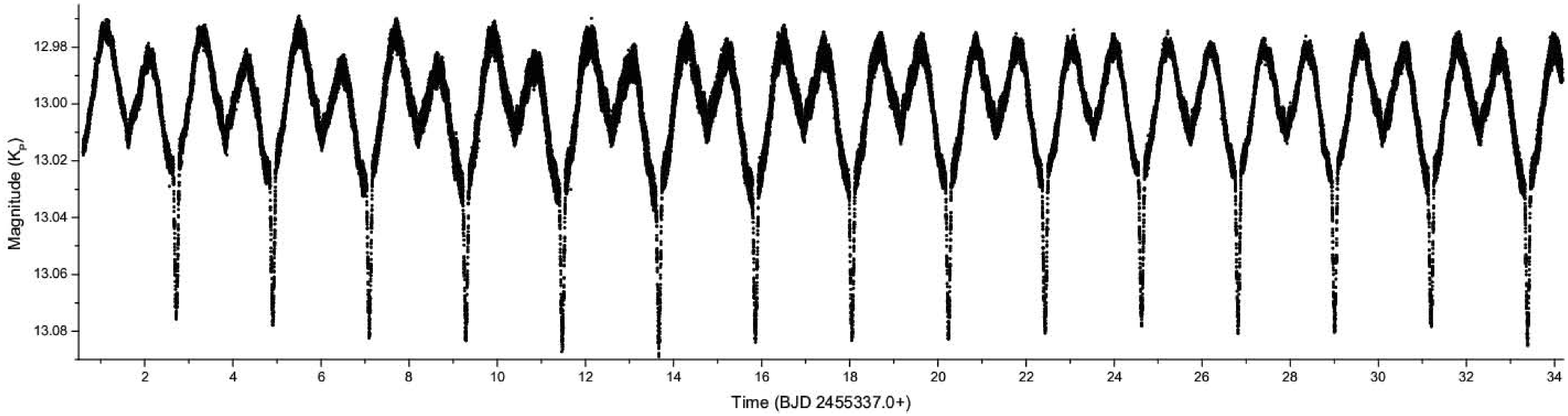}\\
\includegraphics[width=17cm]{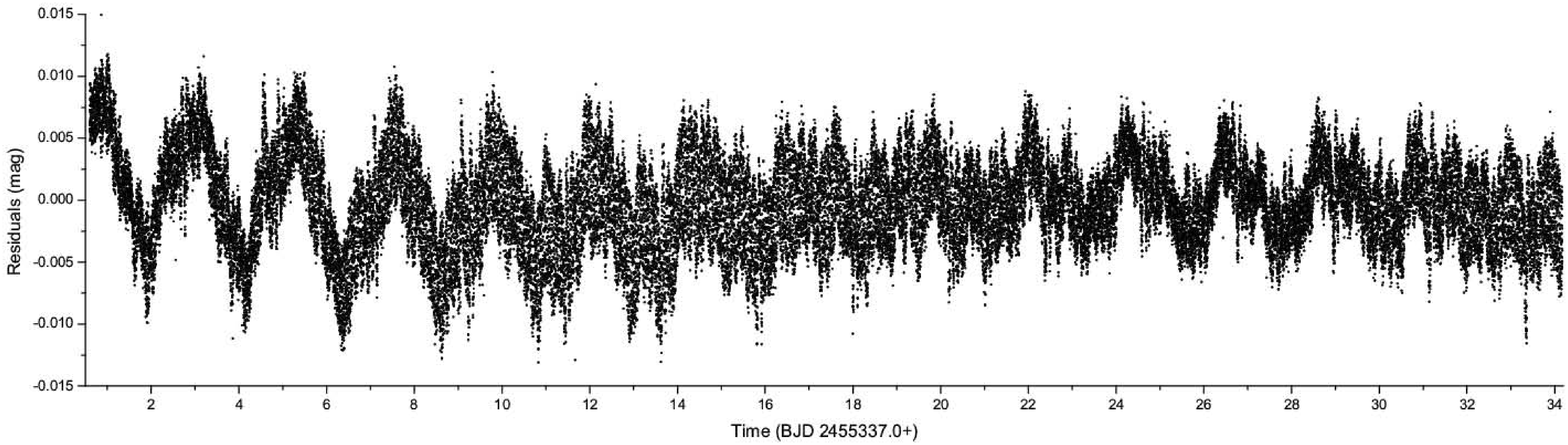}
\end{tabular}
\caption{The same as Fig.~\ref{fig:KIC066LCs_Res}, but for KIC~11175495.}
\label{fig:KIC111LCs_Res}
\end{figure*}

All systems were observed during many quarters of the $Kepler$ mission (see Table~\ref{tab:Obslog}). However, short-cadence data (i.e. time resolution $\sim1$~min) were obtained only during one quarter. Since the objective of this research is the study of the asteroseismic behaviour of the systems' pulsating members, the time resolution as well as the continuous recording are matters of high importance. Therefore, for the following analyses only the short-cadence data, taken from the online version of the $KEBC$ \citep{PRS11}, were used. The total number of data points for each system corresponding to a significant number of fully covered LCs (see Table~\ref{tab:Obslog}) were available at our disposal and they are plotted in the upper panels of Figs~\ref{fig:KIC066LCs_Res}-\ref{fig:KIC111LCs_Res}. It should to be noted that for all systems the light contamination (see Table~\ref{tab:Obslog}) is negligible.

The first step of the analysis was to create a mean LC for each system in order to: (a) eliminate brightness changes due to the pulsations and cool spots (O' Connell effect) and (b) use it for faster calculations for the estimation of mass ratio ($q$) of the EB. For this, the phase diagrams of the LCs folded into the orbital period were used. The ephemerides ($T_{0},~P_{\rm orb}$) for the conversion of the time-dependent LCs into phase folded LCs were taken from the $KEBC$ \citep{PRS11} and are given in Table~\ref{tab:LCmdlAbs}. More specifically, by taking averages every $\sim 80$ data points, one single LC with $\sim 600$ mean points was derived for each system and used only for the `$q$-search' method as is described below.

\begin{figure}
\includegraphics[width=\columnwidth]{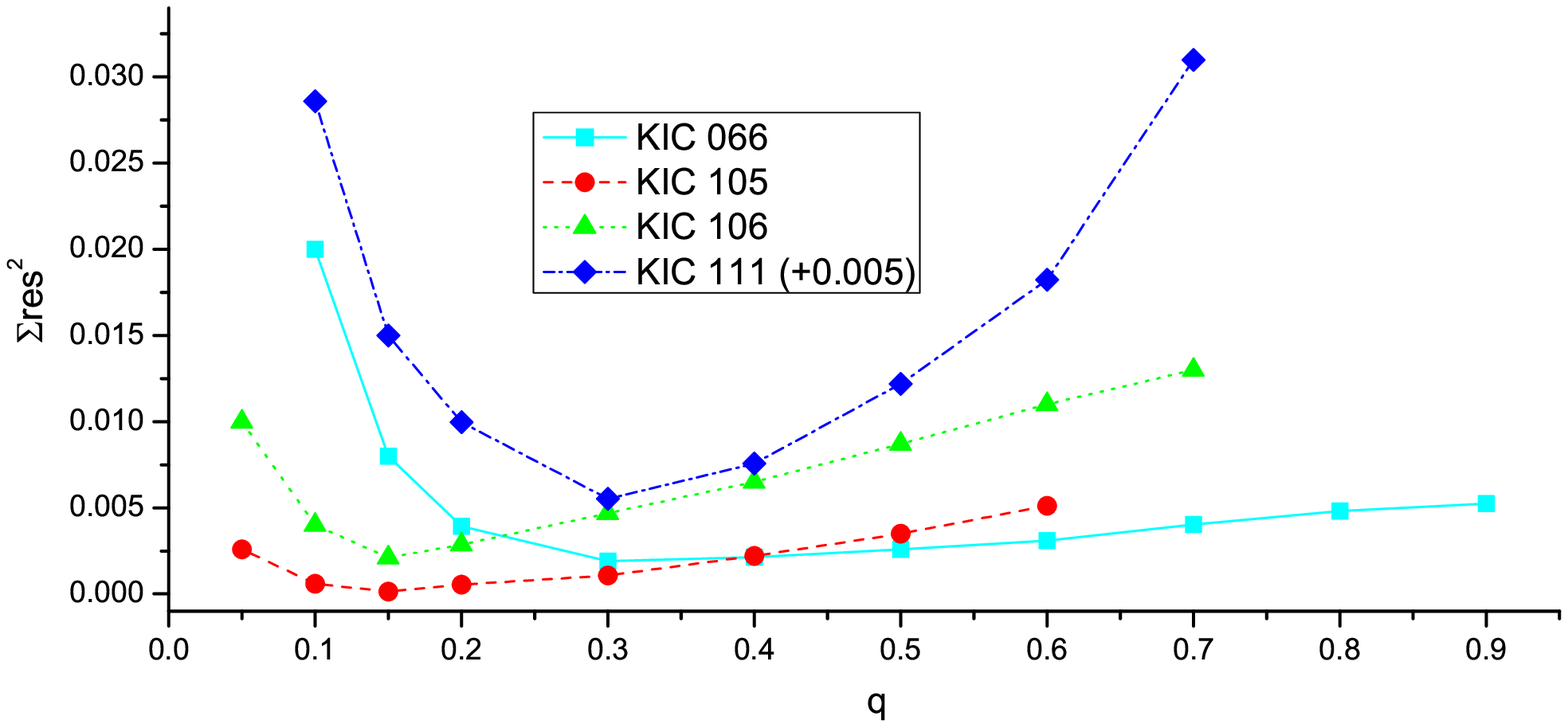}
\caption{$q$-search plots for all systems. The points of KIC~11175495 are shifted vertically for better viewing.}
\label{fig:qs}
\end{figure}

The LCs analysis was made using the \textsc{PHOEBE} v.0.29d software \citep{PRS05} that is based on the 2003 version of the Wilson-Devinney (W-D) code \citep{WIL71, WIL79, WIL90}. In the absence of spectroscopic mass ratios, the `$q$-search' method \citep[for details see][]{LIAN12} was applied in modes 2 (detached system), 4 (semi-detached system with the primary component filling its Roche lobe) and 5 (conventional semi-detached binary) to find feasible (`photometric') estimates of the mass ratio. The step of $q$ change during the search was 0.05-0.1 starting from $q=0.05-0.1$. The effective temperatures of the primaries ($T_{1}$) were given the values derived from the spectral classification (see Section~\ref{sec:sp}) and were kept fixed during the analysis, while the temperatures of the secondaries $T_{2}$ were adjusted. The Albedos, $A_1$ and $A_2$, and gravity darkening coefficients, $g_1$ and $g_2$, were set to generally adopted values for the given spectral types of the components \citep{RUC69, ZEI24, LUC67}. The (linear) limb darkening coefficients, $x_1$ and $x_2$, were taken from the tables of \citet{HAM93}. The dimensionless potentials $\Omega_{1}$ and $\Omega_{2}$, the fractional luminosity of the primary component $L_{1}$, and the system's orbital inclination $i$ were set as adjustable parameters. At this point it should to be noted that since the $Kepler$'s photometer has a spectral response range between approx. 410-910~nm with a peak at $\sim588$~nm, the $R$ filter (Bessell photometric system--range between 550-870~nm and with a transmittance peak at 597~nm) was selected as the best representative for the filter depended parameters (i.e. $x$ and $L$). Moreover, in all systems there is evidence of maxima brightness changes, so parameters of photospheric spots on the secondary's surface were also adjusted. The selection of the magnetically active component was based on the effective temperatures of the systems' members. In all cases the secondaries are clearly cooler than the primaries (i.e. large minima difference), therefore, they host a convective envelope that suits better to magnetically active star. In addition, the hotter ones are candidates for being pulsating stars of $\delta$~Sct type and it is rather rare to present also magnetic activity. For all EBs, except for KIC~111, the third light parameter ($l_{3}$) was also adjusted because the systems are candidate for triplicity (see Sections~\ref{sec:introKIC066}-\ref{sec:introKIC106}). However, during the iterations it resulted in unrealistic values, therefore, it was omitted in the final analysis. Finally, all systems were found to have the minimum $\sum \rm res^{2}$ in mode 5. KIC~066 and KIC~111 have a minimum at $q=0.3$, while KIC~105 and KIC~106 at $q=0.15$. In Fig.~\ref{fig:qs} are shown the respective `$q$-search' plots.

The analyses of the systems confronted three main problems: (a) the spot migration over time, which did not allow to create one simple LC model using all the available points, (b) the widely known `unrealistic' error estimation of \textsc{PHOEBE}, and (c) the LC asymmetries due to pulsations. At this point it should to be noted that the objectives of the LC modelling are the estimation of the systems' parameters (geometrical and absolute) and also the elimination of the binarity influence to the data points in order to be used later for a frequency analysis. A plausible solution to the first two problems was to solve each LC separately, while for the third one, to use again mean points but this time for each LC. Thus, for all systems $\sim 500$ normal points per LC were used. For each LC model as input parameters were used those derived from the `$q$-search' method and following exactly the same assumptions as before and with the addition of $q$ as adjustable parameter, the iterations began. The iterations finished once the parameters could not converge to a solution with less $\sum \rm res^{2}$. This method provides one model per LC i.e. 42, 16, 12, and 15 models for KIC~066, KIC~105, KIC~106, and KIC~111, respectively.

\begin{figure}
\begin{tabular}{c}
\includegraphics[width=\columnwidth]{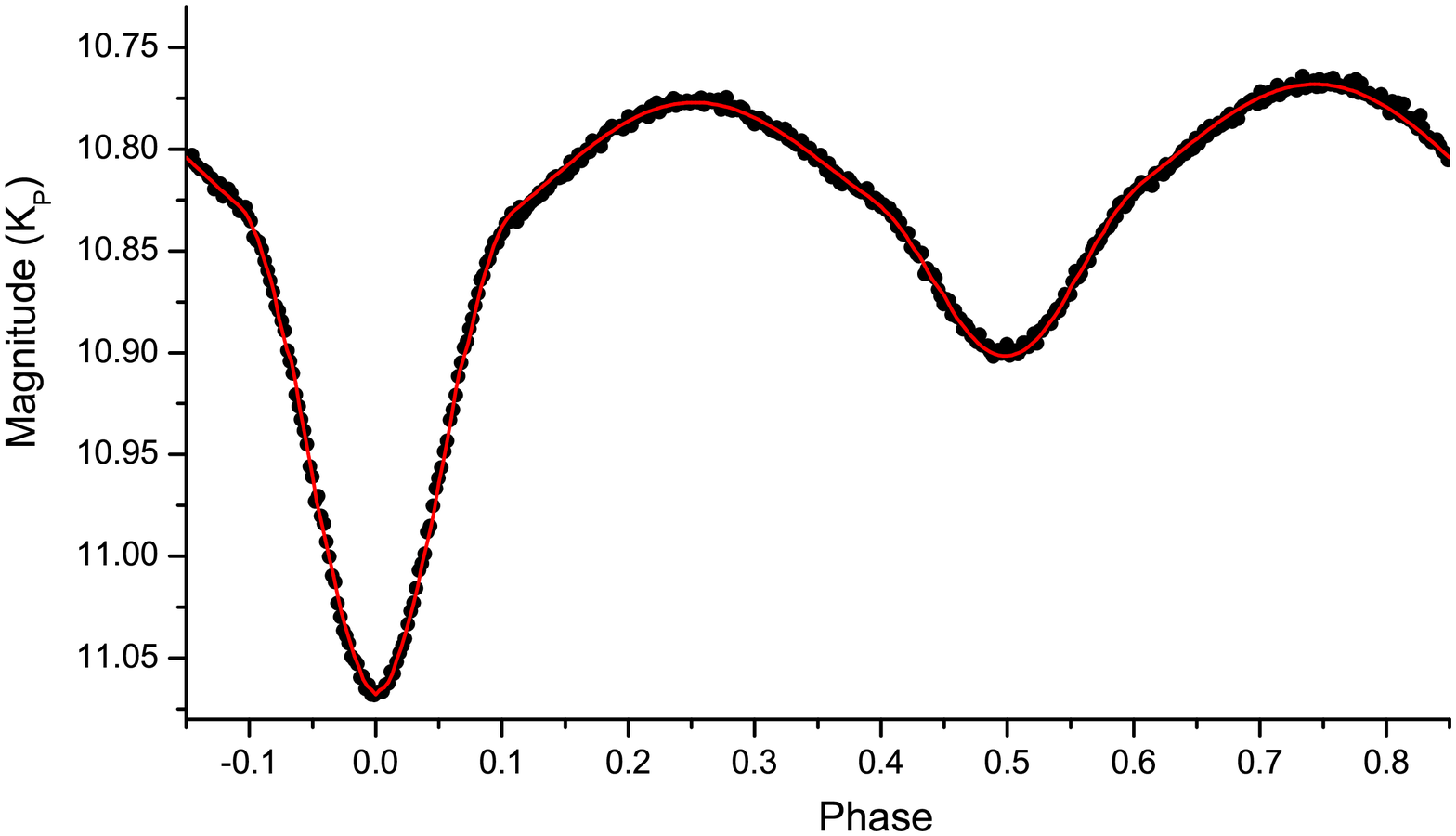}\\
\includegraphics[width=6cm]{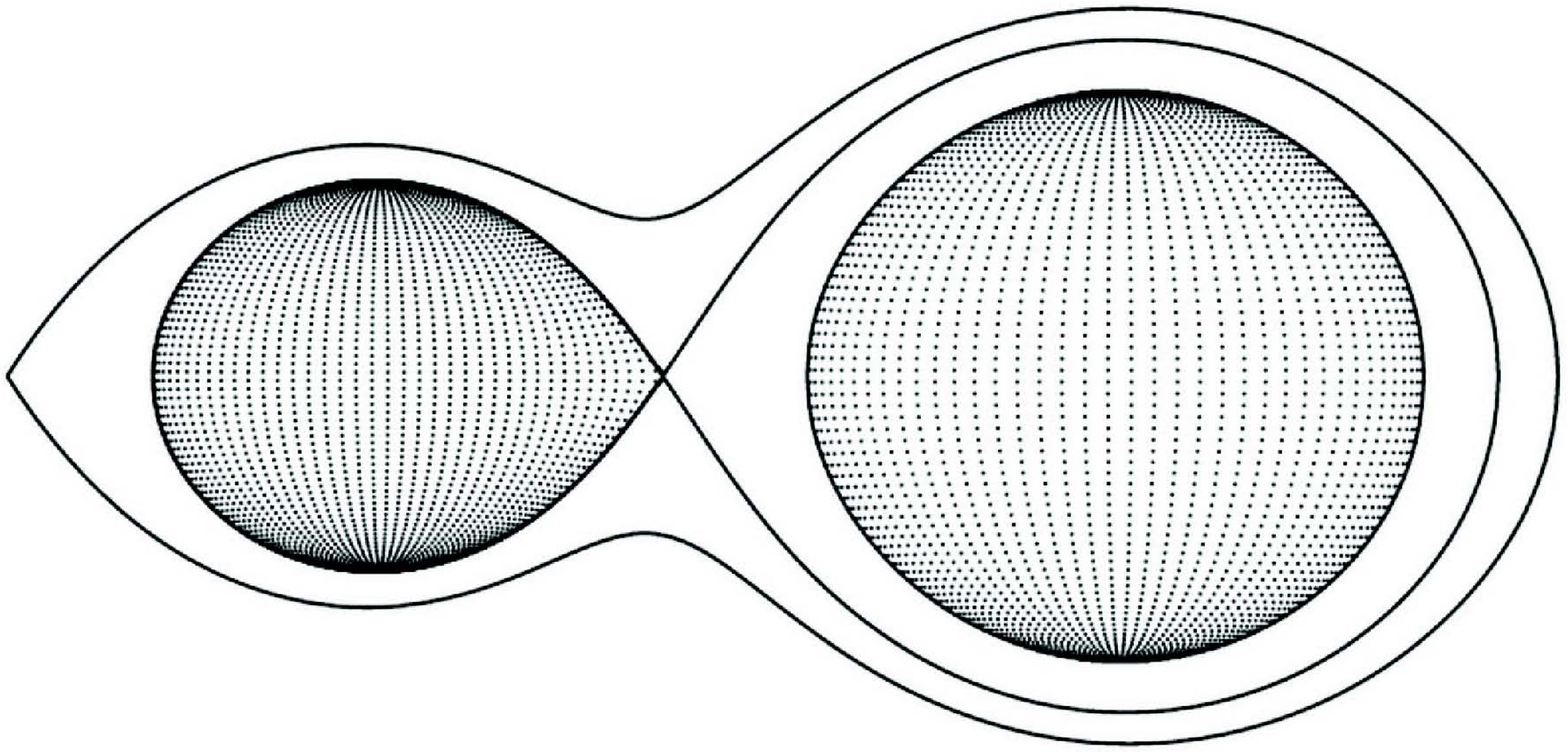}
\end{tabular}
\caption{Upper panel: Theoretical (solid line) over observed (points) LC for KIC~06669809. Lower panel: Three-dimensional representation of the Roche geometry of the system at orbital phase 0.75.}
\label{fig:KIC066LCm3D}
\begin{tabular}{c}
\includegraphics[width=\columnwidth]{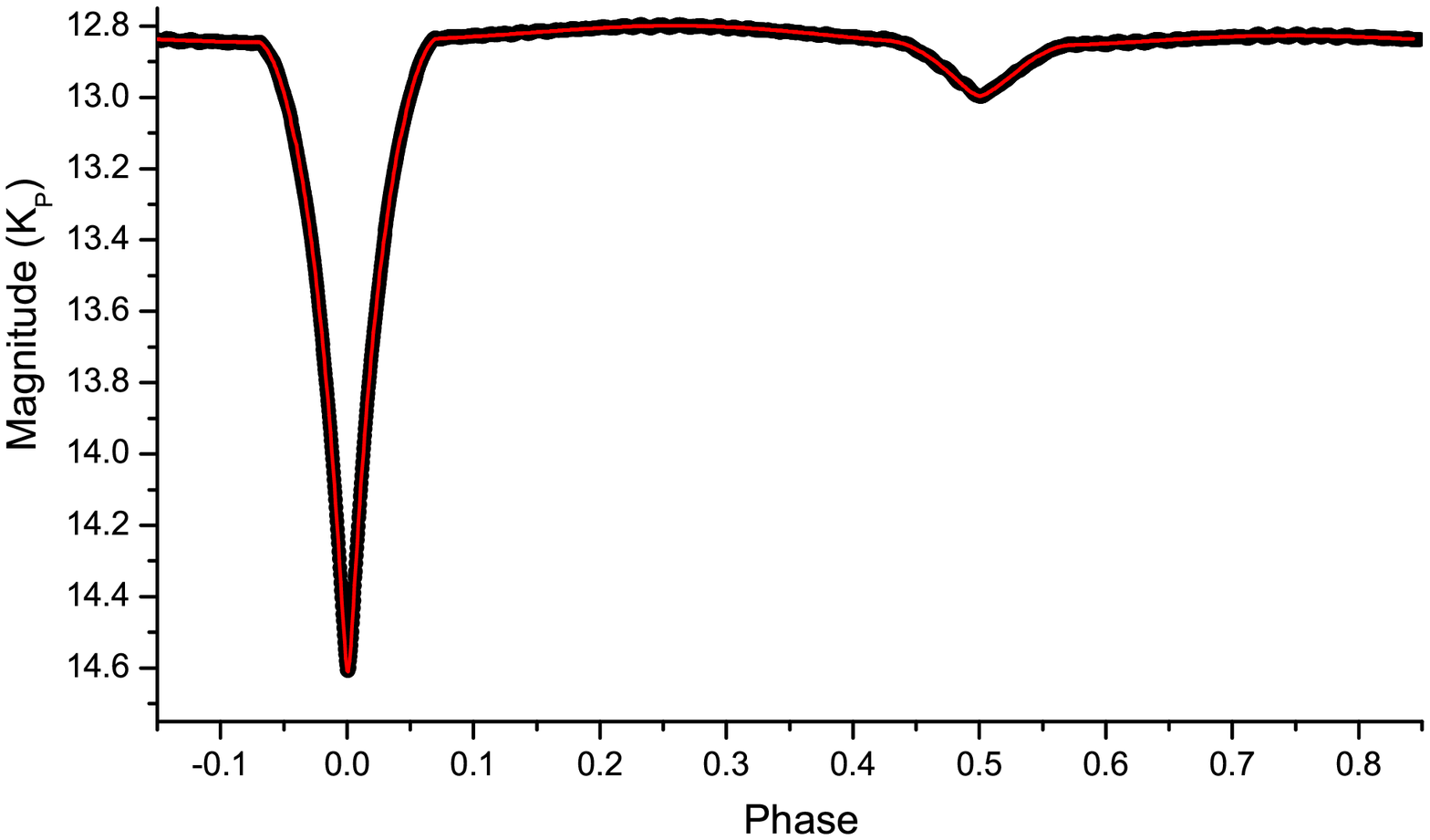}\\
\includegraphics[width=6cm]{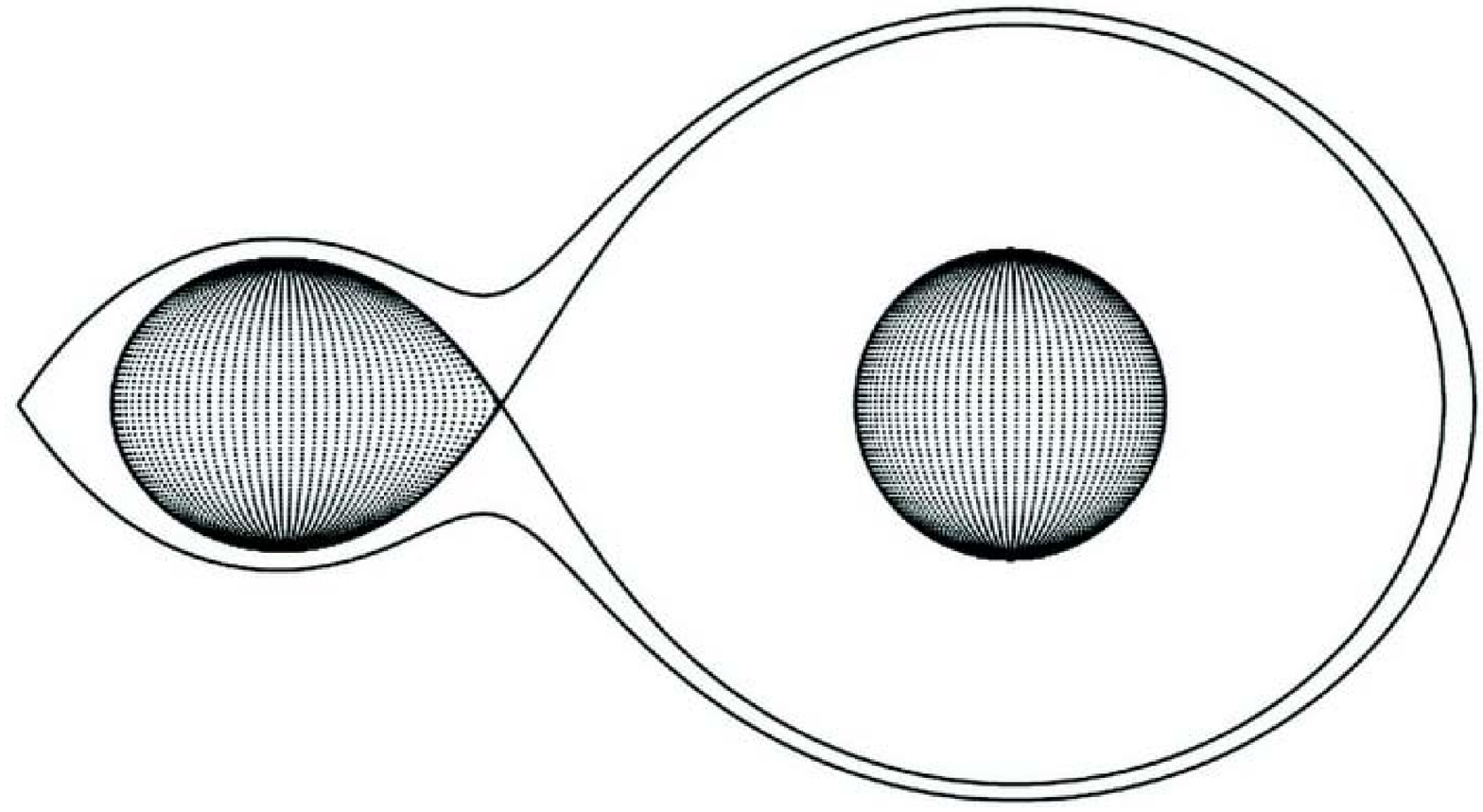}
\end{tabular}
\caption{The same as Fig.~\ref{fig:KIC066LCm3D}, but for KIC~10581918.}
\label{fig:KIC105LCm3D}
\end{figure}

\begin{figure}
\begin{tabular}{c}
\includegraphics[width=\columnwidth]{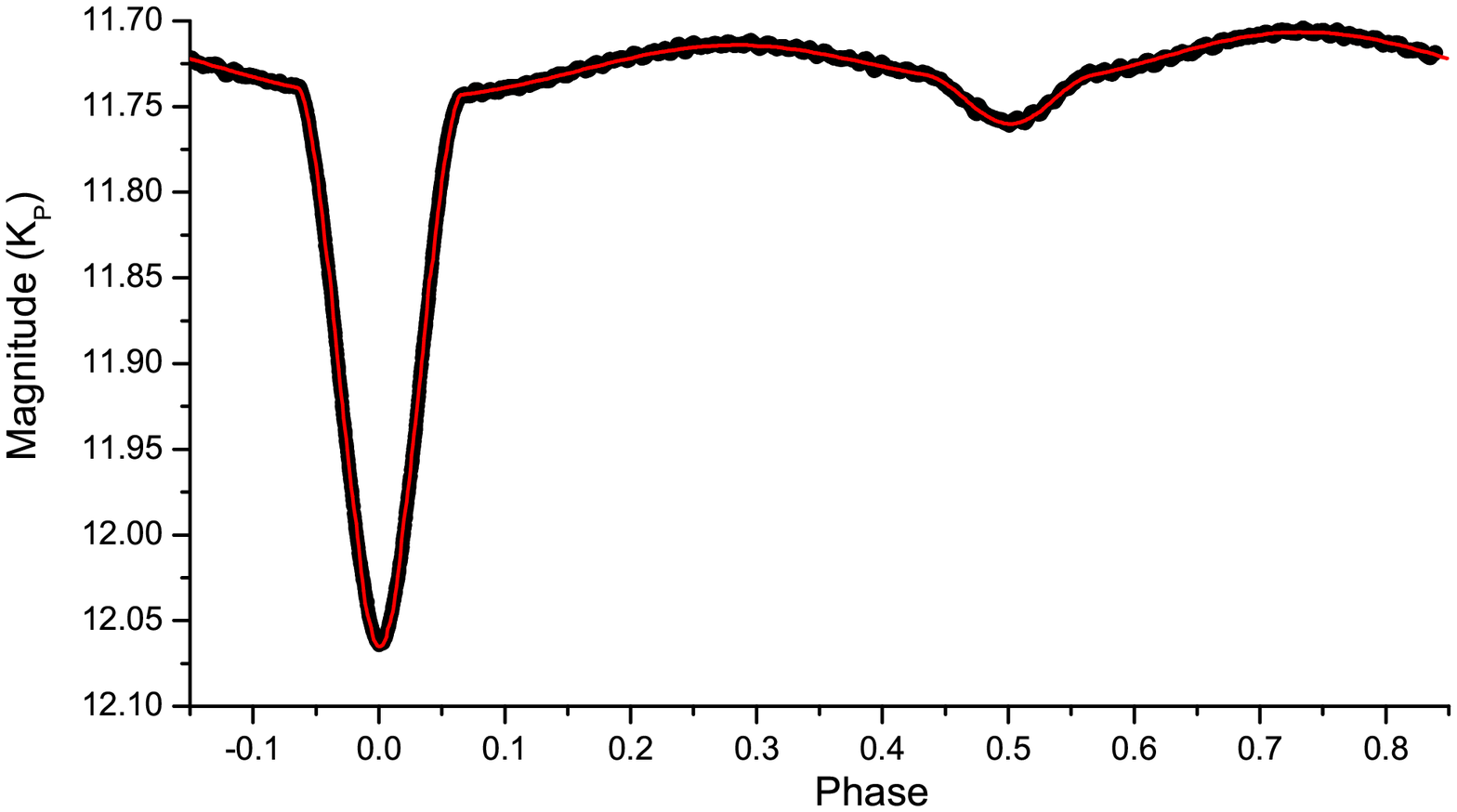}\\
\includegraphics[width=6cm]{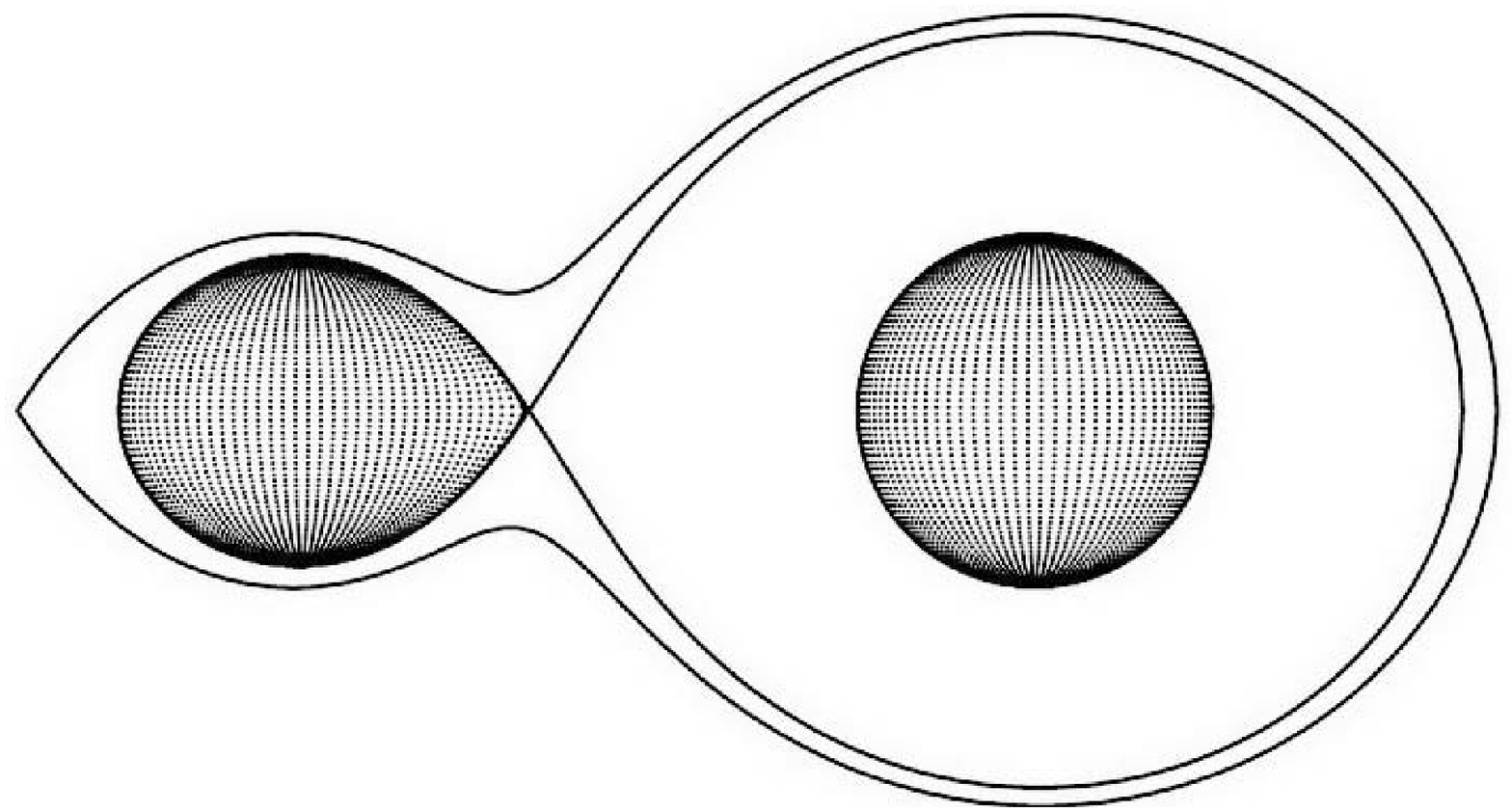}
\end{tabular}
\caption{The same as Fig.~\ref{fig:KIC066LCm3D}, but for KIC~10619109.}
\label{fig:KIC106LCm3D}
\begin{tabular}{c}
\includegraphics[width=\columnwidth]{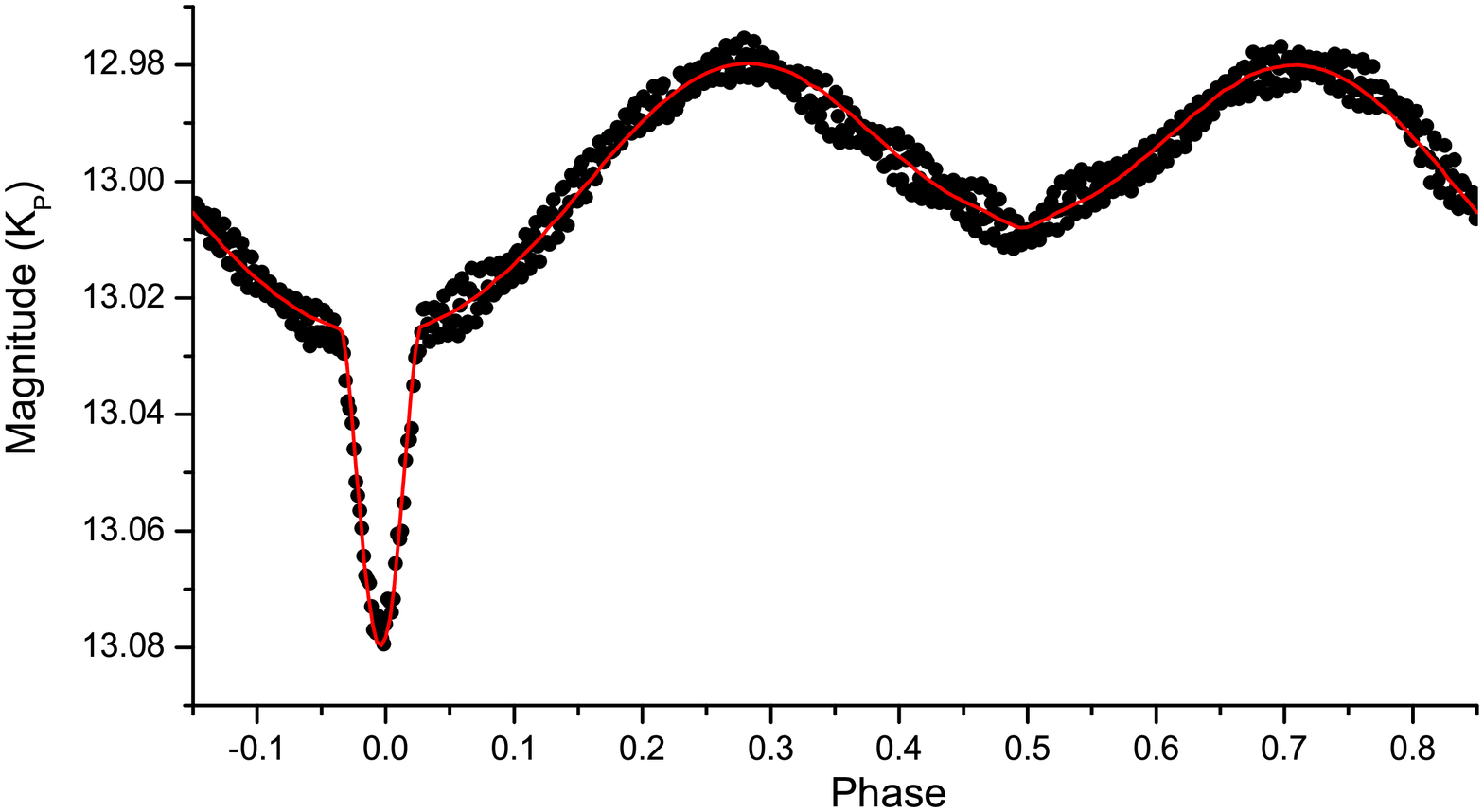}\\
\includegraphics[width=6cm]{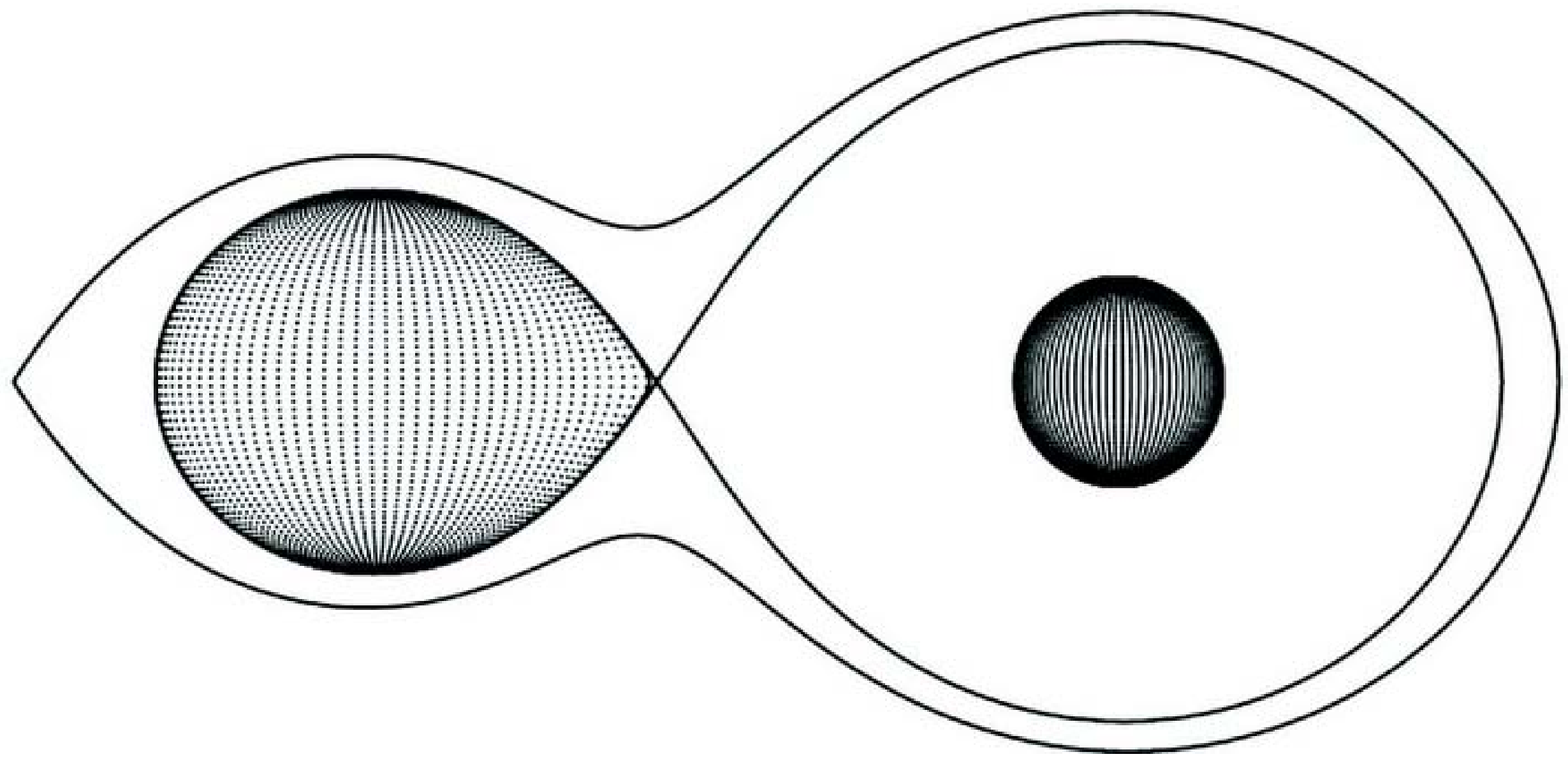}
\end{tabular}
\caption{The same as Fig.~\ref{fig:KIC066LCm3D}, but for KIC~11175495.}
\label{fig:KIC111LCm3D}
\end{figure}

The interesting in these cases is the changes of spot parameters over time. Although the spots are not subject of this study, it was considered useful to show how they migrated and changed during the duration of observations in Appendix~\ref{sec:App2}. In the cases of KIC~066 and KIC~111, one cool photospheric spot was assumed, while two cool spots were used to describe the brightness changes of KIC~105. For KIC~106 two bright spots were assumed on the secondary's surface because the system presents flare activity \citep{GAO16} and wide bright regions are expected. In Table~\ref{tab:spots} are given all the spot parameters, namely co-latitude ($co-lat.$), longitude ($long.$), radius, and temperature factor ($Tf$) for each LC. As representative beginning time of the spot migration is the half value of the EB's period. Then, every successive timing is derived from the previous timing plus the EB's period. The upper parts of Figs~\ref{fig:spotKIC066}-\ref{fig:spotKIC111} show the changes of the parameters of all spots over time for each system, while the lower parts show the spot on the secondaries' surfaces during the first and last days of observations.

\begin{table*}
\centering
\caption{Light curve and absolute parameters for all studied systems. The errors are given in parentheses alongside values and correspond to the last digit(s).}
\label{tab:LCmdlAbs}
\begin{tabular}{l cc cc cc cc}
\hline																	
System:	  &	\multicolumn{2}{c}{KIC 06669809}			  &	\multicolumn{2}{c}{KIC 10581918}			  &	\multicolumn{2}{c}{KIC 10619109}			&	\multicolumn{2}{c}{KIC 11175495}			\\
\hline																	
	                                                             &  \multicolumn{8}{c}{System parameters}																\\
\hline																	
$K_{\rm p}^{\rm a}$~(mag)	  &	\multicolumn{2}{c}{10.76}			  &	\multicolumn{2}{c}{12.80}			  &	\multicolumn{2}{c}{11.70}			&	\multicolumn{2}{c}{12.97}			\\
$T_{0}^{\rm a}$~(BJD)	  &	\multicolumn{2}{c}{2454953.998571}			  &	\multicolumn{2}{c}{2455139.362915}			  &	\multicolumn{2}{c}{2454955.136772}			&	\multicolumn{2}{c}{2454954.005791}		\\
$P_{\rm orb}^{\rm a}$~(days)	  &	\multicolumn{2}{c}{0.733738}			  &	\multicolumn{2}{c}{1.801863}			  &	\multicolumn{2}{c}{ 2.045183}			&	\multicolumn{2}{c}{2.191027}			\\
$q~(m_{2}/m_{1})$ 	  &	\multicolumn{2}{c}{0.31(1)}			  &	\multicolumn{2}{c}{0.12(1)}			  &	\multicolumn{2}{c}{0.14(1)}			&	\multicolumn{2}{c}{0.29(1)}			\\
$i~(\degr)$	  &	\multicolumn{2}{c}{67.5(1)}			  &	\multicolumn{2}{c}{89.0(2)}			  &	\multicolumn{2}{c}{75.0(1)}			&	\multicolumn{2}{c}{68.6(2)}			\\
\hline																	
	                                                     &  \multicolumn{8}{c}{Components parameters}																\\
\hline	
        	  &	     $Prim.$	&	      $Sec.$	  &	     $Prim.$	&	      $Sec.$	&	     $Prim.$	&	      $Sec.$	&	     $Prim.$	&	      $Sec.$	\\
\hline																	
$T_{\rm eff}$~(K)	&	                  7400(100)$^{\rm b}$	&	  4898(57)	&	                  7900(100)$^{\rm b}$	&	  4954(51)	&	7070(75)$^{\rm b}$	&	3949(31)	&	7550(100)$^{\rm b}$	&	  3600(55)	\\
$\Omega$                  	&	        3.04(2)	&	    2.61(1)	&	4.83(2)	&	2.03(1)	&	4.32(1)	&	2.07(2)	&	      7.4(2)	&	 2.41(1)	\\
$A^{\rm c}$ 	&	1	&	0.5	&	1	&	0.5	&	1	&	0.5	&	1	&	0.5	\\
$g^{\rm c}$ 	&	1	&	0.32	&	1	&	0.32	&	1	&	0.32	&	1	&	0.32	\\
$x$	&	0.442	&	0.637	&	0.394	&	0.627	&	0.414	&	0.689	&	0.500	&	0.646	\\
$L/(L_{\rm P}+L_{\rm S})$ 	&	0.905(1)	&	   0.095(3)	&	0.843(4)	&	0.157(1)	&	0.934(7)	&	0.066(2)	&	     0.870(9)	&	0.130(1)	\\
$r_{\rm pole}$	&	     0.385(1)	&	   0.264(1)	&	0.212(1)	&	0.201(1)	&	0.239(1)	&	0.208(1)	&	     0.141(4)	&	0.259(3)	\\
$r_{\rm point}$	&	        0.423(1)  	&	    0.383(1)	&	0.214(1)	&	0.298(1)	&	0.242(1)	&	0.308(1)	&	     0.141(4)	&	0.377(4)	\\
$r_{\rm side}$	&	     0.400(1)	&	   0.275(1)	&	0.214(1)	&	0.209(1)	&	0.241(1)	&	0.217(1)	&	     0.141(4)	&	0.270(4)	\\
$r_{\rm back}$	&	     0.411(1)	&	   0.308(1)	&	0.214(1)	&	0.241(1)	&	0.241(1)	&	0.248(1)	&	     0.141(4)	&	0.303(4)	\\
\hline																	
	                                                   &\multicolumn{8}{c}{Absolute parameters}																\\
\hline																	
$M~$(M$_{\sun})$	&	1.65(17)$^{\rm c}$	&	    0.51(5)	&	1.87(19)$^{\rm c}$	&	0.22(3)	&	1.54(15)$^{\rm c}$	&	0.22(3)	&	  1.70(17)$^{\rm c}$ 	&	   0.49(5)	\\
$R~$(R$_{\sun})$	&	1.81(2)	&	  1.21(1)	&	1.74(3)	&	1.66(3)	&	2.01(3)	&	1.79(3)	&	 1.33(5)	&	    2.48(8)	\\
$L~$(L$_{\sun})$	&	8.6(5)	&	  0.76(6)	&	10.5(6)	&	1.51(8)	&	9.0(5)	&	0.70(6)	&	  5.2(8)	&	     0.93(1) 	\\
$\log g~$(cm~s$^{-2}$)	&	4.14(4)	&	  3.98(5)	&	4.23(5)	&	3.35(6)	&	4.02(5)	&	3.27(6)	&	  4.42(8)	&	     3.34(5) 	\\
\textsl{a}~(R$_{\sun})$	&	1.07(3)	&	 3.46(2)	&	0.87(7)	&	7.29(4)	&	1.03(7)	&	7.34(4)	&	  2.12(6)	&	    7.32(4)	\\
$M_{\rm bol}$~(mag)       	&	2.4(3)	&	 5.1(3)	&	2.2(3)	&	4.3(3)	&	2.4(3)	&	5.1(3)	&	  3.0(4)	&	    4.8(5)	\\
\hline																	
\end{tabular}															
\newline																	
$^{\rm a}$Taken from the $KEBC$ \citep{PRS11}, $^{\rm b}$Results from spectroscopy, $^{\rm c}$assumed, $Prim.$=Primary; $Sec.$=Secondary component
\end{table*}

Each value of the final LC model presented in Table~\ref{tab:LCmdlAbs} is the mean value of the same parameters of the individual LC models, while the errors are the standard deviation of them. In the upper part of Figs~\ref{fig:KIC066LCm3D} - \ref{fig:KIC111LCm3D} are shown for each system the fits to the observed points of an individual observed LC, while in the lower parts are illustrated their Roche models. The residuals of each LC were derived after the subtraction of the respective model from all the observed points of the particular LC. The LC residuals are plotted against time in the lower part of Figs~\ref{fig:KIC066LCs_Res}- \ref{fig:KIC111LCs_Res} for each system.

Although no RVs curves exist for these systems, fair estimates of their absolute parameters can be formed. The mass of each primary was inferred from its spectral type according to the spectral type-mass correlations of \citet{COX00} for main-sequence stars. Mass values of 1.65~M$_{\sun}$, 1.87~M$_{\sun}$, 1.54~M$_{\sun}$, and 1.7~M$_{\sun}$ were assigned for the primaries of KIC~066, KIC~105, KIC~106, and KIC~111, respectively, while a fair error value of 10\% was also adopted. The secondaries' masses follow from the determined mass ratios. The semi-major axes \textsl{a}, which fix the absolute mean radii, can then be derived from Kepler's third law. The luminosities ($L$), the gravity acceleration ($\log g$), and the bolometric magnitudes values ($M_{\rm bol}$) were calculated using the standard definitions. For the calculation of the absolute parameters the software \textsc{AbsParEB} \citep{LIA15} was used in mode 3 and the results are shown in the lower part of Table~\ref{tab:LCmdlAbs}.

\section{Pulsation Frequencies analysis}
\label{sec:Fmdl}
According to the spectroscopic and LCs analyses' results (Sections~\ref{sec:sp} and \ref{sec:LCmdl}), only the primary components of the studied systems have stellar characteristics similar to the $\delta$~Sct type stars (i.e. mass and temperature), therefore the following analyses and results concern only them. Frequency analysis was performed with the software \textsc{PERIOD04} v.1.2 \citep{LEN05} that is based on classical Fourier analysis. Given that typical frequencies for $\delta$~Sct stars range between 4-80~cycle~d$^{-1}$ \citep{BRE00}, the analysis should be made for this range. However, given that $\delta$~Sct stars in binary systems may present $g$-mode pulsations that are connected to their $P_{\rm orb}$ or present hybrid behaviour of $\gamma$~Dor-$\delta$~Sct type the search was extended to 0-80~cycle~d$^{-1}$.

\begin{figure*}[t]
\includegraphics[width=18cm]{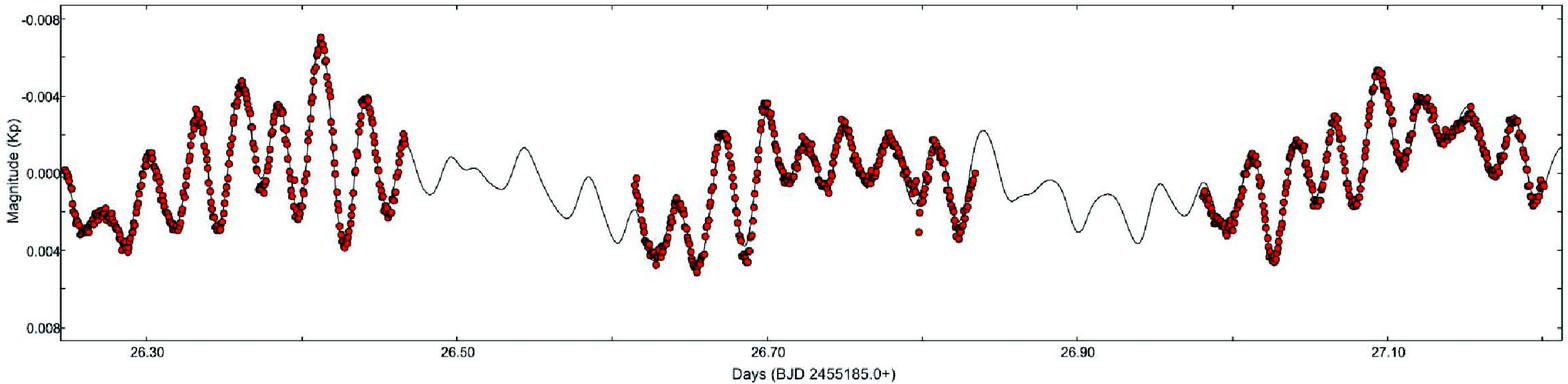}
\caption{Fourier fit (solid line) on various data points for KIC~06669809.}
\label{fig:KIC066FF}
\includegraphics[width=18cm]{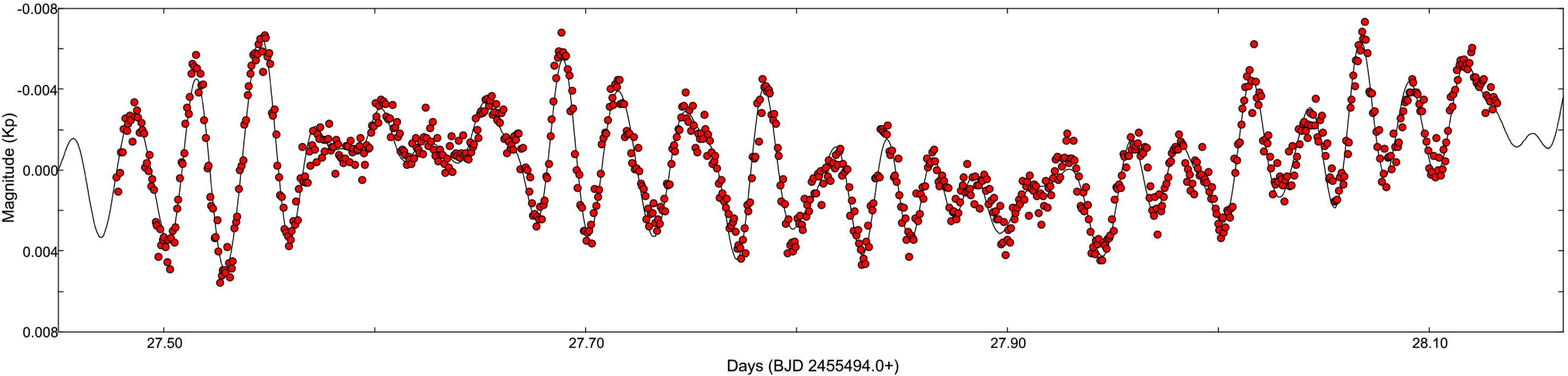}
\caption{The same as Fig.~\ref{fig:KIC066FF}, but for KIC~10581918.}
\label{fig:KIC105FF}
\includegraphics[width=18cm]{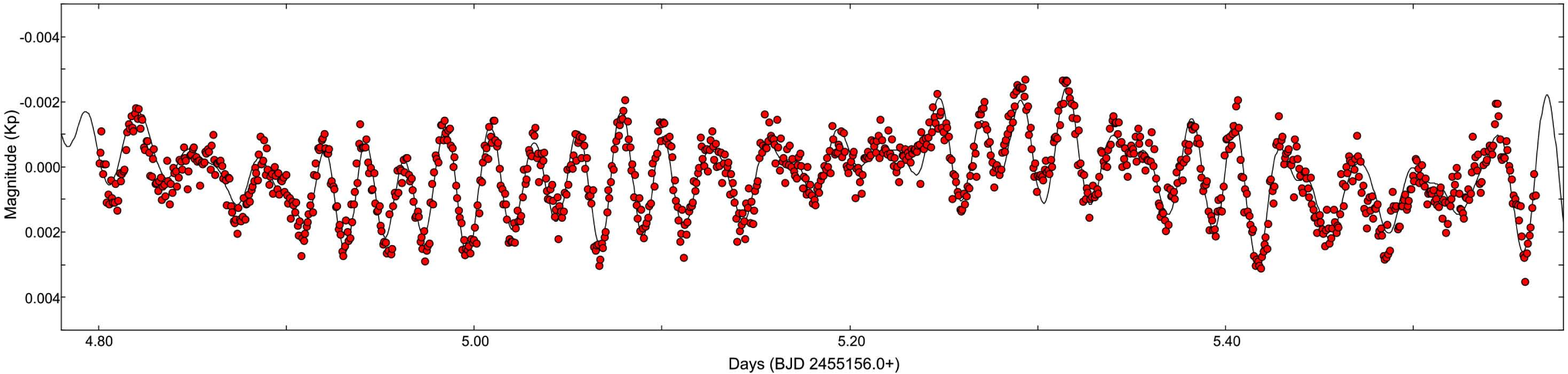}
\caption{The same as Fig.~\ref{fig:KIC066FF}, but for KIC~10619109.}
\label{fig:KIC106FF}
\includegraphics[width=18cm]{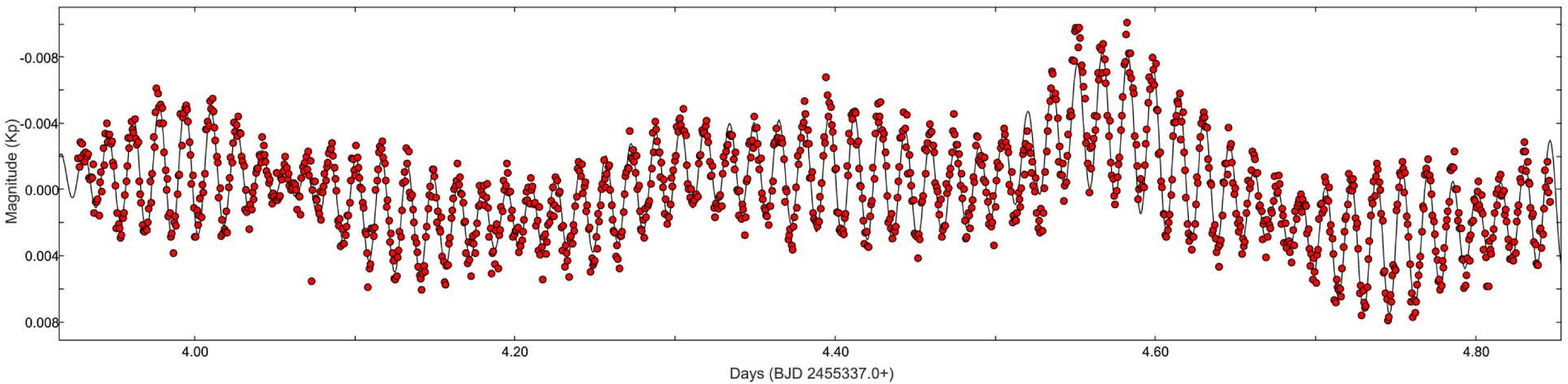}
\caption{The same as Fig.~\ref{fig:KIC066FF}, but for KIC~11175495.}
\label{fig:KIC111FF}
\end{figure*}
According to the method followed for the LC modelling (see Section~\ref{sec:LCmdl}), the binarity influence (e.g. proximity effects) as well as the light variations due to the cool spot existence in both systems were eliminated from the LC residuals. However, given that the total luminosity of a binary varies during the eclipses causing changes to the pulsations as well and, in addition, that the LC fitting is never perfect, only the out-of-eclipse data were used in the subsequent analysis. In particular, the range of orbital phases ($\Phi_{\rm orb}$) of the data that were excluded from the analysis are listed in Table~\ref{tab:PulsPar}.

\begin{table}
\centering
\caption{Pulsation analysis parameters. The columns contain the orbital phases ($\Phi_{\rm orb}$) of the excluded data, the background noise ($bgd$) of each data set, the 4$\upsigma$ detection threshold, the Nyquist number ($Nyq.$) and the frequency resolutions ($\updelta f$). For more details see text.}
\label{tab:PulsPar}
\scalebox{0.94}{
\begin{tabular}{l cc cc cc l cc cc cc}
\hline
System	&	excluded	&	$bgd$	&	4$\upsigma$	&	$Nyq.$	&	$\updelta f$	\\
	    &	($\Phi_{\rm orb}$)	&	($\upmu$mag)	&	(mmag)	&		&	cycle~d$^{-1}$	\\
\hline											
KIC066	&	0.90-0.10; 	&	8.4	&	0.034	&	727.1	&	0.0483	\\
	    &	0.40-0.60	&		&		    &		    &		    \\
KIC105	&	0.93-0.07; 	&	12.2&	0.049	&	720.9	&	0.0518	\\
	    &	0.44-0.56	&		&		    &		    &		    \\
KIC106	&	0.93-0.07; 	&	7.1	&	0.028	&	721.9	&	0.0581	\\
	    &	0.44-0.57	&		&		    &		    &		    \\
KIC111	&	0.97-0.03; 	&	9.4	&	0.038	&	723.6	&	0.0487	\\
	    &	0.43-0.57	&		&		    &		    &		    \\
\hline																											
\end{tabular}}
\end{table}

Before the pulsation frequency search, another significant problem, which, in general, arises when very accurate photometric data such as those of $Kepler$ are analysed, had to be confronted. Because of the high accuracy signals, frequencies can be found very close to each other and very often the calculation of their S/N may be problematic and unrealistic. More specifically, the software calculates the S/N of a given frequency based on its amplitude and the background noise around it. The problem is the calculation of the background noise itself. In particular, in $Kepler$'s data there might be many other frequencies of lower amplitudes around a given one. Although it is possible to increase the window for the background noise calculation around the detected frequency in order to minimize the contribution of the closest ones, the results remain unrealistic, since it is very often for the frequencies of multi periodic pulsators to occupy wide `bands' in the periodogram. This problem is more obvious when reaching the last significant frequencies. In order to solve this problem and provide more realistic S/N for the detected frequencies, the background noise ($bgd$) of each data set was calculated and listed in Table~\ref{tab:PulsPar} in regions where no frequencies seem to exist with a spacing of 2~cycle~d$^{-1}$ and a box size of 2. The software has a critical limit of 4$\upsigma$ (i.e. S/N$>4$) to detect reliable frequencies, so the same threshold was adopted and is given for each system in Table~\ref{tab:PulsPar}. In the same table, the Nyquist number ($Nyq.$) and the frequency resolutions ($\updelta f$) according to the Rayleigh-Criterion (i.e. 1.5/$T$, where $T$ is the observations time range in days) are also given for each data set. Finally, after the first frequency computation the residuals were subsequently pre-whitened for the next one until the detected frequency had S/N$\sim$4. Table~\ref{tab:IndF} contains the values of only the independent frequencies found for each system. The errors of all values listed in Tables~\ref{tab:IndF}, and \ref{tab:DepFreqKIC066}-\ref{tab:DepFreqKIC111} were calculated using analytical simulations \citep[for details see][]{LEN05}.

\begin{table*}
\centering
\caption{Independent oscillation frequencies for the pulsating components of all systems. The columns include for each system the frequency values $f_{\rm i}$ (where $i$ is an increasing number), the semi-amplitudes $A$, the phases $\Phi$, S/N, the pulsation constants $Q$, and the pulsation-orbital period ratios ($P_{\rm pul}/P_{\rm orb}$). The errors are given in parentheses alongside values and correspond to the last digit(s).}
\label{tab:IndF}
\begin{tabular}{l cc cc cc cc}
\hline																	
$i$	&	  $f_{\rm i}$	&	$A$	&	  $\Phi$	&	S/N	  & 	$Q$	&	         $P_{\rm pul}/P_{\rm orb}$$^{\rm a}$	&	$l$-degree	&	Mode	\\
	&	     (cycle~d$^{-1}$)	&	(mmag)	&	$(\degr)$	&		  & 	(d)     	&		&		&		\\
\hline																	
                                        \multicolumn{9}{c}{KIC 06669809}																	\\
\hline																	
2	&	    35.28312(7)	&	0.829(3)  	&	260.9(2)	&	102.3	&	0.0150(14)	&	0.0386	&	2	&	NR $p_{4}$	\\
3	&	    23.54355(9)	&	0.687(3)  	&	317.9(3)	&	84.7	&	0.0225(21)	&	0.0579	&	1	&	NR $p_{2}$	\\
8	&	    38.61245(13)	&	0.448(3)  	&	300.7(4)	&	55.3	&	0.0173(13)	&	0.0353	&	2	&	NR $p_{5}$	\\
\hline																	
                    \multicolumn{9}{c}{KIC 10581918}																	\\
\hline																	
1	&	34.0784(1)	&	1.194(6)	&	14.0(3)	&	98.1	&	0.0175(16)	&	0.0163	&	2	&	NR $p_{3}$	\\
2	&	36.4615(1)	&	1.048(6)	&	160.0(3)	&	86.1	&	0.0164(15)	&	0.0152	&	3	&	NR $p_{3}$	\\
\hline																	
                    \multicolumn{9}{c}{KIC 10619109}																	\\
\hline																	
1	&	42.8022(1)	&	0.661(4)	&	135.3(3)	&	93.7	&	0.0103(9)	&	0.0114	&	0	&	R $7H$	\\
2	&	32.0947(2)	&	0.509(4)	&	15.5(4)	&	72.1	&	0.0137(12)	&	0.0152	&	1	&	NR $p_{5}$	\\
3	&	31.1941(3)	&	0.273(4)	&	343.0(8)	&	38.6	&	0.0141(13)	&	0.0157	&	3	&	NR $p_{4}$	\\
6	&	35.7549(3)	&	0.238(4)	&	212.7(9)	&	33.7	&	0.0123(11)	&	0.0137	&	3	&	NR $p_{5}$	\\
7	&	45.9953(3)	&	0.232(4)	&	109.8(9)	&	32.8	&	0.0096(9)	&	0.0106	&	3	&	NR $p_{7}$	\\
\hline																	
                    \multicolumn{9}{c}{KIC 11175495}																	\\
\hline																	
1	&	64.44272(3)	&	3.151(5)	&	129.0(1)	&	336.2	&	 0.0132(45) 	&	0.0071	&	1	&	NR $p_{5}$	\\
4	&	57.73592(9)	&	0.947(5)	&	48.7(3)	&	101	&	0.0147(50)	&	0.0079	&	2	&	NR $p_{2}$	\\
5	&	56.68742(14)	&	0.643(5)	&	98.3(5)	&	68.6	&	0.0150(51)	&	0.0081	&	2	&	NR $p_{2}$	\\
\hline																	
\end{tabular}
\newline
$^{\rm a}$Error values are of 10$^{-7}$-10$^{-8}$ order of magnitude; R=Radial, NR=Non Radial pulsation
\end{table*}

\begin{figure}
\includegraphics[width=\columnwidth]{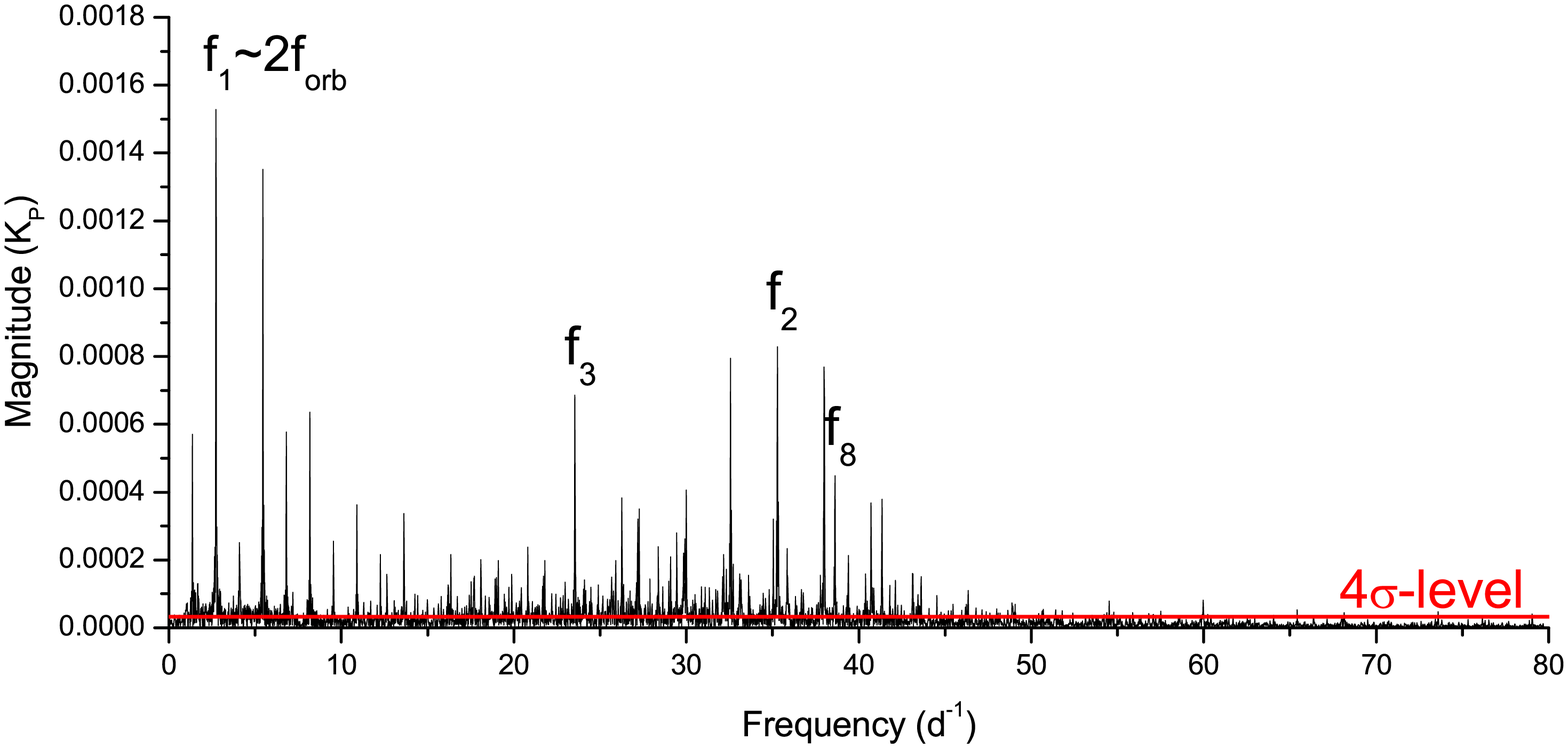}
\caption{Periodogram for KIC~06669809. The independent frequencies, the strong frequencies that are connected to the $P_{\rm orb}$, as well as the significance level are also indicated.}
\label{fig:KIC066FS}
\includegraphics[width=\columnwidth]{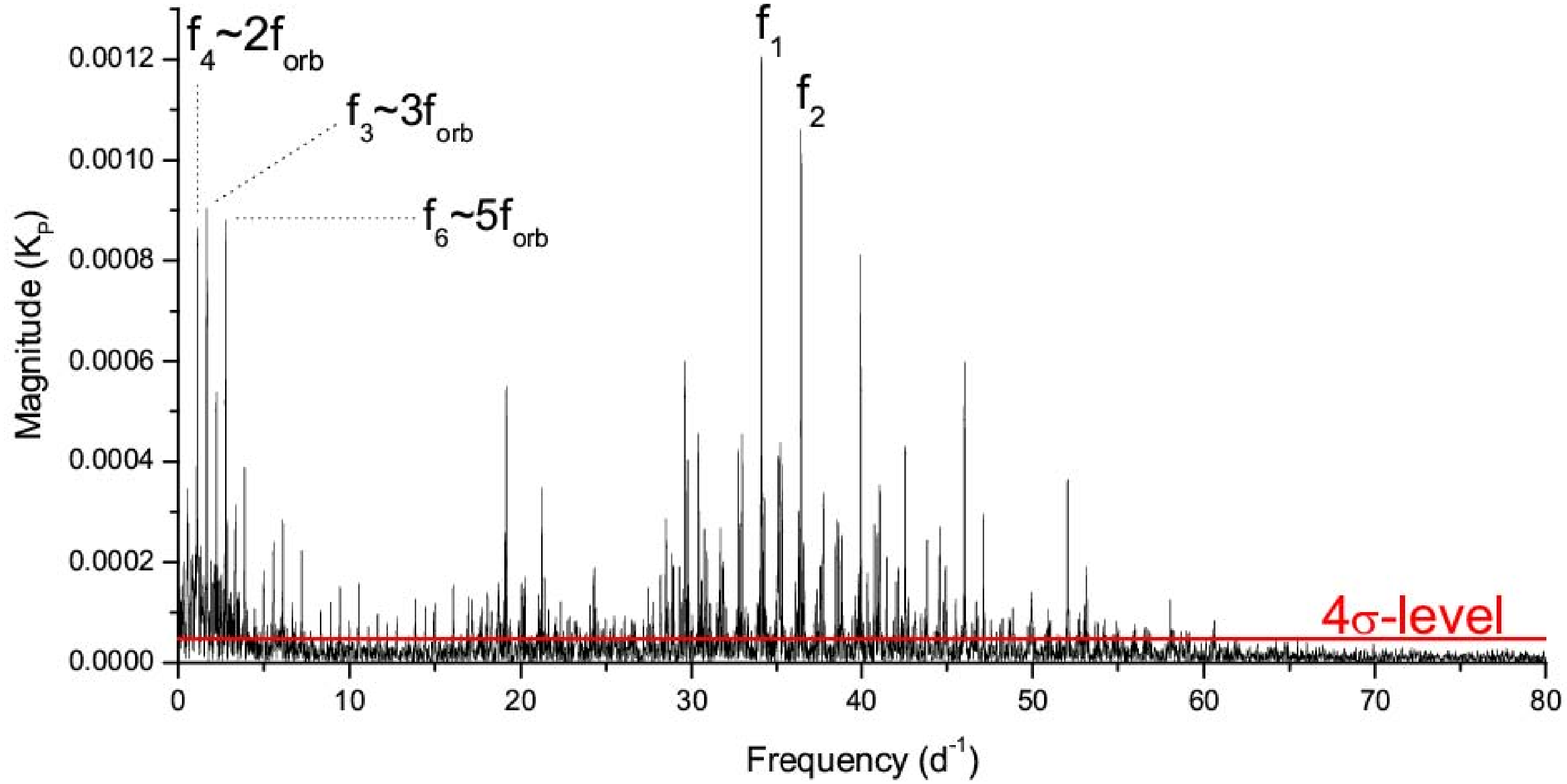}
\caption{The same as Fig.~\ref{fig:KIC066FS}, but for KIC~10581918.}
\label{fig:KIC105FS}
\includegraphics[width=\columnwidth]{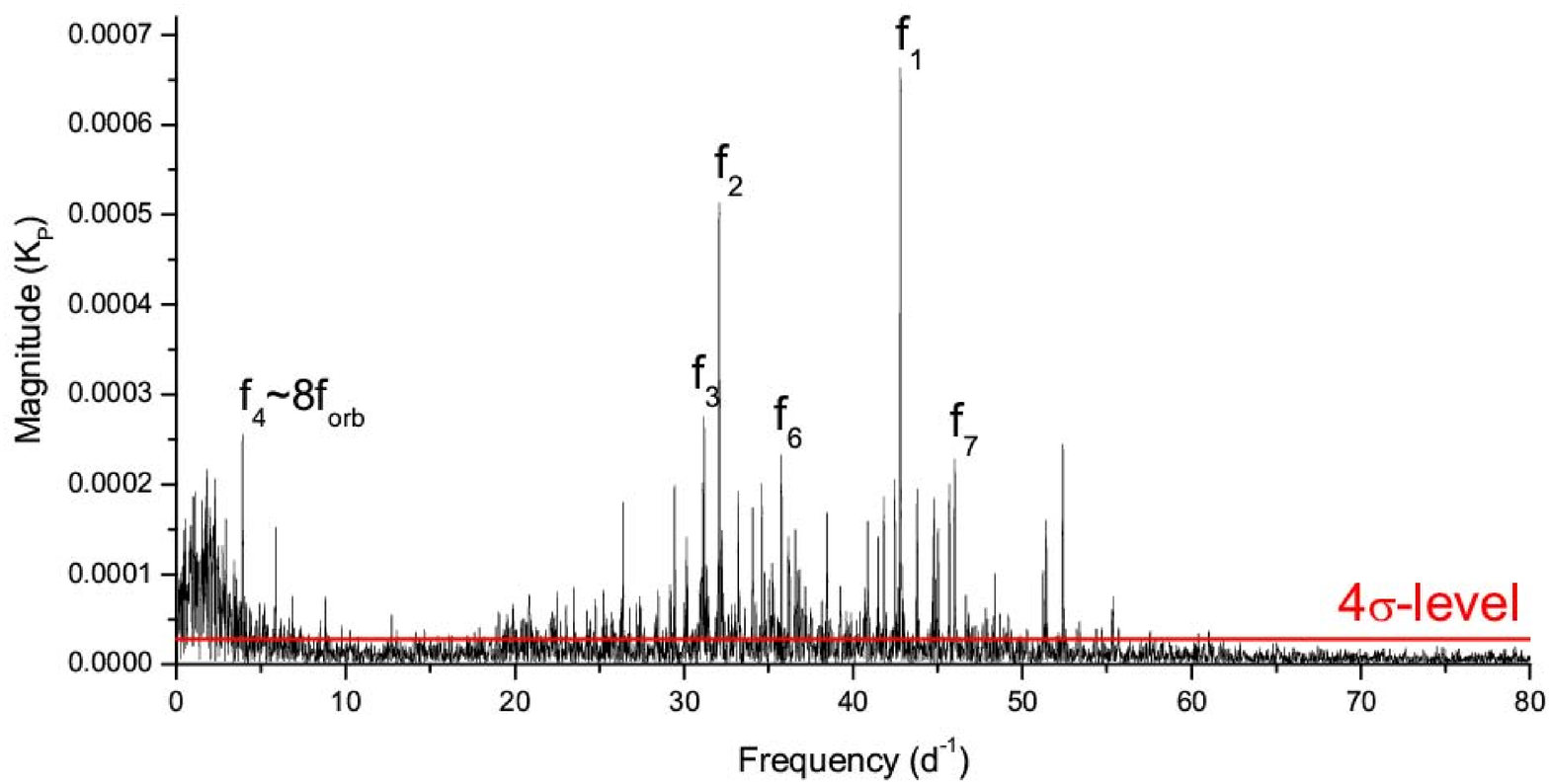}
\caption{The same as Fig.~\ref{fig:KIC066FS}, but for KIC~10619109.}
\label{fig:KIC106FS}
\includegraphics[width=\columnwidth]{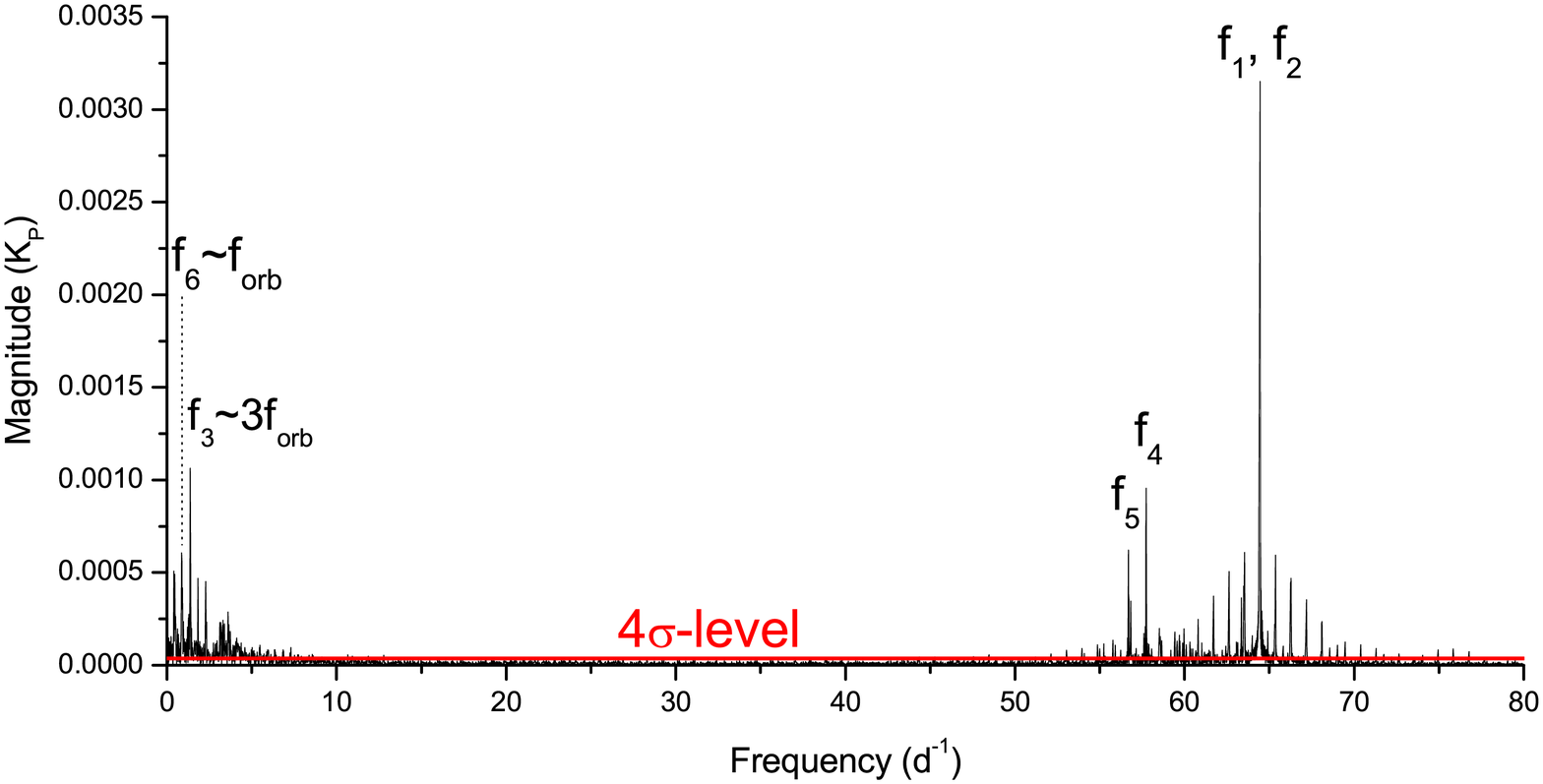}
\caption{The same as Fig.~\ref{fig:KIC066FS}, but for KIC~11175495.}
\label{fig:KIC111FS}
\end{figure}

\subsection{KIC 06669809}
\label{sec:FourierResultsKIC066}

KIC~066 was found to pulsate in three independent frequencies, namely $f_{2}$, $f_{3}$, and $f_{8}$ in the range 23-38~cycle~d$^{-1}$, while the corresponding frequency to the orbital period of the system is $f_{\rm orb}$=1.36288~cycle~d$^{-1}$. The most powerful frequency ($f_{1}$) is the 2nd harmonic of $f_{\rm orb}$. In total, 255 frequencies were detected, but the 252 of them, which are listed in Table~\ref{tab:DepFreqKIC066}, are either combinations or harmonics of the others. The fourier fit on individual data points is shown in Fig.~\ref{fig:KIC066FF}, while the periodogram, in which the level of significance as well as the independent frequencies are indicated, is illustrated in Fig.~\ref{fig:KIC066FS}. In the periodogram it is clear that frequencies' distribution show uniformity in all over the spectrum, but there are two main bands between 0-15~cycle~d$^{-1}$ and 23-45~cycle~d$^{-1}$, where their majority is concentrated. The slowest frequency is the $f_{54}\sim0.037$~cycle~d$^{-1}$ and the fastest the $f_{187}\sim68.145$~cycle~d$^{-1}$, but both of them are combinations of others.

\subsection{KIC 10581918}
\label{sec:FourierResultsKIC105}
Two independent oscillation frequencies (see Table~\ref{tab:IndF}) between 34-36.5~cycle~d$^{-1}$ and other 205 depended ones, listed in Table~\ref{tab:DepFreqKIC105}, were detected for KIC~105. As shown in the periodogram (Fig.~\ref{fig:KIC105FS}), the majority of the frequencies spread in the range 18-55~cycle~d$^{-1}$, while the range 0-5~cycle~d$^{-1}$ is dominated by the frequencies that are connected to the orbital period ($f_{\rm orb}$=0.55498~cycle~d$^{-1}$). In Fig.~\ref{fig:KIC105FF} is illustrated the Fourier fit on a selected data set.

\subsection{KIC 10619109}
\label{sec:FourierResultsKIC106}
The pulsating star of KIC~106 pulsates in 208 frequencies. In particular, $f_{1},~f_{2},~f_{3},~f_{6}$ and $f_{7}$ are the five independent ones, lie in the range 31-46~cycle~d$^{-1}$ (see Table~\ref{tab:IndF}), while the other 203 are combined ones and are listed in Table~\ref{tab:DepFreqKIC106}. The periodogram (Fig.~\ref{fig:KIC106FS}) shows four main frequency concentration ranges, one between 22-38~cycle~d$^{-1}$, another between 41-48~cycle~d$^{-1}$, a third one between 50-54~cycle~d$^{-1}$, and a last one between 0-6~cycle~d$^{-1}$. The latter range has the highest peak at 3.9135~cycle~d$^{-1}$, which corresponds to the 8th harmonic of the $f_{\rm orb}$ (=0.48895~cycle~d$^{-1}$) of the system. In Fig.~\ref{fig:KIC106FF} is shown the Fourier fit on individual observations.

\subsection{KIC 11175495}
\label{sec:FourierResultsKIC111}
The periodogram of KIC~111 (Fig.~\ref{fig:KIC111FS}) shows clearly two main frequencies concentrations, one between 0-12~cycle~d$^{-1}$ and a second between 55-70~cycle~d$^{-1}$, with only one frequency to be detected in the range 12-55~cycle~d$^{-1}$, namely $f_{156}$. The dominant frequency is the $f_{1}\sim64.44$~cycle~d$^{-1}$ with an amplitude of $\sim3.1$~mmag, while the other two independent ones ($f_{4},~f_{5}$) were found in the range 56.5-57.7~cycle~d$^{-1}$ (see Table~\ref{tab:IndF}). The rest 153 detected frequencies are either combinations or harmonics of the others and they are listed in Table~\ref{tab:DepFreqKIC111}. The Fourier fit on a selected day of observations is given in Fig.~\ref{fig:KIC111FF}.

In order to check the most possible oscillations modes for the independent frequencies, the pulsation constants ($Q$), based on the equation of \citet{BRE00} were calculated:
\begin{equation}
\log Q = -\log f + 0.5 \log g + 0.1M_{\rm bol} + \log T_{\rm eff} - 6.456
\end{equation}
where $f$ is the frequency of the pulsation, and $\log g$, $M_{\rm bol}$, and $T_{\rm eff}$ denote the standard quantities (see Section~\ref{sec:LCmdl}). Together with the $Q$ values, the $P_{\rm pul}/P_{\rm orb}$ ratios were also calculated and all of them are listed in Table~\ref{tab:IndF}. At this point, it should to be noted that the quantity $P_{\rm pul}$ refers to the dominant pulsation mode (i.e. the frequency with the highest amplitude). Firstly, according to \citet{ZHA13} if $P_{\rm pul}/P_{\rm orb}$ of a frequency is less than 0.07, then this frequency potentially belongs to the $p$-mode region. For all independent frequencies it was found that their $P_{\rm pul}/P_{\rm orb}$ ratios are indeed smaller than this value. Secondly, the $Q$ of each frequency was compared with the theoretical models of \citet{FIT81} for $M$=1.5~M$_{\sun}$ for all stars except for the primary of KIC~105, whose $Q$ values were compared with the models for $M$=2~M$_{\sun}$. The $l$-degrees and the type of each oscillation frequency are listed in Table~\ref{tab:IndF}. All stars were found to pulsate in non radial $p$-modes, except for the primary of KIC~106, which pulsates in both radial and non-radial modes. In addition, for the latter, the ratio $f_{2}/f_{1}$ is $\sim0.75$, that is the typical ratio for the radial fundamental to the first overtone mode \citep{STE79, OAS06}, and this can be an alternative explanation for its first two pulsating modes. Finally, it should to be also noted that the range of $Q$ errors allow also other possible mode determination, but the closest ones to the mean values were finally chosen.

The frequency search results of all stars show that the dominant independent pulsations occur in frequency ranges that are typical for the $\delta$~Sct type stars. Therefore, and in combination with the spectroscopic results regarding their spectral types (Section~\ref{sec:sp}), they can be plausibly considered of this type of pulsators. However, the distributions of their frequencies in the periodograms (Figs~\ref{fig:KIC066FS}-\ref{fig:KIC111FS}), i.e. existence of slow ($<5$ cycle~d$^{-1}$) and fast pulsations, may reveal the possibility of potential $\delta$~Sct-$\gamma$~Dor hybrid behaviour. This issue is discussed in the last section.

\section{Discussion and Conclusions}
\label{sec:Dis}

In this paper, detailed LCs and frequency analyses for four $Kepler$ EBs were presented. Spectroscopic observations of the systems permitted to estimate the spectral types of their primaries with an error of half sub-class. The absolute parameters of the systems' components as well as their evolutionary stages were also estimated, while the pulsational characteristics of their primaries were determined.

\begin{figure}
\includegraphics[width=\columnwidth]{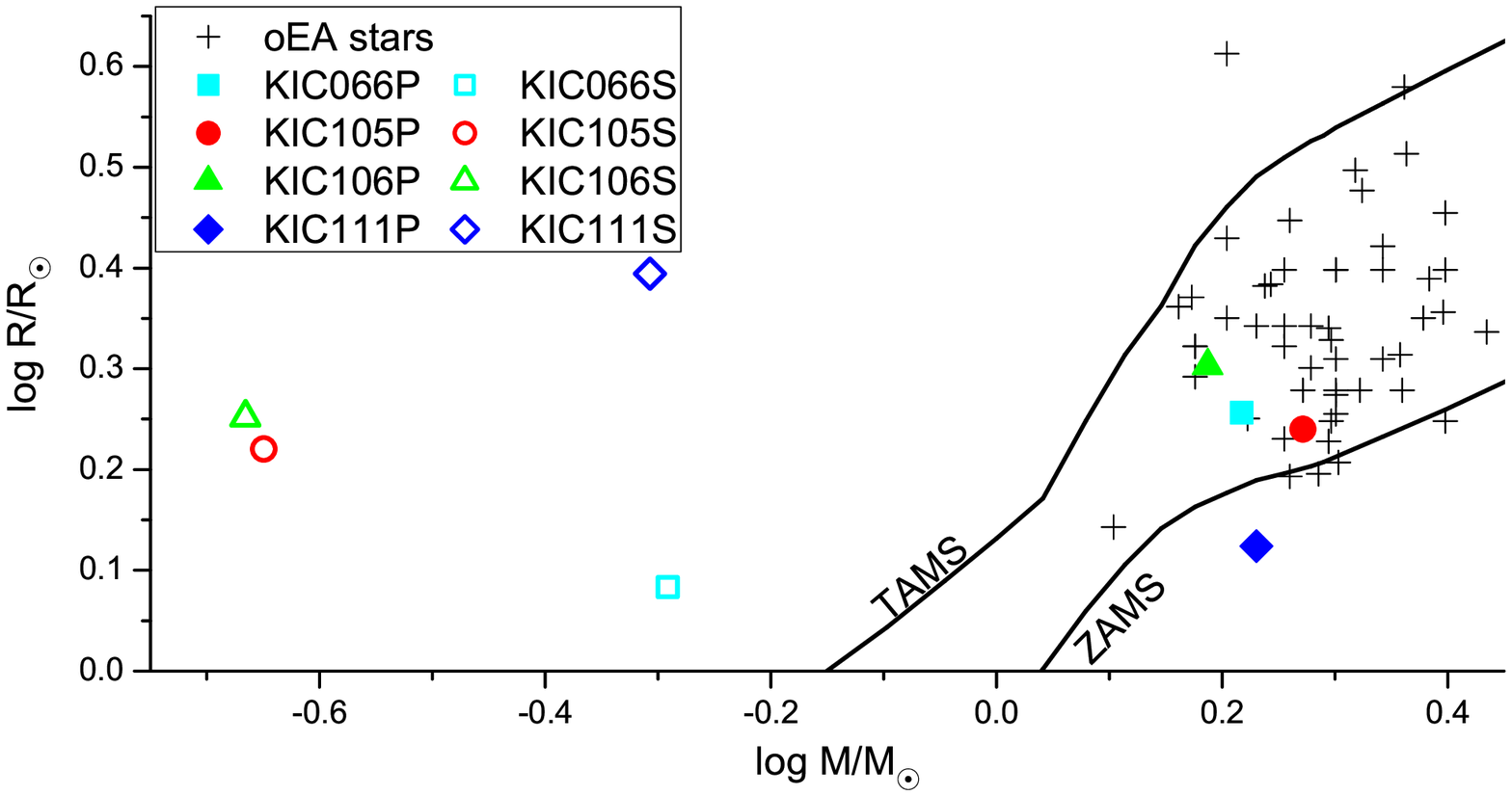}
\caption{Location of the primary (filled symbols) and secondary (empty symbols) components of KIC~06669809 (square), KIC~10581918 (circle),  KIC~10619109 (triangle)  and KIC~11175495 (diamond) within the Mass-Radius diagram. Crosses denote the $\delta$~Sct components of other oEA stars, while the solid lines the boundaries of main-sequence.}
\label{fig:MR}
\includegraphics[width=\columnwidth]{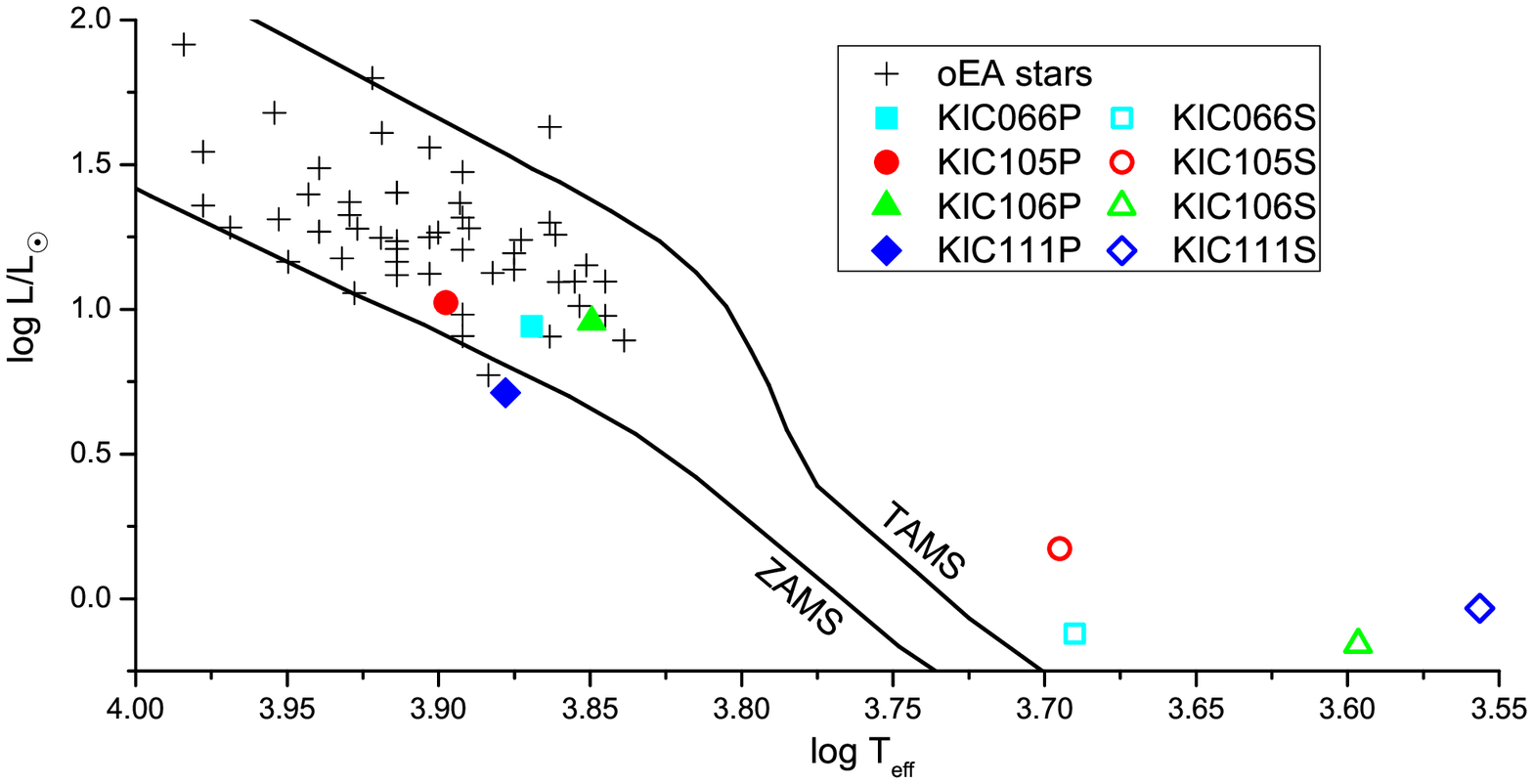}
\caption{Location of the components of all studied systems within the H-R diagram. Symbols and lines have the same meaning as in Fig.~\ref{fig:MR}.}
\label{fig:HR}
\end{figure}

All systems were found to be in a conventional semi-detached configuration, with their less massive components filling their Roche lobes and, therefore, can be categorized as classical Algol-type stars regarding their evolutionary status. The positions of the systems' components in the mass-radius ($M-R$) and Hertzsprung-Russell ($HR$) evolutionary diagrams are given in Figs~\ref{fig:MR} and \ref{fig:HR}, respectively. Zero and Terminal Age (ZAMS and TAMS, respectively) main-sequence lines (for solar metallicity composition, i.e. $Z$=0.019) for these diagrams were taken from \citet{GIR00}, while the positions of the other $\delta$~Scuti stars of oEA systems from \citet{LIAN17}. The secondaries of all systems have left main-sequence and they are in the subgiant evolutionary phase. Moreover, these components were found to be magnetically active, with their spots to migrate in both latitude and longitude and, in addition, to change radii in only $\sim$30 days. On the other hand, all primaries, except for that of KIC~111, are well inside the main-sequence boundaries. In particular, the primaries of KIC~066 and KIC~105 are closer to the ZAMS, while the one of KIC~106 is approximately in the middle of main-sequence band. The primary of KIC~111 was found to evolve slightly before ZAMS and it is currently the youngest of all other stars of the same type.

Although none of the systems exhibit total eclipses, which can be potentially used for the determination of the pulsating component (i.e. occurrence/absence of pulsations during the flat-phase part of the total eclipse), the primaries of all systems are A-F type stars according to the present spectral classification, thus, they fit better to a profile of a $\delta$~Scuti star in comparison with the secondaries, which are much cooler. Therefore, it is plausibly concluded that, in combination with the previous results for their evolutionary stages, all systems can be considered by definition as oEA stars. Three independent $p$-mode frequencies in the range 23-38~cycle~d$^{-1}$ were found for the primary of KIC~066 with the dominant at 35.283~cycle~d$^{-1}$ and an amplitude of $\sim0.83$~mmag, while the primary of KIC~105 pulsates in two main non radial pressure modes in the range 34-36.5~cycle~d$^{-1}$. The main component of KIC~106 exhibits five independent oscillation frequencies, with the dominant one (42.8~cycle~d$^{-1}$) to be radial mode and the other four non radial $p$-modes. The primary of KIC~111 was also found to oscillate in three frequencies in the range 56-65~cycle~d$^{-1}$ and it is the fastest pulsator in binary systems that has ever been found \citep[cf.][]{LIAN17}.

\begin{figure}
\includegraphics[width=\columnwidth]{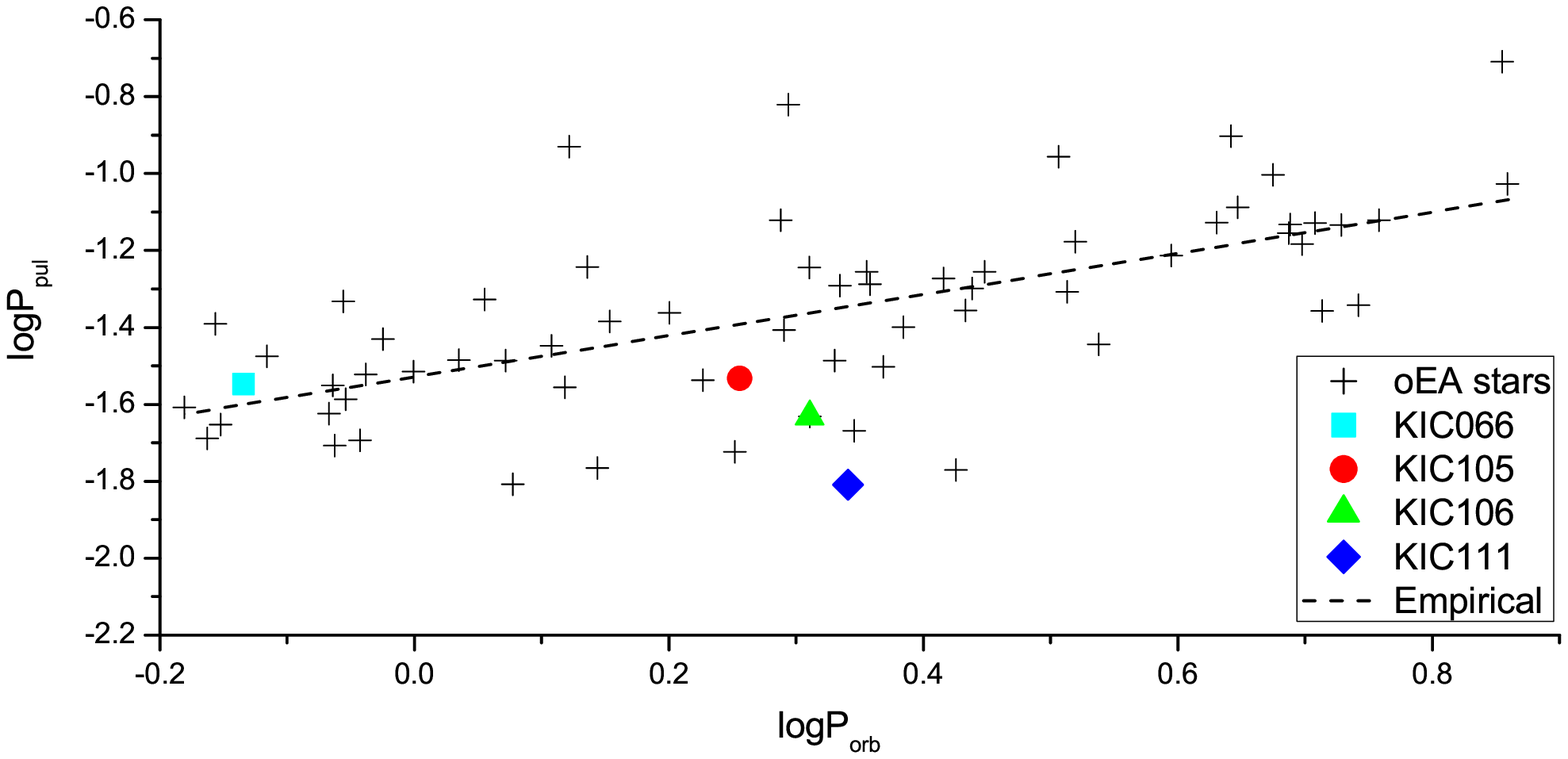}
\caption{Location of the pulsating components of all studied systems among other $\delta$~Sct stars (crosses) of oEA systems within the $P_{\rm orb}-P_{\rm pul}$ diagram. The dashed line represents the empirical relation for oEA stars of \citet{LIAN17}. Symbols have the same meaning as in Fig.~\ref{fig:MR}.}
\label{fig:PP}
\includegraphics[width=\columnwidth]{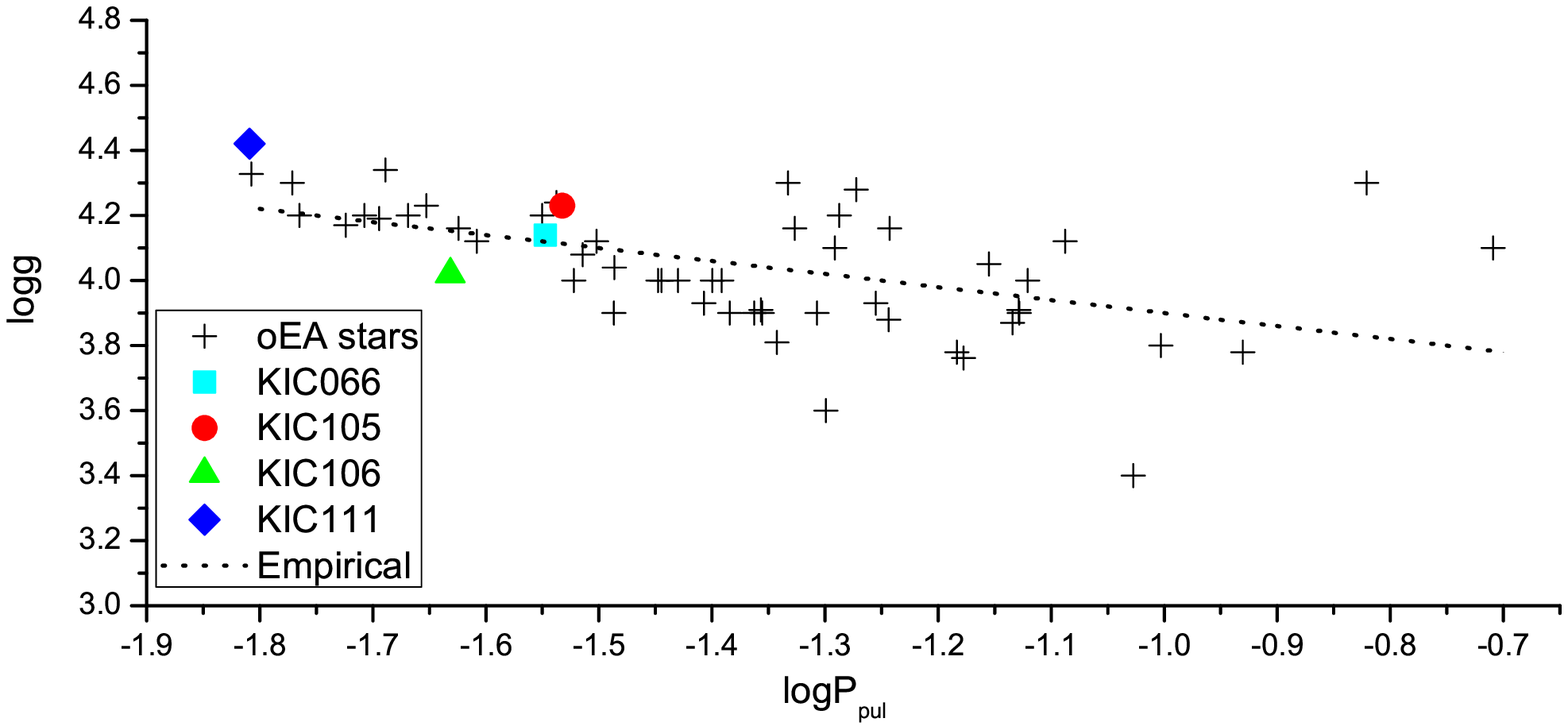}
\caption{Location of the pulsating components of all studied systems within the $\log g-P_{\rm pul}$ diagram. Symbols have the same meaning as in Fig.~\ref{fig:MR} and the line denotes the present empirical correlation for oEA stars with a $\delta$~Sct component.}
\label{fig:gP}
\end{figure}

A direct comparison between the pulsators of these systems and the $\delta$~Scuti components of other oEA stars is presented in the $P_{\rm orb}-P_{\rm pul}$ diagram (Fig.~\ref{fig:PP}). Although the present analysis for these systems is much more detailed than the preliminary one of \citet{LIAN17}, the new derived dominant frequencies are close enough to the previous values. Therefore, and given that these systems are the $\sim$6\% of the total available sample of oEA stars with $\delta$~Scuti component, the current correlation \citep[][eq.~1]{LIAN17} remains as it is. The primary of KIC~066 is among the first five oEA stars with the shortest $P_{\rm orb}$ values and is very close to the empirical fit. The pulsators of KIC~105 and KIC~106 follow well the empirical fit and the distribution of the other stars. On the other hand, the $\delta$~Scuti component of KIC~111 is the fastest pulsator of the sample but is also one of the most deviated cases from the fit. However, the latter cannot characterize it as an extreme case since, as mentioned by \citet{LIAN17}, there is a large scatter between 1.2~d$<P_{\rm orb}<$2.6~d (i.e. $0.08<\log P_{\rm orb}<0.42$) and 6.4~cycle~d$^{-1}<f_{\rm dom}<64.5$~cycle~d$^{-1}$ (i.e. $-1.81<\log P_{\rm pul}<-0.81$) for unknown reasons so far.

In Fig.~\ref{fig:gP} the locations of the primaries of the studied systems and other of oEA stars within the $\log g - \log P_{\rm pul}$ diagram are also shown. Contrary to the previous empirical fit, in this case the contribution of the systems to a new empirical relation between $\log g - \log P_{\rm pul}$ is quite significant. The previous fit \citep[][eq.~7]{LIAN17} included all the known systems with a $\delta$~Scuti component, whose absolute parameters are known, regardless their Roche geometry. In the present study and for the following empirical correlation, only the oEA stars are taken into account. The four studied systems consist the $\sim7$\% of the whole sample of oEA stars (56 in total) and they contribute for the first time in this fit with updated and more accurate absolute parameters (Table~\ref{tab:LCmdlAbs}). Therefore, the new empirical correlation between $\log g - \log P_{\rm pul}$ for the semi-detached EBs with a $\delta$~Scuti component (i.e. oEA stars) is the following:
\begin{equation}
\log g =3.5(1) - 0.40(9) \log P_{\rm pul}
\end{equation}
and is drawn in Fig.~\ref{fig:gP} as well. In this case, all the studied stars are relatively close enough to the empirical fit. This diagram shows also that the $\delta$~Scuti member of KIC~111 is the youngest of the sample, while the rest three cases are among the 20 youngest stars of this type.

The frequencies distributions of all systems may arise possible hybrid $\delta$~Sct-$\gamma$~Dor scenarios. It is clear from Figs~\ref{fig:KIC066FS}-\ref{fig:KIC111FS} that there are bands of frequencies of high and low values. In order to decide whether a star presents indeed an hybrid behaviour, the criteria proposed by \citet{UYT11} should be checked. According to that research work, the $\delta$~Sct-$\gamma$~Dor hybrids have to fulfil the following: a) present frequencies in both $\delta$~Sct and $\gamma$~Dor domains, b) the amplitudes in the two domains must be either comparable or the amplitudes should not differ more than a factor of 5-7, and c) at least two independent frequencies are detected in both regimes with amplitudes higher than 100~ppm. For all systems, except for KIC~111, no independent pulsation frequencies were detected in the $\gamma$-Dor regime, thus the hybrid scenario can be directly rejected for these three EBs. On the other hand, for KIC~111 two separate concentrations of frequencies exist (Fig.~\ref{fig:KIC111FS}), one in the $\delta$~Sct frequency range and another in the $\gamma$~Dor region (i.e. $<5$~cycle~d$^{-1}$). The first frequency found in the latter region is the $f_{6}=0.8668$~cycle~d$^{-1}$. In this case, the first and second criteria are indeed satisfied i.e. there are many frequencies in both regions and the amplitude of $f_{6}$ (0.59~mmag) is $\sim5.3$ times less than the amplitude of $f_{1}$. Regarding the third criterion, it is found that no other independent frequencies exist in the $\gamma$~Dor region. However, if we compare the $f_{6}$ with the second harmonic of the orbital frequency (2$f_{\rm orb}=0.9128$~cycle~d$^{-1}$), it could be noticed that they differ only by 0.046~cycle~d$^{-1}$, with the frequency resolution for this sample to be 0.049~cycle~d$^{-1}$. Therefore, it can be plausibly assumed that $f_{6}$ is probably the $2f_{\rm orb}$. Moreover, another possible reason for declining $f_{6}$ as a true $\gamma$~Dor frequency is its $P_{\rm pul}/P_{\rm orb}$ ratio, that has a value 0.526. By comparing this value with others of known hybrids \citep[e.g.][]{ZHA13}, it could be seen that it is outside their range. Therefore, it is concluded that $f_{6}$ is probably the second harmonic of $f_{\rm orb}$ instead of an independent frequency and that the pulsator of KIC~111 is not an hybrid $\delta$~Sct-$\gamma$~Dor star.

The ETV analysis for KIC~066 by \citet{BOR16} suggested the presence of a third body with a minimum mass of $m_{3, \rm min}\sim0.32$~M$_{\sun}$, based on the assumption that the total mass of the binary is 2~M$_{\sun}$. For reasons of completeness, the $m_{3, \rm min}$ value is recalculated according to the currently derived mass values (Table~\ref{tab:LCmdlAbs}) and the formula of the mass function of a third body \citep{MAY90}:
\begin{equation}
f(m_{3})=\frac{(m_{3}~\sin i_{3})^{3}}{(m_{1}+m_{2}+m_{3})^{2}}
\end{equation}
where $m$ are the masses of the components, and $i_{3}$ the inclination of the tertiary body's orbit. For $i_{3}=90~\degr$ the orbits of the EB and the third body's are coplanar and the minimum mass of the third component $m_{3, \rm min}$ can be derived. The above calculation results in $m_{3, \rm min} \sim$0.34~M$_{\sun}$. Assuming that the potential third body is a main-sequence star and by following the formalism of \citet{LIA11}, the luminosity contribution to the total light, using the \textsc{InPeVEB} software \citep{LIA15}, is found to be $\sim0.25$\%. Therefore, the absence of a third light in the LC solution seems very reasonable. For KIC~105 the orbital period analysis of \citet{WOL15} and \citet{ZAS15} resulted in a cyclic modulation of the $P_{\rm orb}$ with a period of $\sim$14 yr, which was attributed as prepense of a third companion with a mass function of $f(m_{3})$=0.00027(9)~M$_{\sun}$. Following the same method as in the previous case, a minimum mass of $m_{3, \rm min}\sim0.1$~M$_{\sun}$ and a luminosity contribution of less than 0.05\% are derived. Therefore, also in this case, the light of the potential third body is too weak to be detected either spectroscopically or photometrically. Both systems host magnetically active components, therefore the Applegate's mechanism \citep{APP92} has to be tested as a potential orbital period modulator. The quadrupole moment variation $\Delta Q$ is calculated based on the formula of \citet{LAN02} using the \textsc{InPeVEB} software \citep{LIA15}:
\begin{equation}
\Delta Q=-\frac{1}{9} \frac{\Delta P_{\rm orb}}{P_{\rm orb}} M {\rm a}^{2}
\end{equation}
where $\Delta P_{\rm orb}$ is the period of the variation of $P_{\rm orb}$, $M$ the mass of the star, and \textsl{a} the semi-major axis of the orbit. The $\Delta Q$ values are found less than $10^{51}$~gr~cm$^{2}$ for both systems, thus, according to the criterion of \citet{LAN02}, the magnetic activity cannot stand as possible explanation for the cyclic changes of their orbital periods. However, it should to be noted that the amplitudes of the periodic terms of the ETV fitting functions are quite small ($\sim0.0014$~d and $\sim0.0021$~d for KIC~066 and KIC~105, respectively) and given that the secondary components are magnetically active stars, these cyclic variations of the $P_{\rm orb}$ could be caused due to the spot's visibility \citep{KAL02, TRA13}. The ETV diagrams of KIC~106 and KIC~111 \citep{CON14, GIE15} seem that cannot provide any other useful information for the systems so far, except for some irregularities probably due to the presence of pulsations and spots.

Regarding any future studies on these systems, it could be mentioned that, RVs measurements will certify or modify the absolute parameters values. For this, telescopes with high resolution spectrographs and with diameters more than $\sim$2~m for KIC~066 and $\sim$4~m for the rest three systems are needed in order to detect signal from both components, given that the primaries dominate (more than 84\%) the spectra especially in bluer wavelength bands. In any case, the pulsation models, which were the objectives of this study, are not expected to be dramatically changed. Follow-up observations for minima timings are also welcome, in a sense that many years later they may offer the opportunity to study in detail the ETV diagrams for any possible orbital period modulating mechanisms. In general, detailed LCs and pulsation analyses of other $Kepler$ oEA stars is highly recommended, since the sample of these systems is still quite small but at the same time it is very promising regarding our knowledge for pulsations in binary systems.

\begin{acknowledgements}
The author acknowledges financial support by the European Space Agency (ESA) under the Near Earth object Lunar Impacts and Optical TrAnsients (NELIOTA) programme, contract no. 4000112943, and wishes to thank Mrs Maria Pizga for proofreading the text. The `Aristarchos' telescope is operated on Helmos Observatory by the Institute for Astronomy, Astrophysics, Space Applications and Remote Sensing of the National Observatory of Athens. The author wishes to thank the reviewer for the valuable suggestions. This research has made use of NASA's Astrophysics Data System Bibliographic Services, the SIMBAD, the Mikulski Archive for Space Telescopes (MAST), and the $Kepler$ Eclipsing Binary Catalog data bases. To the memory of Lans.
\end{acknowledgements}

%
%


\begin{appendix}

\section{Lists of combined frequencies}
\label{sec:App1}

Tables~\ref{tab:DepFreqKIC066}-\ref{tab:DepFreqKIC111} contain the values $f_{\rm i}$ (where $i$ is an increasing number), semi-amplitudes $A$, phases $\Phi$, S/N and the combinations of the depended frequencies of the systems. Details can be found in Section~\ref{sec:Fmdl}.

\begin{table*}
\centering
\caption{Combined frequencies of KIC~06669809. The errors are given in parentheses alongside values and correspond to the last digit(s).}
\label{tab:DepFreqKIC066}
\scalebox{0.85}{
\begin{tabular}{cc cc cl cc cc cl}
\hline																													
$i$	&	$f_{\rm	i}$	&	$A$	&	$\Phi$	&	S/N	&	Combination				&	$i$	&	$f_{\rm	i}$	&	$A$	&	$\Phi$	&	S/N	&	Combination		\\
	&	(cycle~d$^{-1}$)	&	(mmag)	&	$(\degr)$	&	&						&		&	(cycle~d$^{-1}$)	&	(mmag)	&	$(\degr)$	&	&				\\
\hline																													
1	&	2.72399(4)	&	1.529(3)	&	280.7(1)	&	188.6	&	2$f_{\rm	orb}$				&	76	&	24.54653(77)	&	0.078(3)	&	101.7(2.5)	&	9.6	&	2$f_{62}$			\\
4	&	37.98941(10)	&	0.605(3)	&	122.8(3)	&	74.6	&	$f_{1}$+$f_{2}$					&	77	&	24.45155(76)	&	0.079(3)	&	119.7(2.5)	&	9.7	&	$f_{11}$-$f_{1}$			\\
5	&	6.80998(10)	&	0.578(3)	&	345.4(3)	&	71.3	&	5$f_{\rm	orb}$				&	78	&	24.85403(78)	&	0.077(3)	&	87.8(2.5)	&	9.5	&	$f_{63}$+$f_{7}$			\\
6	&	1.36683(11)	&	0.548(3)	&	258.8(4)	&	67.6	&	$f_{5}$-2$f_{1}$	($\sim	f_{\rm	orb}$)		&	79	&	1.48918(78)	&	0.076(3)	&	137.9(2.5)	&	9.4	&	$f_{12}$-$f_{31}$			\\
7	&	5.44799(11)	&	0.537(3)	&	53.3(4)	&	66.2	&	2$f_{1}$	($\sim	4f_{\rm	orb}$)		&	80	&	29.79167(79)	&	0.076(3)	&	233.5(2.6)	&	9.3	&	$\sim	f_{47}$		\\
9	&	32.56879(14)	&	0.438(3)	&	123.3(4)	&	54.1	&	$f_{2}$-$f_{1}$					&	81	&	31.50785(79)	&	0.076(3)	&	85.4(2.5)	&	9.3	&	$f_{9}$-$f_{73}$			\\
\hline																																											
\end{tabular}}
\end{table*}

\begin{table*}
\scalebox{0.85}{
\begin{tabular}{cc cc cl cc cc cl}
\hline																												
$i$	&	$f_{\rm	i}$	&	$A$	&	$\Phi$	&	S/N	&	Combination				&	$i$	&	$f_{\rm	i}$	&	$A$	&	$\Phi$	&	S/N	&	Combination		\\
	&	(cycle~d$^{-1}$)	&	(mmag)	&	$(\degr)$	&	&						&		&	(cycle~d$^{-1}$)	&	(mmag)	&	$(\degr)$	&	&				\\
\hline																													
149	&	34.32039(117)	&	0.051(3)	&	89.0(3.8)	&	6.3	&	$f_{2}$-$f_{112}$					&	203	&	36.28289(146)	&	0.041(3)	&	116.4(4.7)	&	5.1	&	2$f_{133}$			\\
150	&	32.45127(118)	&	0.051(3)	&	7.6(3.8)	&	6.3	&	$f_{16}$-$f_{19}$					&	204	&	22.06242(146)	&	0.041(3)	&	33.6(4.7)	&	5.0	&	$f_{3}$-$f_{79}$			\\
151	&	2.12832(119)	&	0.050(3)	&	11.8(3.8)	&	6.2	&	$f_{4}$-$f_{22}$					&	205	&	5.63152(147)	&	0.041(3)	&	46.8(4.7)	&	5.0	&	$f_{12}$-$f_{13}$			\\
152	&	2.17340(118)	&	0.051(3)	&	56.5(3.8)	&	6.3	&	$\sim	f_{151}$				&	206	&	13.29798(148)	&	0.040(3)	&	152.1(4.8)	&	5.0	&	$f_{113}$+$f_{5}$			\\
153	&	45.45785(120)	&	0.050(3)	&	21.9(3.9)	&	6.2	&	2$f_{51}$					&	207	&	2.22814(147)	&	0.041(3)	&	110.5(4.8)	&	5.0	&	$f_{2}$-$f_{34}$			\\
154	&	43.52111(120)	&	0.050(3)	&	260.9(3.9)	&	6.2	&	$f_{53}$+$f_{7}$					&	208	&	37.09590(150)	&	0.040(3)	&	357.2(4.9)	&	4.9	&	2$f_{84}$			\\
\hline																													
\end{tabular}}

\vspace{1cm}
\centering
\caption{Combined frequencies of KIC~10581918. The errors are given in parentheses alongside values and correspond to the last digit(s).}
\label{tab:DepFreqKIC105}
\scalebox{0.85}{
\begin{tabular}{cc cc cl cc cc cl}
\hline																													
$i$	&	$f_{\rm	i}$	&	$A$	&	$\Phi$	&	S/N	&	Combination				&	$i$	&	$f_{\rm	i}$	&	$A$	&	$\Phi$	&	S/N	&	Combination		\\
	&	(cycle~d$^{-1}$)	&	(mmag)	&	$(\degr)$	&	&						&		&	(cycle~d$^{-1}$)	&	(mmag)	&	$(\degr)$	&	&				\\
\hline
3	&	1.6666(1)	&	0.906(6)	&	288.2(0.4)	&	74.4	&	3$f_{\rm orb}$	&	16	&	52.0507(3)	&	0.362(6)	&	197.3(0.9)	&	29.7	&	2$f_{7}-f_{5}$	\\
4	&	1.1111(1)	&	0.857(6)	&	89.0(0.4)	&	70.4	&	2$f_{\rm orb}$	&	17	&	0.5366(3)	&	0.353(6)	&	193.6(1.0)	&	29.0	&	$f_{3}-f_{4}~(\sim f_{\rm orb})$	\\
5	&	39.9446(1)	&	0.807(6)	&	248.3(0.4)	&	66.3	&	$f_{4}+2f_{2}-f_{1}$	&	18	&	37.7467(3)	&	0.339(6)	&	107.5(1.0)	&	27.8	&	2$f_{1}-f_{10}$	\\
6	&	2.7742(2)	&	0.682(6)	&	300.2(0.5)	&	56.0	&	$f_{3}+f_{4}~(\sim 5f_{\rm orb}$)	&	19	&	21.2481(3)	&	0.337(6)	&	137.0(1.0)	&	27.7	&	$f_{1}+f_{14}-f_{11}$	\\
7	&	46.0158(2)	&	0.600(6)	&	148.1(0.6)	&	49.2	&	$f_{5}+f_{6}+2f_{3}$	&	20	&	31.6962(4)	&	0.307(6)	&	350.6(1.1)	&	25.2	&	2$f_{1}-f_{2}$	\\
8	&	29.6242(2)	&	0.579(6)	&	213.1(0.6)	&	47.6	&	$f_{1}-f_{3}-f_{6}$	&	21	&	35.0967(4)	&	0.289(6)	&	283.7(1.2)	&	23.7	&	$f_{13}+f_{2}-f_{1}$	\\
\hline																													
\end{tabular}}
\end{table*}

\begin{table*}
\scalebox{0.85}{
\begin{tabular}{cc cc cl cc cc cl}
\hline																												
$i$	&	$f_{\rm	i}$	&	$A$	&	$\Phi$	&	S/N	&	Combination				&	$i$	&	$f_{\rm	i}$	&	$A$	&	$\Phi$	&	S/N	&	Combination		\\
	&	(cycle~d$^{-1}$)	&	(mmag)	&	$(\degr)$	&	&						&		&	(cycle~d$^{-1}$)	&	(mmag)	&	$(\degr)$	&	&				\\
\hline
29	&	38.5144(5)	&	0.221(6)	&	247.0(1.5)	&	18.2	&	2$f_{1}-f_{8}$	&	103	&	26.6016(12)	&	0.094(6)	&	232.7(3.6)	&	7.7	&	$f_{102}+f_{4}$	\\
30	&	0.7574(5)	&	0.217(6)	&	246.8(1.6)	&	17.8	&	$f_{10}-f_{8}$	&	104	&	49.9494(13)	&	0.089(6)	&	342.8(3.8)	&	7.3	&	$\sim f_{45}$	\\
31	&	0.5003(5)	&	0.206(6)	&	143.3(1.7)	&	16.9	&	$\sim f_{17}$	&	105	&	3.1572(13)	&	0.088(6)	&	136.9(3.9)	&	7.2	&	2$f_{56}$	\\
32	&	28.8582(6)	&	0.203(6)	&	256.8(1.7)	&	16.7	&	$f_{13}-f_{15}$	&	106	&	44.5321(13)	&	0.088(6)	&	226.5(3.9)	&	7.2	&	$\sim f_{23}$	\\
33	&	3.3004(6)	&	0.200(6)	&	78.0(1.7)	&	16.4	&	2$f_{3}~(\sim 6f_{\rm orb}$)	&	107	&	60.6200(13)	&	0.086(6)	&	219.2(3.9)	&	7.1	&	$f_{36}+f_{9}$	\\
34	&	24.3294(6)	&	0.191(6)	&	230.5(1.8)	&	15.7	&	2$f_{1}-f_{25}$	&	108	&	24.4157(13)	&	0.085(6)	&	96.8(4.0)	&	7.0	&	$f_{102}-f_{4}$	\\
\hline																													
\end{tabular}}
\end{table*}

\begin{table*}
\scalebox{0.85}{
\begin{tabular}{cc cc cl cc cc cl}
\hline																												
$i$	&	$f_{\rm	i}$	&	$A$	&	$\Phi$	&	S/N	&	Combination				&	$i$	&	$f_{\rm	i}$	&	$A$	&	$\Phi$	&	S/N	&	Combination		\\
	&	(cycle~d$^{-1}$)	&	(mmag)	&	$(\degr)$	&	&						&		&	(cycle~d$^{-1}$)	&	(mmag)	&	$(\degr)$	&	&				\\
\hline
177	&	25.6061(21)	&	0.055(6)	&	141.3(6.2)	&	4.5	&	$f_{8}-f_{97}$	&	193	&	23.3115(22)	&	0.052(6)	&	357.1(6.5)	&	4.3	&	$f_{23}-f_{19}$	\\
178	&	1.7235(21)	&	0.055(6)	&	113.2(6.2)	&	4.5	&	$\sim f_{68}$	&	194	&	5.5484(21)	&	0.054(6)	&	336.2(6.3)	&	4.4	&	2$f_{6}~(\sim 10 f_{\rm orb})$	\\
179	&	1.8063(20)	&	0.057(6)	&	244.2(6.0)	&	4.6	&	$\sim f_{68}$	&	195	&	5.1119(21)	&	0.053(6)	&	340.3(6.5)	&	4.3	&	2$f_{84}$	\\
180	&	5.1947(21)	&	0.055(6)	&	77.7(6.2)	&	4.5	&	2$f_{84}$	&	196	&	6.8216(22)	&	0.052(6)	&	139.0(6.5)	&	4.3	&	$f_{2}-f_{8}$	\\
181	&	58.6722(21)	&	0.055(6)	&	347.5(6.2)	&	4.5	&	$f_{47}+f_{5}$	&	197	&	59.0086(22)	&	0.052(6)	&	70.2(6.5)	&	4.3	&	$f_{27}+f_{5}$	\\
182	&	36.4527(21)	&	0.054(6)	&	232.1(6.3)	&	4.4	&	$\sim f_{2}$	&	198	&	40.8953(22)	&	0.052(6)	&	273.2(6.5)	&	4.3	&	$\sim f_{24}$	\\
\hline																													
\end{tabular}}

\vspace{1cm}


\centering
\caption{Combined frequencies of KIC~10619109. The errors are given in parentheses alongside values and correspond to the last digit(s).}
\label{tab:DepFreqKIC106}
\scalebox{0.85}{
\begin{tabular}{cc cc cl cc cc cl}
\hline																													
$i$	&	$f_{\rm	i}$	&	$A$	&	$\Phi$	&	S/N	&	Combination				&	$i$	&	$f_{\rm	i}$	&	$A$	&	$\Phi$	&	S/N	&	Combination		\\
	&	(cycle~d$^{-1}$)	&	(mmag)	&	$(\degr)$	&	&						&		&	(cycle~d$^{-1}$)	&	(mmag)	&	$(\degr)$	&	&				\\
\hline
4	&	3.9135(3)	&	0.255(4)	&	358.7(0.8)	&	36.2	&	8$f_{\rm orb}$	&	56	&	41.4925(10)	&	0.079(4)	&	359.3(2.7)	&	11.2	&	$f_{1}-f_{32}$	\\
5	&	52.3919(3)	&	0.246(4)	&	250.7(0.9)	&	34.8	&	2$f_{2}-3f_{4}$	&	57	&	30.4380(10)	&	0.079(4)	&	54.2(2.7)	&	11.2	&	$f_{2}-f_{24}$	\\
8	&	1.7930(4)	&	0.218(4)	&	92.2(1.0)	&	31.0	&	2$f_{2}-2f_{3}$	&	58	&	22.4908(10)	&	0.078(4)	&	197.1(2.8)	&	11.1	&	$f_{17}-f_{4}$	\\
9	&	45.6890(4)	&	0.209(4)	&	288.0(1.0)	&	29.7	&	$f_{7}+2f_{8}-f_{4}$	&	59	&	40.7268(10)	&	0.078(4)	&	9.9(2.8)	&	11.1	&	$f_{1}-f_{33}$	\\
10	&	2.9347(4)	&	0.207(4)	&	166.1(1.0)	&	29.3	&	$f_{9}-f_{1}~(\sim 6f_{\rm orb}$)	&	60	&	25.2452(10)	&	0.078(4)	&	115.8(2.8)	&	11.1	&	$f_{17}-f_{47}$	\\
11	&	33.2060(4)	&	0.205(4)	&	132.8(1.1)	&	29.0	&	2$f_{1}-f_{5}$	&	61	&	45.0164(10)	&	0.077(4)	&	146.6(2.8)	&	11.0	&	2$f_{58}$	\\
\hline																																											
\end{tabular}}
\end{table*}

\begin{table*}
\scalebox{0.85}{
\begin{tabular}{cc cc cl cc cc cl}
\hline																												
$i$	&	$f_{\rm	i}$	&	$A$	&	$\Phi$	&	S/N	&	Combination				&	$i$	&	$f_{\rm	i}$	&	$A$	&	$\Phi$	&	S/N	&	Combination		\\
	&	(cycle~d$^{-1}$)	&	(mmag)	&	$(\degr)$	&	&						&		&	(cycle~d$^{-1}$)	&	(mmag)	&	$(\degr)$	&	&				\\
\hline
106	&	38.3485(16)	&	0.049(4)	&	348.9(4.4)	&	7.0	&	$f_{31}+f_{8}$	&	158	&	46.9567(22)	&	0.036(4)	&	355.8(5.9)	&	5.2	&	$f_{21}+f_{7}$	\\
107	&	22.1147(17)	&	0.049(4)	&	184.3(4.4)	&	6.9	&	$f_{1}-f_{105}$	&	159	&	17.2165(22)	&	0.036(4)	&	100.6(6.0)	&	5.1	&	$f_{5}-f_{30}$	\\
108	&	19.0463(17)	&	0.048(4)	&	157.4(4.5)	&	6.8	&	$f_{92}-f_{8}$	&	160	&	6.6583(23)	&	0.035(4)	&	68.6(6.1)	&	5.0	&	$\sim f_{155}$	\\
109	&	37.2262(17)	&	0.047(4)	&	288.9(4.6)	&	6.6	&	$f_{19}+f_{6}$	&	161	&	48.8020(23)	&	0.035(4)	&	346.3(6.1)	&	5.0	&	$f_{4}+f_{45}$	\\
110	&	44.9835(17)	&	0.047(4)	&	224.2(4.6)	&	6.6	&	$\sim f_{61}$	&	162	&	40.1608(23)	&	0.035(4)	&	49.4(6.1)	&	5.0	&	$f_{16}-f_{12}$	\\
111	&	24.4311(17)	&	0.047(4)	&	211.9(4.6)	&	6.6	&	$f_{17}-f_{27}$	&	163	&	0.6455(23)	&	0.035(4)	&	73.7(6.2)	&	4.9	&	$\sim f_{122}$	\\
\hline																																											
\end{tabular}}

\vspace{1cm}
\centering
\caption{Combined frequencies of KIC~11175495. The errors are given in parentheses alongside values and correspond to the last digit(s).}
\label{tab:DepFreqKIC111}
\scalebox{0.85}{
\begin{tabular}{cc cc cl cc cc cl}
\hline																												
$i$	&	$f_{\rm	i}$	&	$A$	&	$\Phi$	&	S/N	&	Combination				&	$i$	&	$f_{\rm	i}$	&	$A$	&	$\Phi$	&	S/N	&	Combination		\\
	&	(cycle~d$^{-1}$)	&	(mmag)	&	$(\degr)$	&	&						&		&	(cycle~d$^{-1}$)	&	(mmag)	&	$(\degr)$	&	&				\\
\hline																													
2	&	64.40521(7)		&	1.304(5)	&	304.8(2)	&	139.1	&	$\sim f_{1}$			&	19	&	3.38379(41)		&	0.212(5)	&	92.2(1.4)	&	22.6	&	$f_{1}+f_{15}-f_{2}$		\\
3	&	1.36722(8)		&	1.063(5)	&	217.8(3)	&	113.4	&	$\sim 3f_{\rm orb}$			&	20	&	0.52127(43)		&	0.203(5)	&	111.0(1.5)	&	21.7	&	$\sim f_{10}$		\\
6	&	0.86680(15)		&	0.590(5)	&	49.5(5)	&	63.0	&	$\sim 2f_{\rm orb}$			&	21	&	3.25720(39)		&	0.222(5)	&	349.4(1.4)	&	23.7	&	$f_{1}+f_{14}-f_{2}$		\\
7	&	1.82891(18)		&	0.492(5)	&	243.2(6)	&	52.4	&	2$f_{3}-f_{6}~(\sim 4f_{\rm orb}$)			&	22	&	3.12018(43)		&	0.202(5)	&	190.7(1.5)	&	21.6	&	$f_{18}+f_{6}$		\\
8	&	0.40957(20)		&	0.443(5)	&	241.3(7)	&	47.3	&	2$f_{6}-f_{3}$			&	23	&	59.96398(43)		&	0.202(5)	&	0.2(1.5)	&	21.6	&	$f_{15}+f_{5}$		\\
9	&	63.35075(26)		&	0.338(5)	&	158.8(9)	&	36.0	&	$f_{1}+f_{5}-f_{4}$			&	24	&	3.65783(46)		&	0.191(5)	&	79.0(1.6)	&	20.4	&	2$f_{7}~(\sim 8f_{\rm orb}$)		\\
\hline																																											
\end{tabular}}
\end{table*}

\begin{table*}
\scalebox{0.85}{
\begin{tabular}{cc cc cl cc cc cl}
\hline																												
$i$	&	$f_{\rm	i}$	&	$A$	&	$\Phi$	&	S/N	&	Combination				&	$i$	&	$f_{\rm	i}$	&	$A$	&	$\Phi$	&	S/N	&	Combination		\\
	&	(cycle~d$^{-1}$)	&	(mmag)	&	$(\degr)$	&	&						&		&	(cycle~d$^{-1}$)	&	(mmag)	&	$(\degr)$	&	&				\\
\hline																															
34	&	0.68659(61)		&	0.143(5)	&	92.9(2.1)	&	15.3	&	$f_{28}+f_{8}$			&	96	&	4.31165(141)		&	0.062(5)	&	321.5(4.9)	&	6.6	&	$\sim 2f_{65}$		\\
35	&	1.25552(63)		&	0.139(5)	&	10.7(2.2)	&	14.9	&	$f_{6}+f_{8}$			&	97	&	1.52062(144)		&	0.061(5)	&	289.8(5.0)	&	6.5	&	$\sim f_{83}$		\\
36	&	2.95933(65)		&	0.135(5)	&	31.7(2.3)	&	14.4	&	$f_{31}-f_{5}$			&	98	&	0.32617(139)		&	0.063(5)	&	144.2(4.8)	&	6.7	&	$\sim 2f_{69}$		\\
37	&	3.35549(65)		&	0.134(5)	&	347.1(2.3)	&	14.3	&	$\sim f_{19}$			&	99	&	3.53124(147)		&	0.060(5)	&	193.1(5.1)	&	6.4	&	$\sim f_{38}$		\\
38	&	3.49251(65)		&	0.135(5)	&	325.7(2.3)	&	14.4	&	$f_{22}+f_{8}$			&	100	&	63.86457(147)		&	0.060(5)	&	241.0(5.1)	&	6.4	&	$f_{1}-f_{39}$		\\
39	&	0.61957(69)		&	0.127(5)	&	59.6(2.4)	&	13.6	&	$f_{6}-f_{28}$			&	101	&	59.46952(147)		&	0.060(5)	&	265.9(5.1)	&	6.4	&	$f_{4}+f_{48}$		\\
\hline																																																									\end{tabular}}
\end{table*}

\section{Spot migration}
\label{sec:App2}
This appendix includes information for the spots migration in time for all systems. Details can be found in Section~\ref{sec:LCmdl}.

\begin{table*}
\centering
\caption{Spot parameters for all systems.}
\label{tab:spots}
\scalebox{0.95}{
\begin{tabular}{c cc cc c cc cc}
\hline			
Time	&	Co-lat	&	Long	&	Radius	&	Tf ($\frac{T_{\rm spot}}{T_{\rm eff}})$	&	Time	&	Co-lat	&	Long	&	Radius	&	Tf ($\frac{T_{\rm spot}}{T_{\rm eff}})$	\\
(days)	&	($\degr$)	&	($\degr$)	&	($\degr$)	&		&	(days)	&	($\degr$)	&	($\degr$)	&	($\degr$)	&		\\
\hline
                \multicolumn{5}{c}{KIC 06669809}        &	21.645	&	26.5	&	42.8	&	25.4	&	0.67	\\
\cline{1-5}		
0.367	&	59.6	&	27.1	&	20.5	&	0.80	&	22.379	&	25.7	&	44.8	&	25.7	&	0.64	\\
1.101	&	58.7	&	26.0	&	20.4	&	0.80	&	23.113	&	25.4	&	45.5	&	25.3	&	0.63	\\
1.834	&	59.2	&	25.9	&	19.6	&	0.78	&	23.846	&	25.3	&	46.5	&	25.2	&	0.62	\\
2.568	&	57.7	&	27.4	&	19.6	&	0.77	&	24.580	&	24.6	&	47.4	&	25.3	&	0.61	\\
3.302	&	56.3	&	27.5	&	19.9	&	0.77	&	25.314	&	23.7	&	48.3	&	25.7	&	0.63	\\
\hline																			
\end{tabular}}
\end{table*}

\begin{figure}[h]
\begin{tabular}{cc}
\multicolumn{2}{c}{\includegraphics[width=7.5cm]{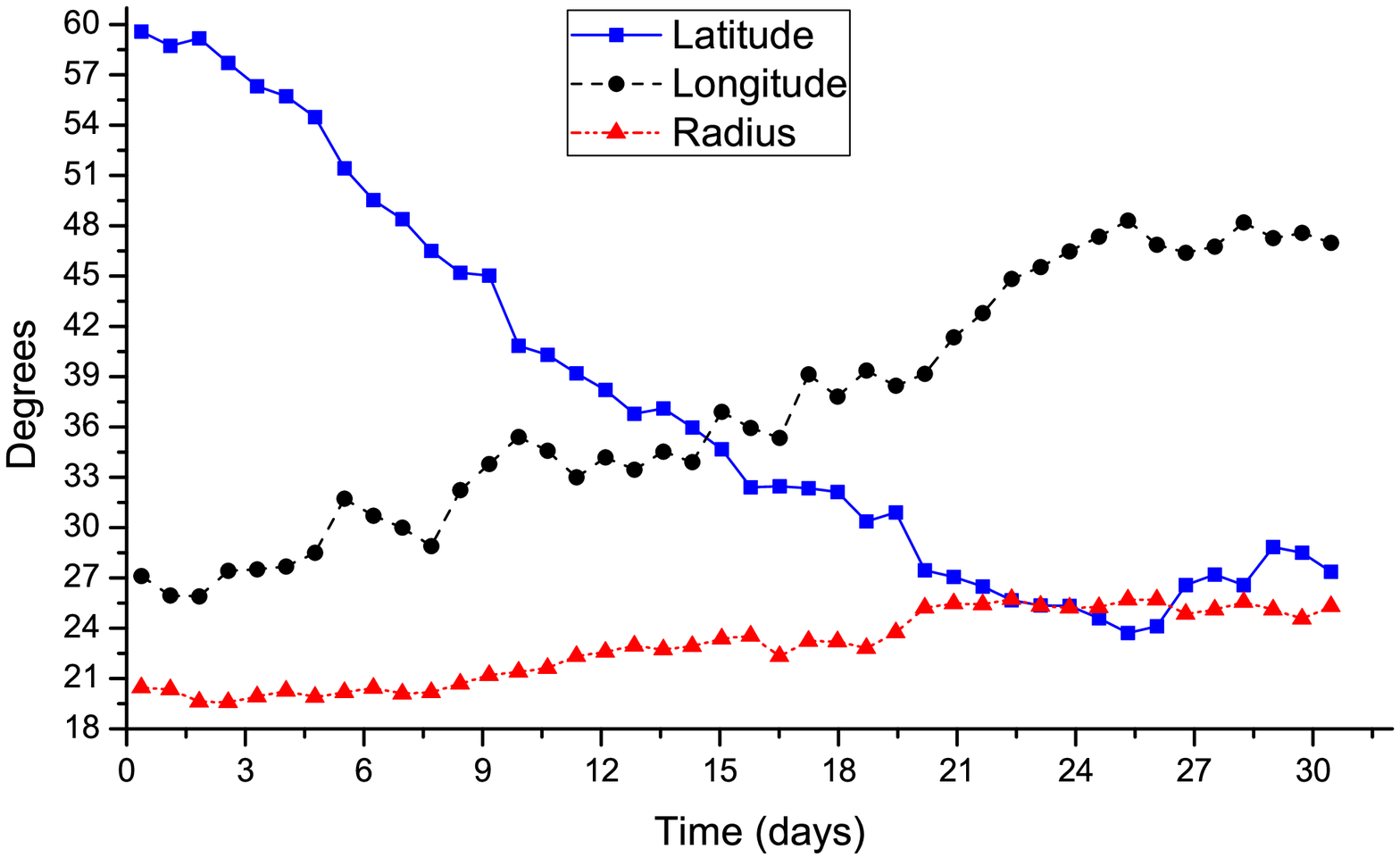}}\\
\includegraphics[width=3.7cm]{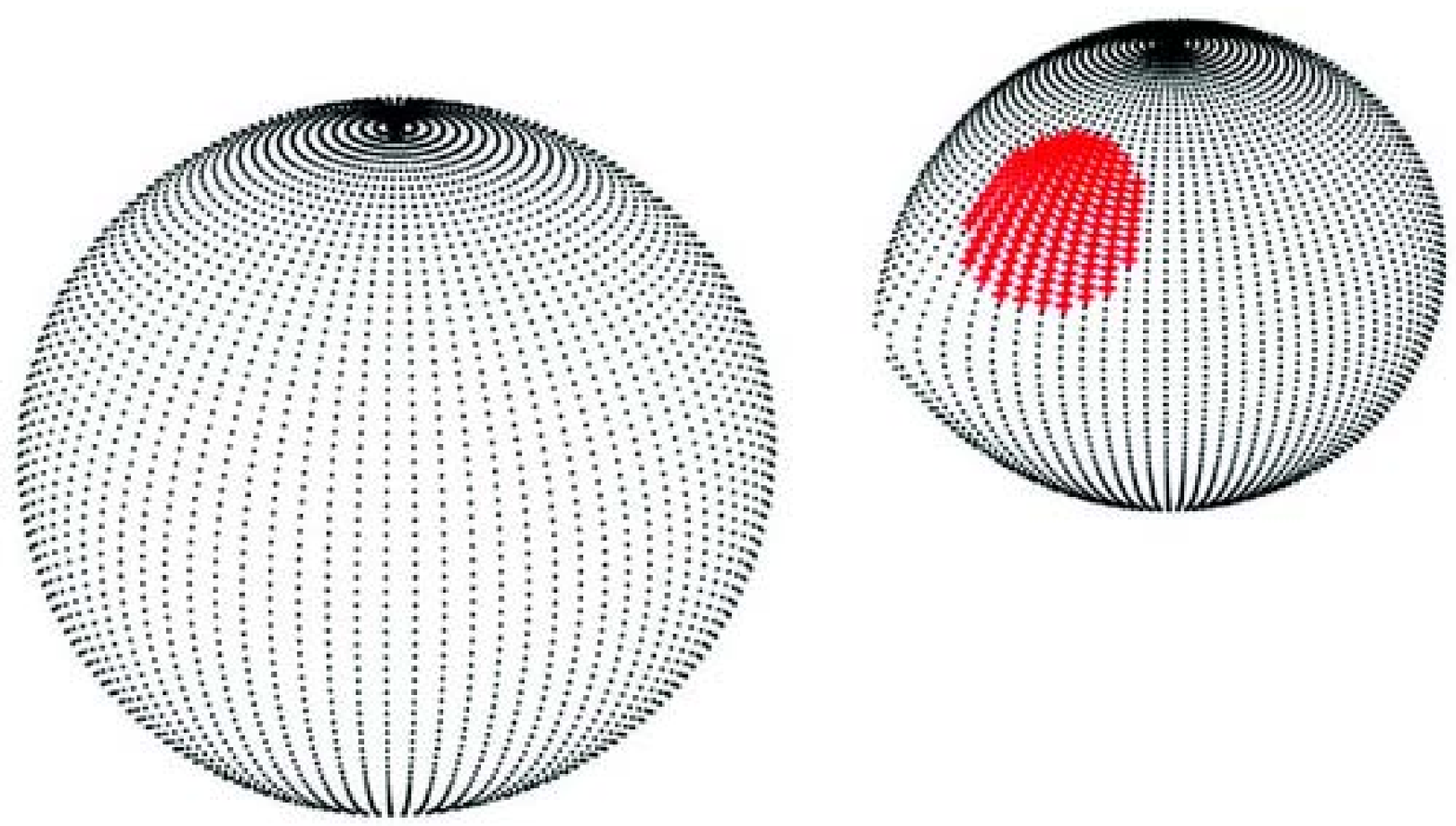}&\hspace{0.5cm}\includegraphics[width=3.7cm]{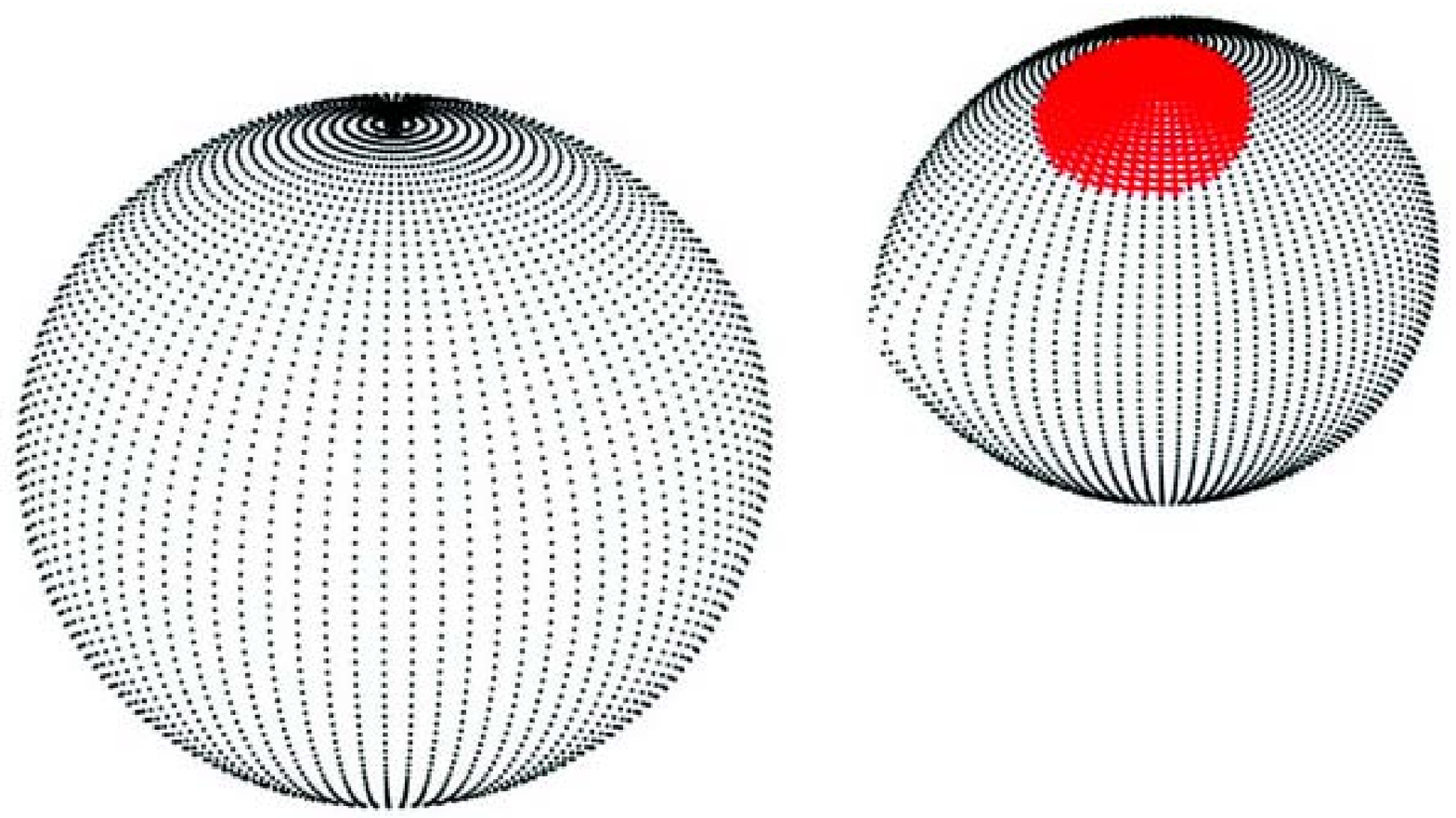}\\
\end{tabular}
\caption{Upper panel: Spot migration diagram for KIC~06669809. Lower panels: The location of the spot (crosses) on the surface of the secondary component during the first day (left) and the last day (right), when the system is at orbital phase 0.34.}
\label{fig:spotKIC066}

\begin{tabular}{cc}
\multicolumn{2}{c}{\includegraphics[width=7.5cm]{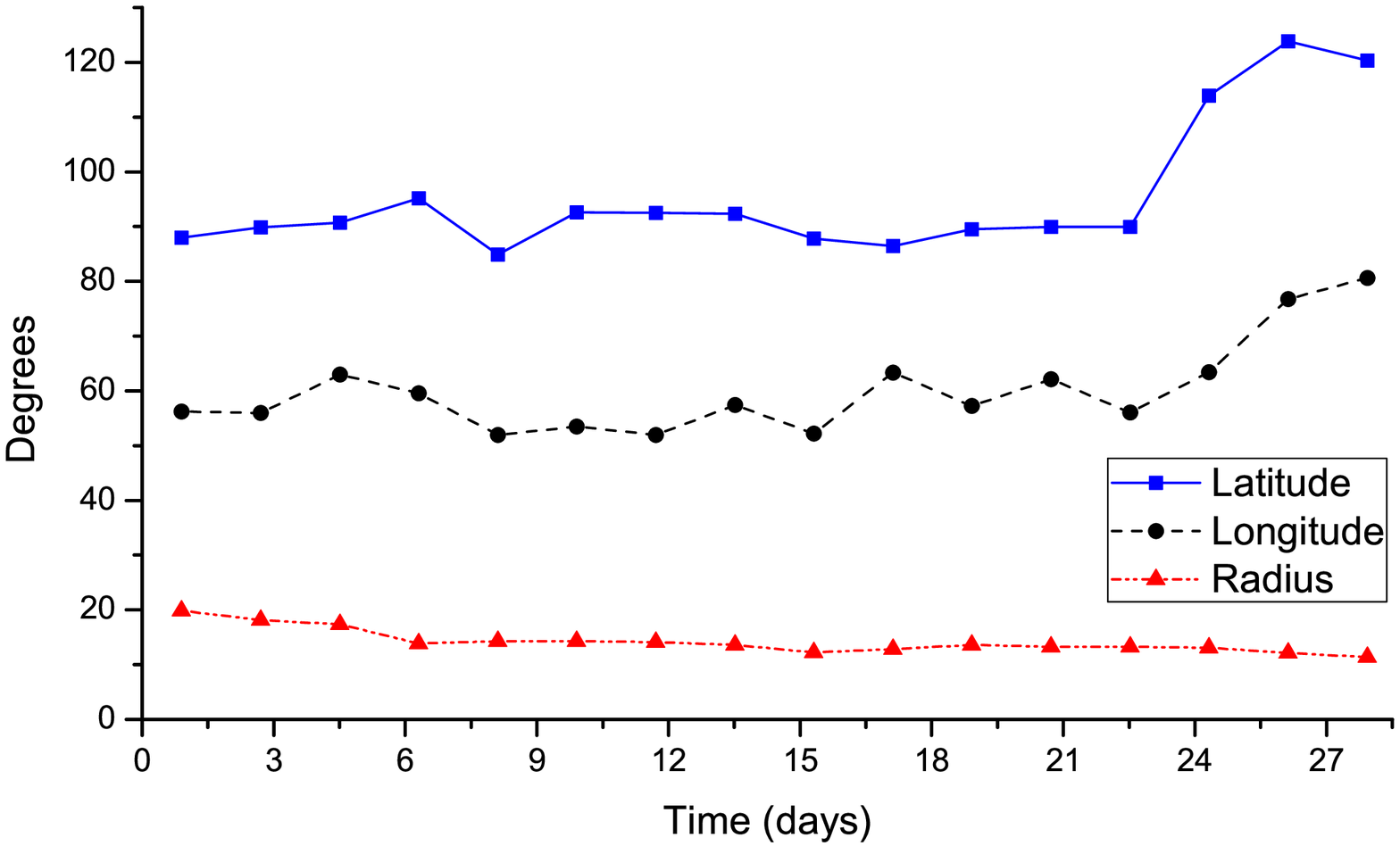}}\\
\includegraphics[width=3.7cm]{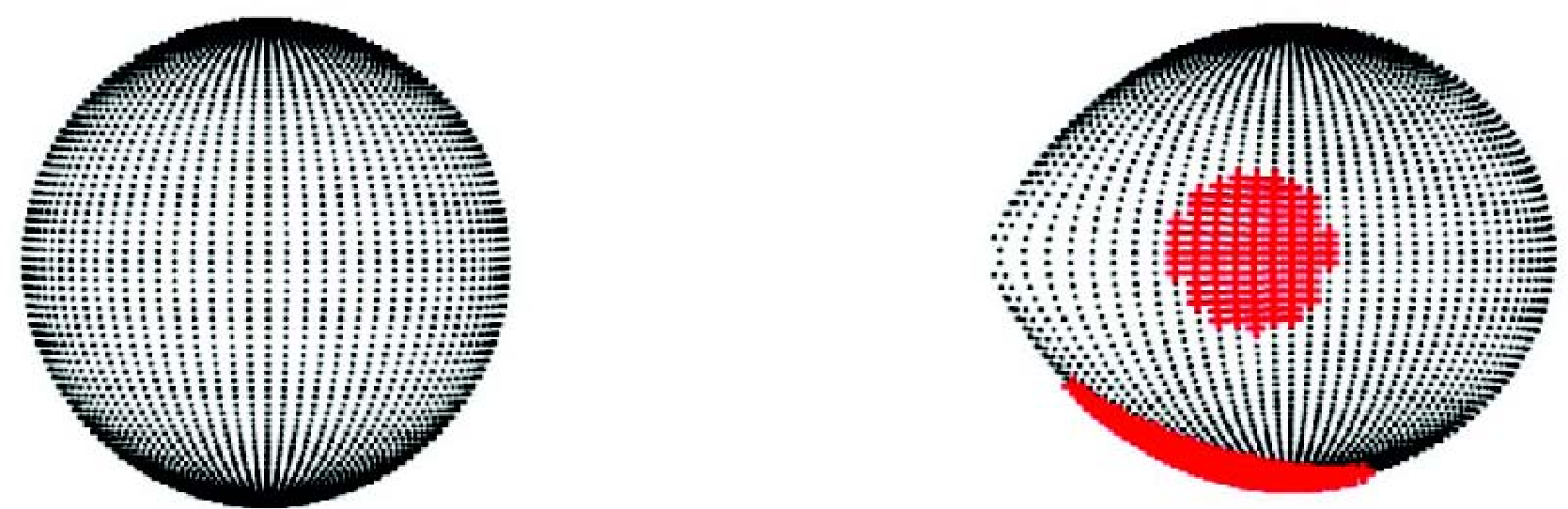}&\hspace{0.5cm}\includegraphics[width=3.7cm]{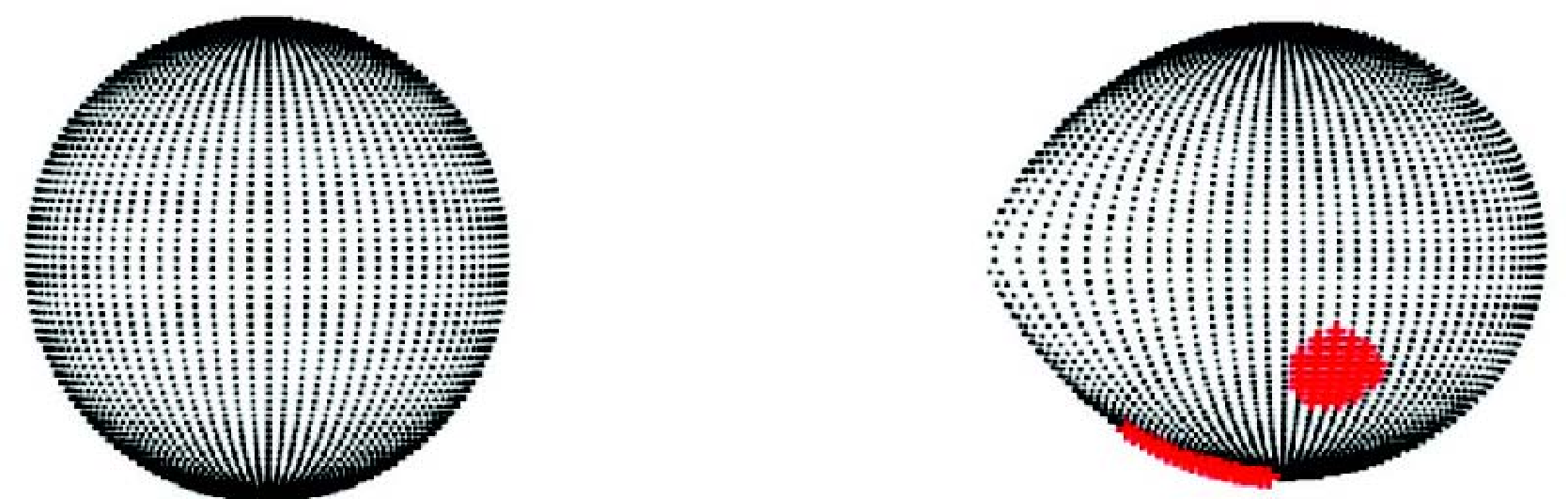}\\
\end{tabular}
\caption{The same as Fig.~\ref{fig:spotKIC066}, but for the first spot of KIC~ 10581918, when the system is at orbital phase 0.32.}
\label{fig:spotKIC105s1}

\begin{tabular}{cc}
\multicolumn{2}{c}{\includegraphics[width=7.5cm]{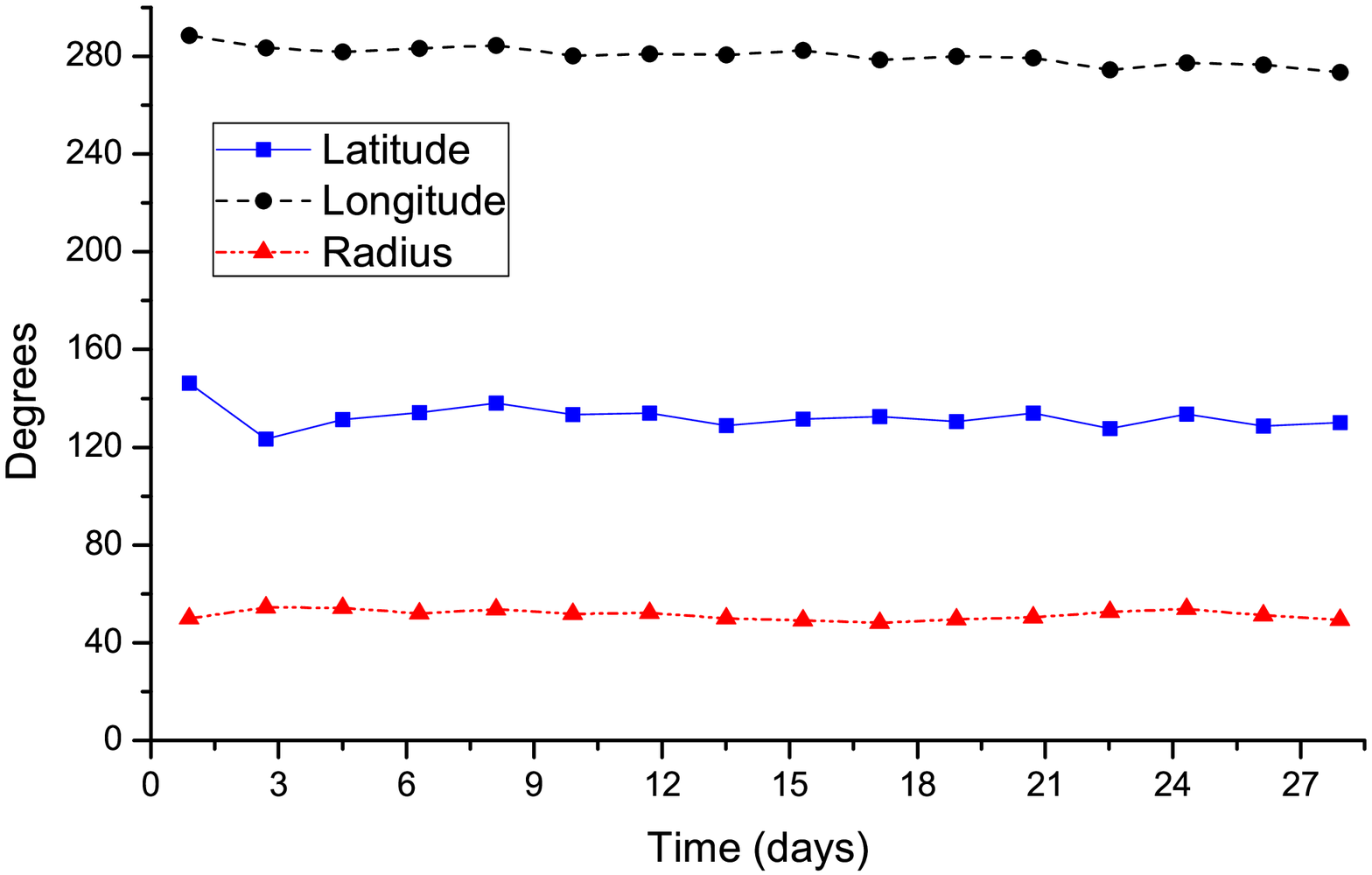}}\\
\includegraphics[width=3.7cm]{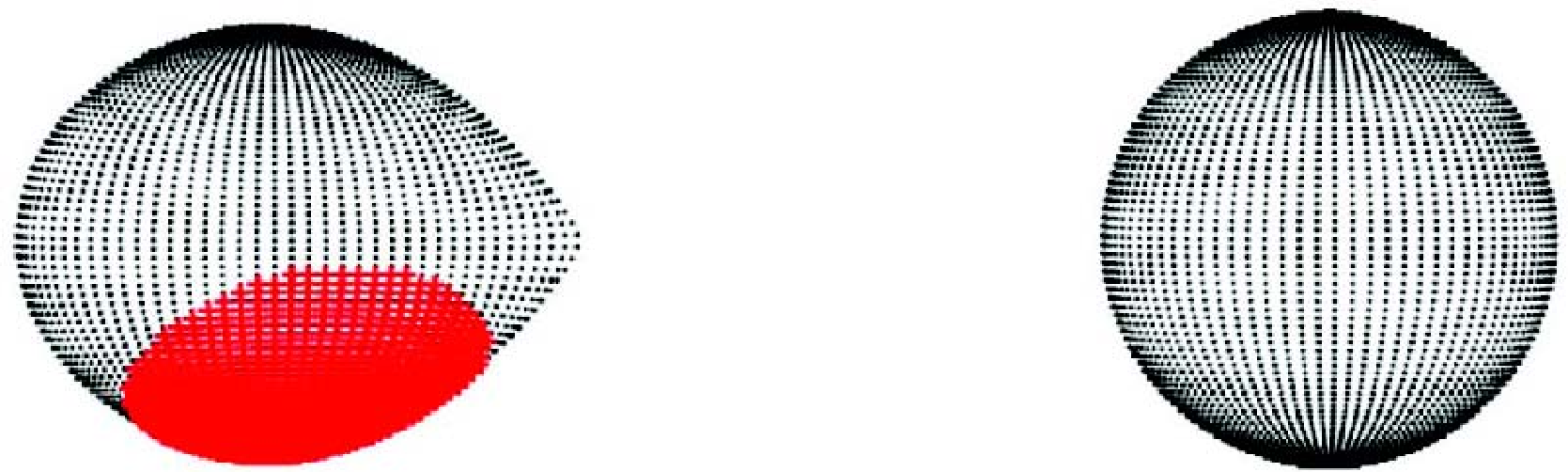}&\hspace{0.5cm}\includegraphics[width=3.7cm]{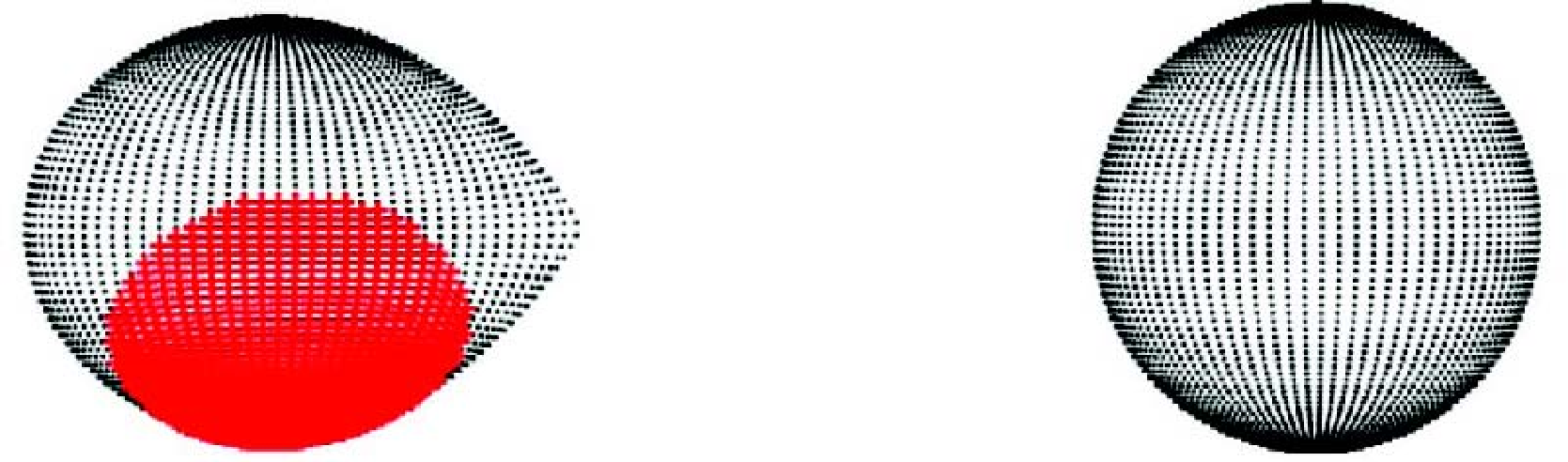}\\
\end{tabular}
\caption{The same as Fig.~\ref{fig:spotKIC066}, but for the second spot of KIC~10581918, when the system is at orbital phase 0.76.}
\label{fig:spotKIC105s2}

\end{figure}

\begin{figure}[h]

\begin{tabular}{cc}
\multicolumn{2}{c}{\includegraphics[width=7.5cm]{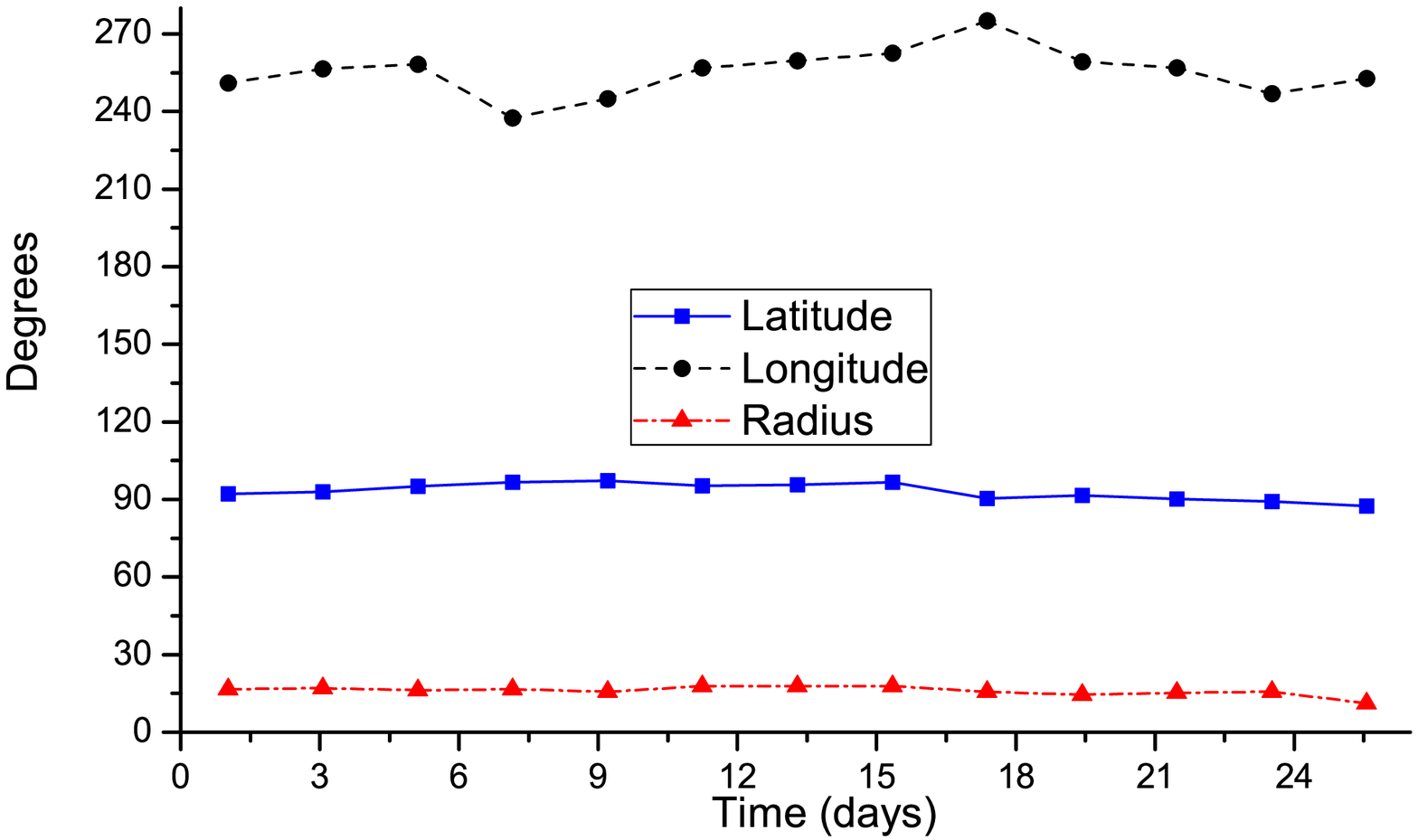}}\\
\includegraphics[width=3.7cm]{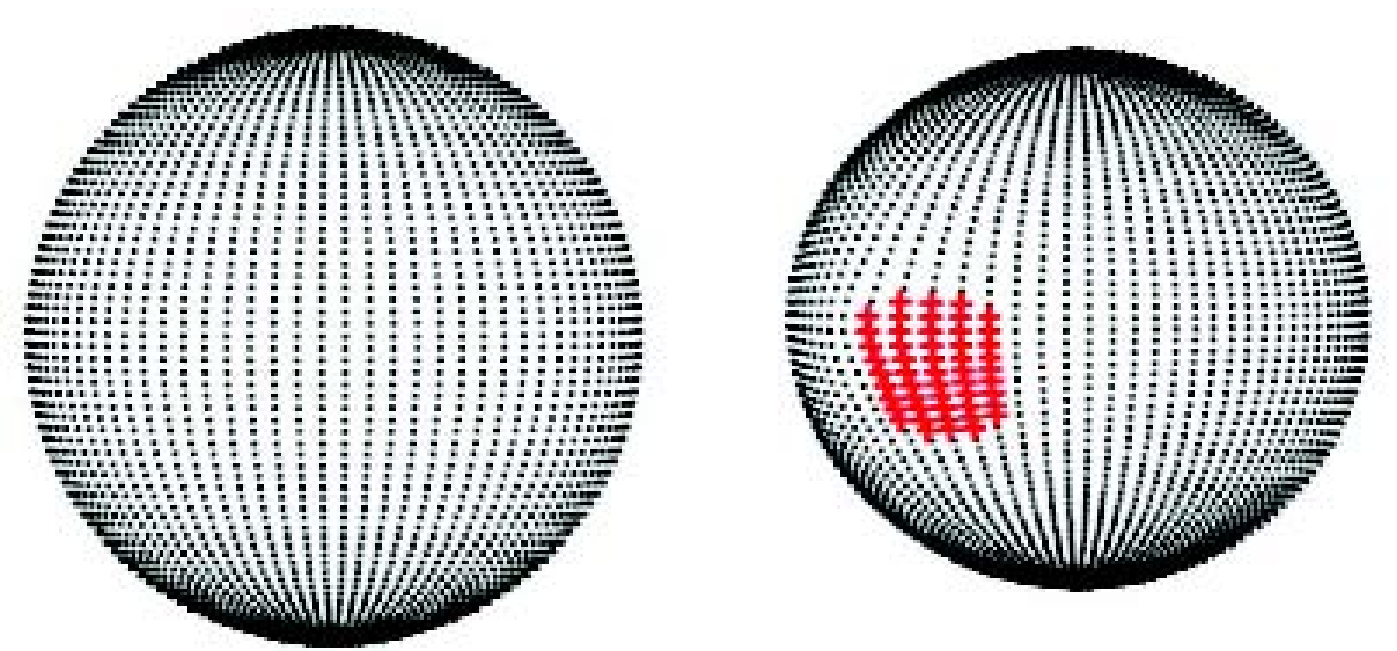}&\hspace{0.5cm}\includegraphics[width=3.7cm]{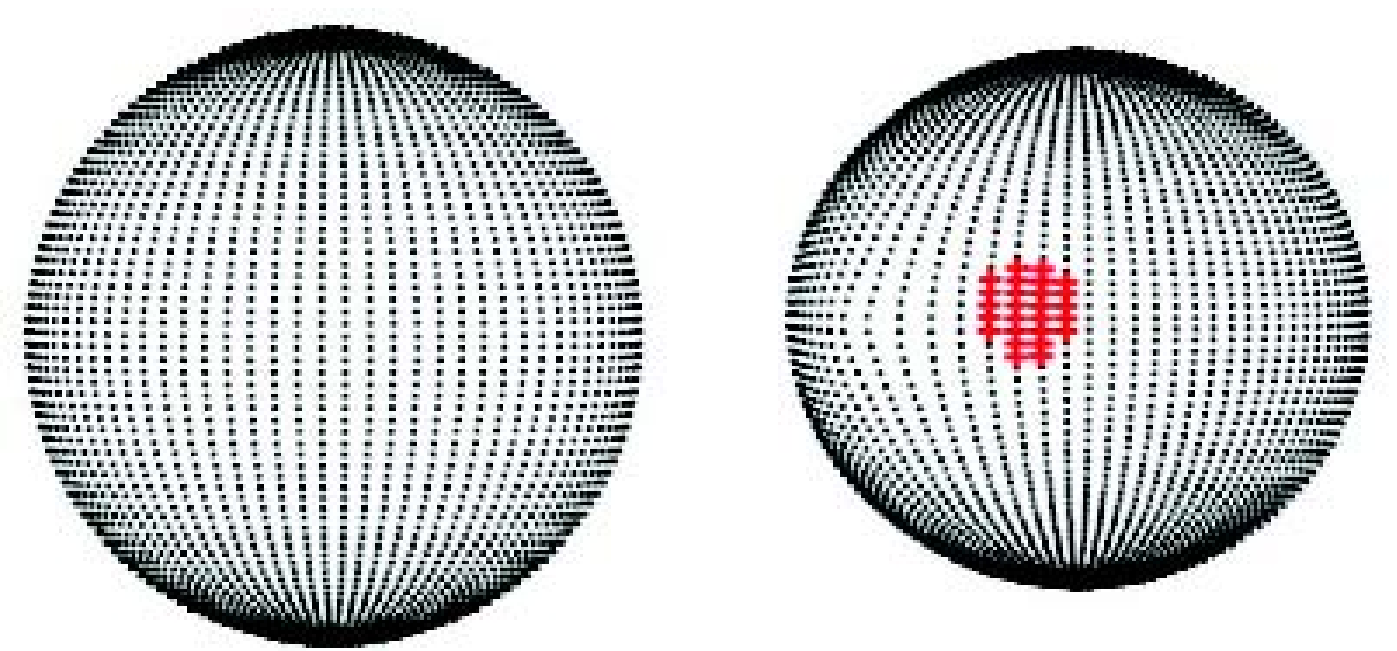}\\
\end{tabular}
\caption{The same as Fig.~\ref{fig:spotKIC066}, but for the first spot of KIC~10619109, when the system is at orbital phase 0.40.}
\label{fig:spotKIC106s1}

\begin{tabular}{cc}
\multicolumn{2}{c}{\includegraphics[width=7.5cm]{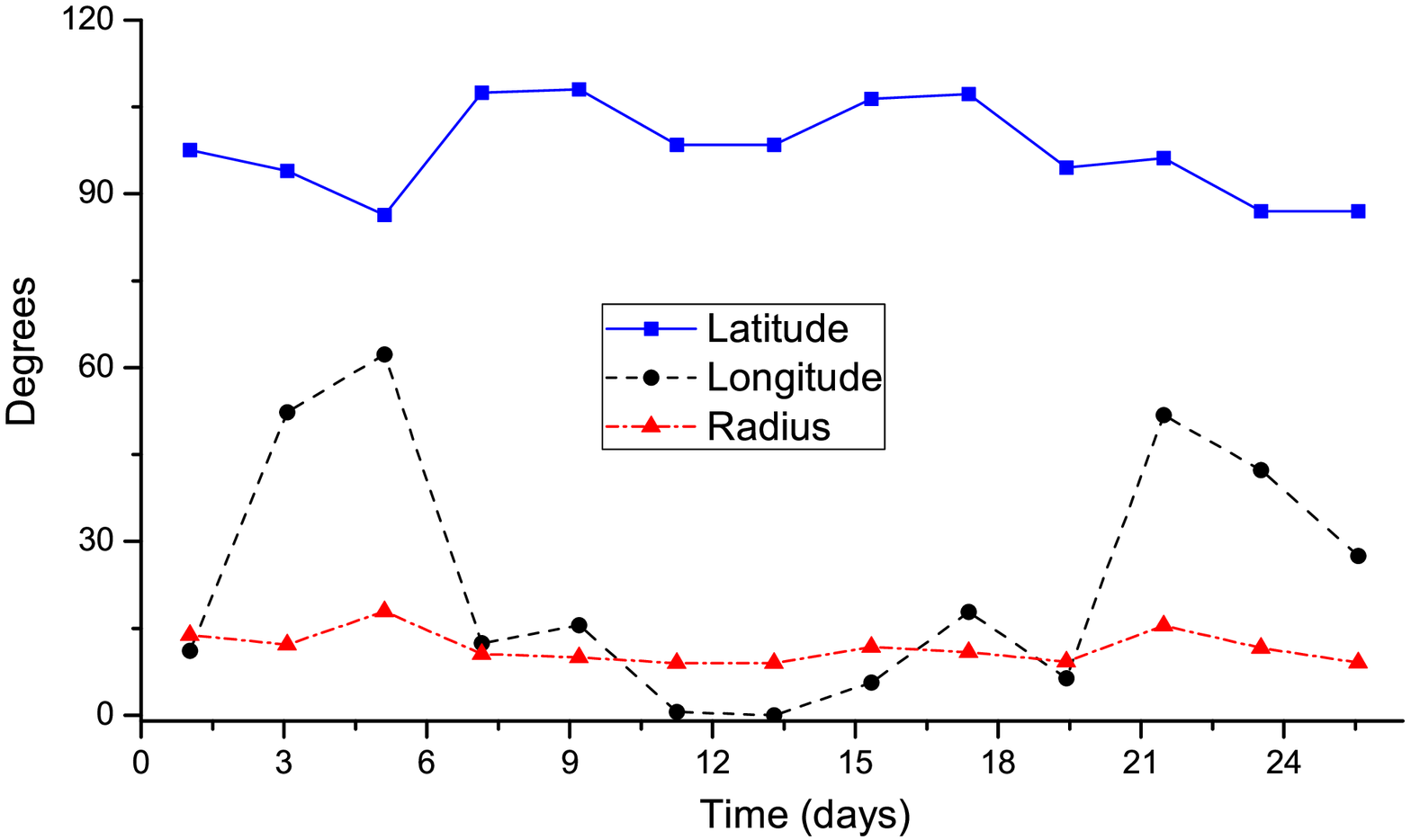}}\\
\includegraphics[width=3.7cm]{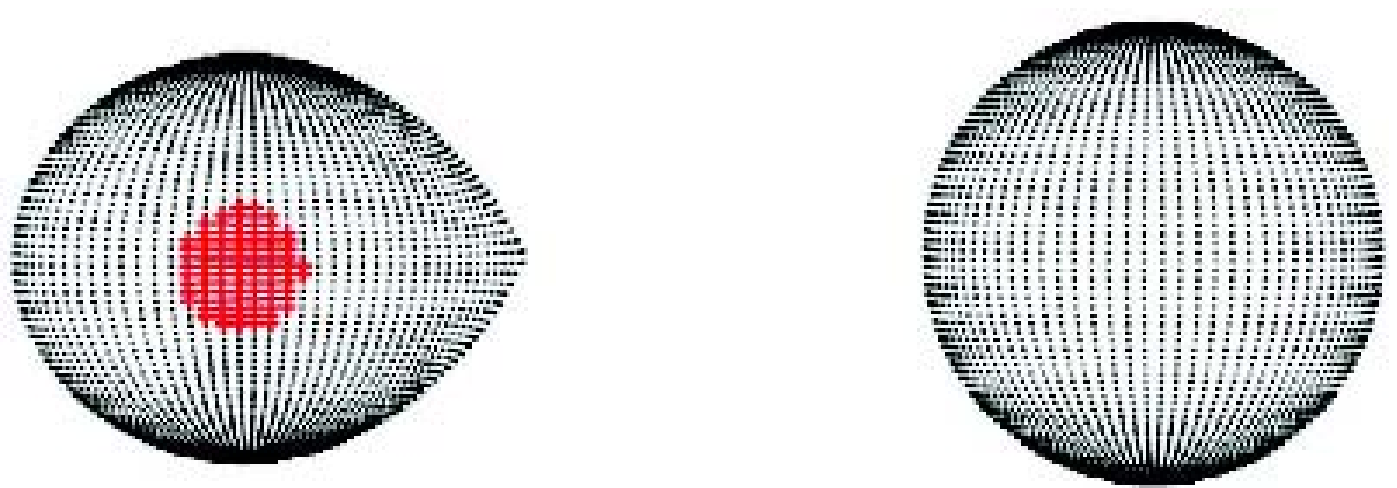}&\hspace{0.5cm}\includegraphics[width=3.7cm]{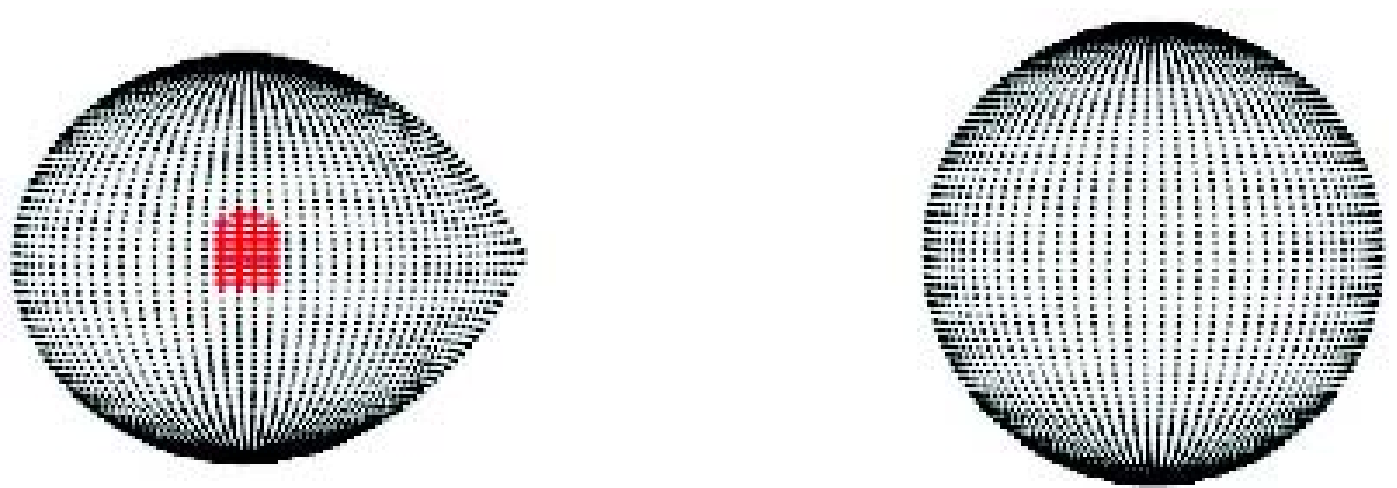}\\
\end{tabular}
\caption{The same as Fig.~\ref{fig:spotKIC066}, but for the second spot of KIC~10619109, when the system is at orbital phase 0.80.}
\label{fig:spotKIC106s2}

\begin{tabular}{cc}
\multicolumn{2}{c}{\includegraphics[width=7.5cm]{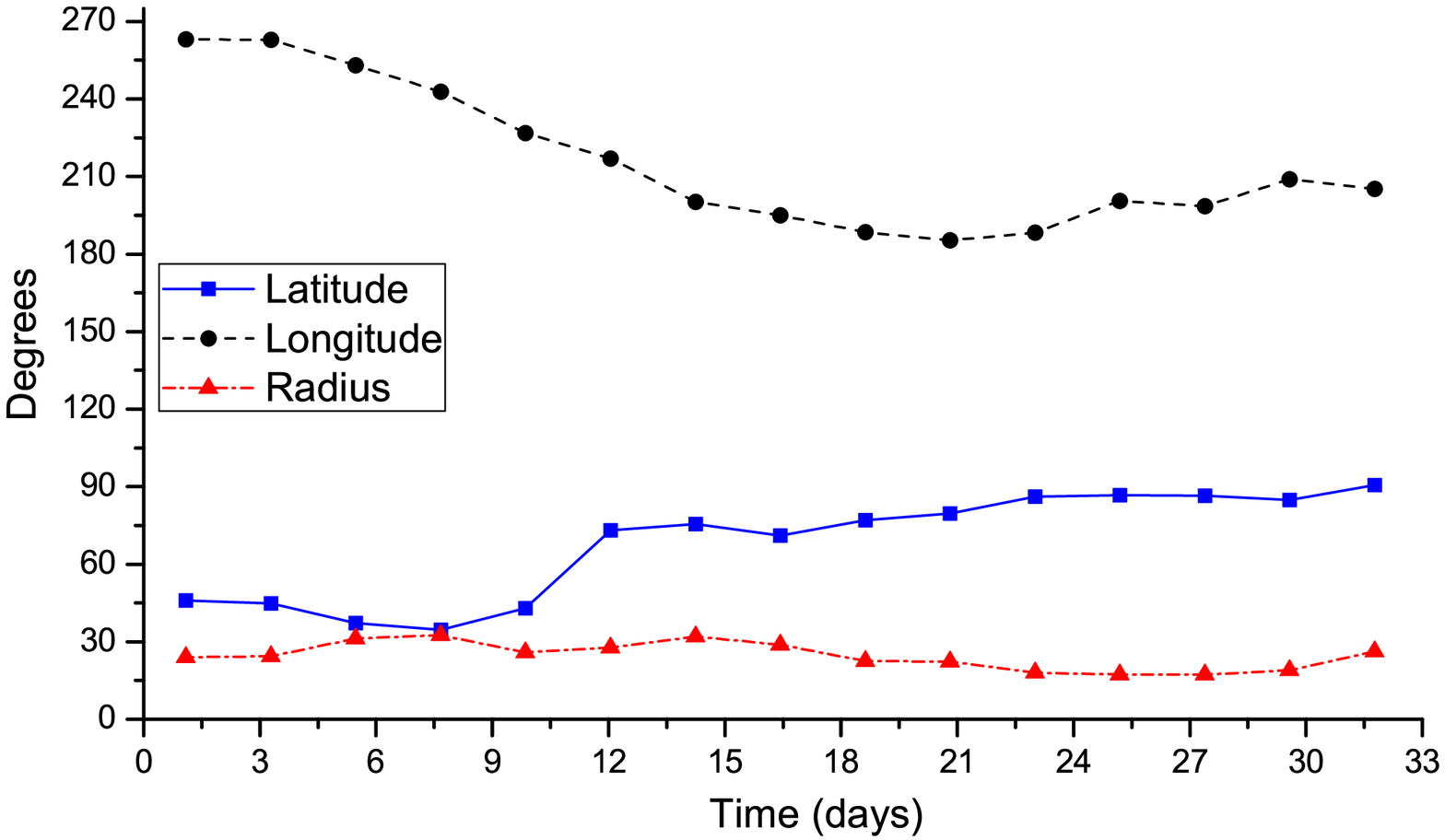}}\\
\includegraphics[width=3.7cm]{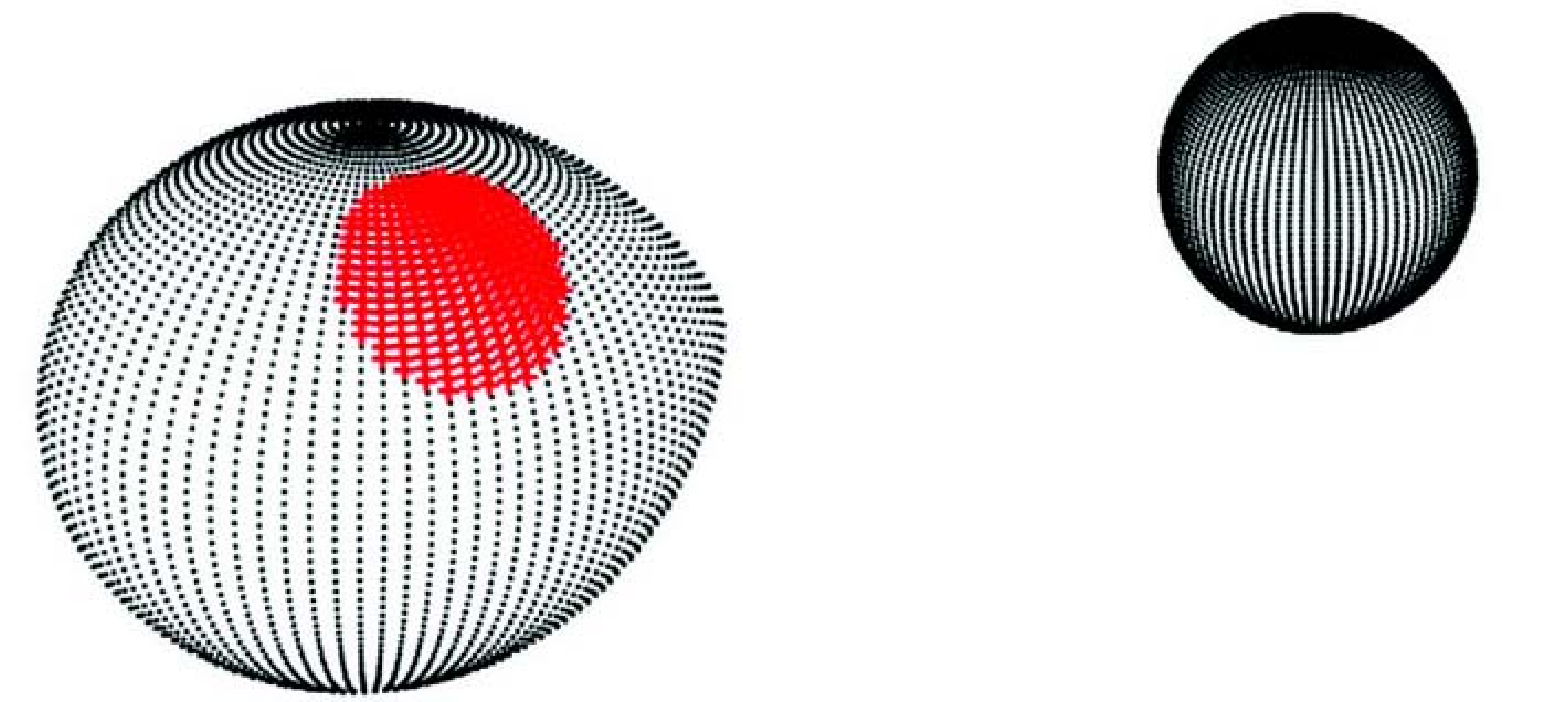}&\hspace{0.5cm}\includegraphics[width=3.7cm]{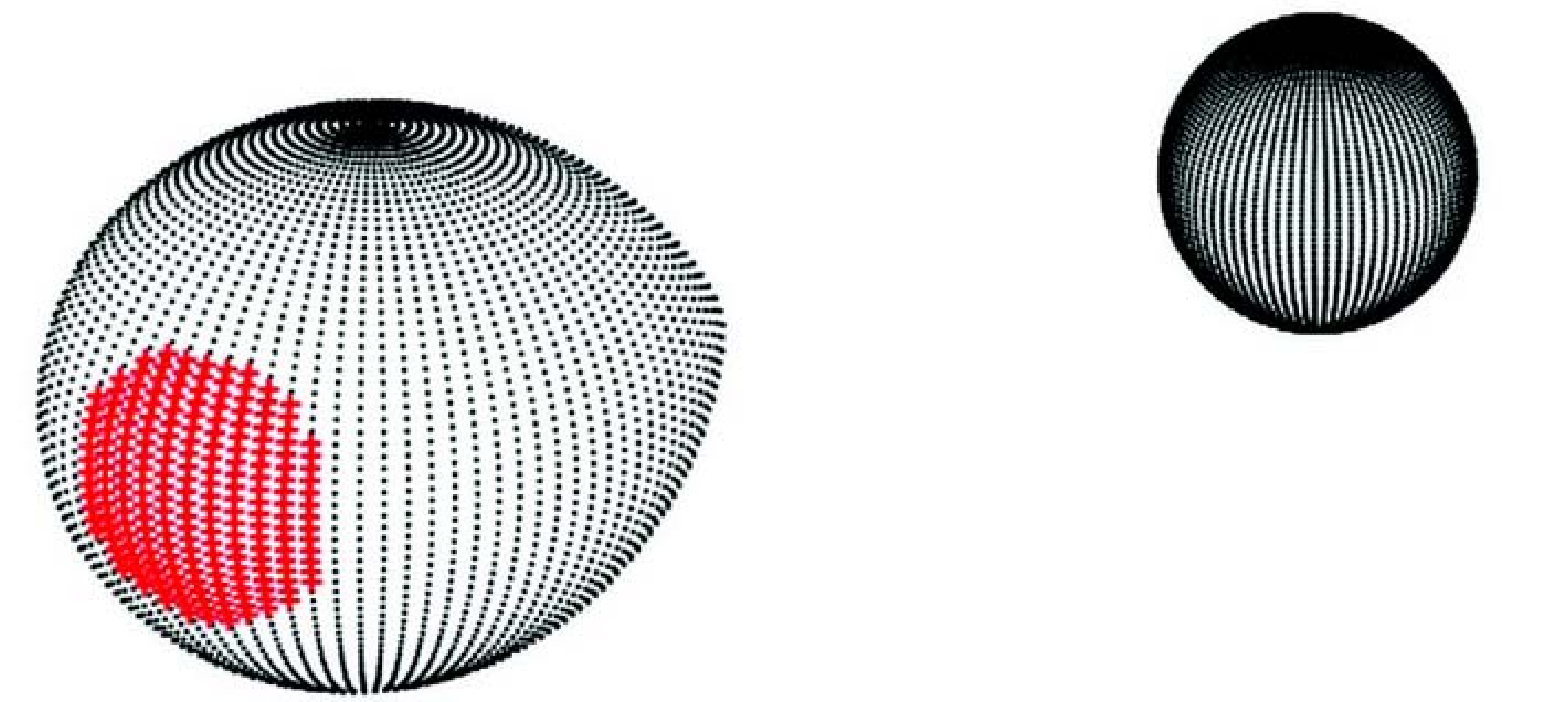}\\
\end{tabular}
\caption{The same as Fig.~\ref{fig:spotKIC066}, but for KIC~11175495, when the system is at orbital phase 0.84.}
\label{fig:spotKIC111}
\end{figure}	
\end{appendix}

\end{document}